\documentclass[fleqn,usenatbib]{mnras}

\usepackage{newtxtext,newtxmath}
\usepackage[T1]{fontenc}

\DeclareRobustCommand{\VAN}[3]{#2}
\let\VANthebibliography\thebibliography
\def\thebibliography{\DeclareRobustCommand{\VAN}[3]{##3}\VANthebibliography}

\usepackage{graphicx}	% Including figure files
\usepackage{amsmath}	% Advanced maths commands

\usepackage[dvipsnames]{xcolor}
\usepackage{siunitx}
\usepackage{listings}

\usepackage{soul} % For highlighting for reviewer

\DeclareSIUnit \h {\ensuremath{\mathit{h}}}
\DeclareSIUnit \parsec {pc}
\DeclareSIUnit \msol {\ensuremath{M_{\odot}}}

\newcommand{\myvec}[1]{\boldsymbol{#1}}

\newcommand{\Tid}[0]{\mathbfss{T}}

\newcommand{\rvir}[0]{r_{\rm{200c}}}
\newcommand{\rhvir}[0]{r_{\rm{200c,h}}}
\newcommand{\mvir}[0]{M_{\rm{200c}}}
\newcommand{\lvir}[0]{\lambda_{\rm{200c}}}
\newcommand{\tvir}[0]{t_{\rm{200c}}}

\newcommand{\rev}[1]{#1}
\newcommand{\revcom}[1]{}

\defcitealias{amorisco_2021}{A21}

\title[Tidal Stripping in the Adiabatic Limit]{Tidal Stripping in the Adiabatic Limit}

\author[J. St\"ucker et al.]{
Jens St\"ucker$^{1}$\thanks{E-mail: jstuecker@dipc.org},
Go Ogiya$^{2,3,4}$,
Raul E. Angulo$^{1,5}$,
Alejandra Aguirre-Santaella$^{6,7}$ and
\newauthor \-  Miguel A. Sánchez-Conde$^{6,7}$
\\
$^{1}$Donostia International Physics Center (DIPC), Paseo Manuel de Lardizabal 4, 20018 Donostia-San Sebastian, Spain.\\
$^{2}$Institute for Astronomy, School of Physics, Zhejiang University, Hangzhou 310027, China \\
$^{3}$Waterloo Centre for Astrophysics, University of Waterloo, Waterloo, ON N2L 3G1, Canada \\
$^{4}$Department of Physics and Astronomy, University of Waterloo, 200 University Avenue West, Waterloo, Ontario N2L 3G1, Canada \\
$^{5}$IKERBASQUE, Basque Foundation for Science, E-48013, Bilbao, Spain.\\
$^{6}$ Instituto de F\'isica Te\'orica UAM-CSIC, Universidad Aut\'onoma de Madrid, C/ Nicol\'as Cabrera, 13-15, 28049 Madrid, Spain\\
$^{7}$ Departamento de F\'isica Te\'orica, M-15, Universidad Aut\'onoma de Madrid, E-28049 Madrid, Spain
}

\date{Accepted XXX. Received YYY; in original form ZZZ}

\pubyear{2015}

\begin{document}
\label{firstpage}
\pagerange{\pageref{firstpage}--\pageref{lastpage}}
\maketitle

% Abstract of the paper
\begin{abstract}
We present a model for the remnants of  haloes that have gone through an adiabatic tidal stripping process. We show that this model exactly reproduces the remnant of an NFW halo that is exposed to a slowly increasing isotropic tidal field and approximately for an  anisotropic tidal field. The model can be used to predict the asymptotic mass loss limit for orbiting subhaloes, solely as a function of the initial structure of the subhalo and the value of the tidal field at pericentre.% -- under the assumption of (1) a smooth tidal field (2) with a negligible time-dependence (3) in the distant tide approximation. 
Predictions can easily be made for differently concentrated host-haloes with and without baryonic components, which differ most notably in their relation between pericentre radius and tidal field. The model correctly predicts several empirically measured relations such as the `tidal track' and the `orbital frequency relation' that was reported by Errani \& Navarro (2021) for the case of an isothermal sphere. Further, we \rev{propose applications of} the `structure-tide' degeneracy which implies that increasing the concentration of a subhalo has exactly the same impact on tidal stripping as reducing the amplitude of the tidal field. Beyond this, we find that simple relations hold for the bound mass, truncation radius, WIMP annihilation luminosity and tidal ratio of tidally stripped NFW haloes in relation to quantities measured at the radius of maximum circular velocity. Finally, we note that NFW haloes cannot be completely disrupted when exposed adiabatically to tidal fields of arbitrary magnitudes. We provide an open-source implementation of our model and suggest that it can be used to improve predictions of dark matter annihilation.
\end{abstract}

% Select between one and six entries from the list of approved keywords.
% Don't make up new ones.

%%%%%%%%%%%%%%%%%%%%%%%%%%%%%%%%%%%%%%%%%%%%%%%%%%

%%%%%%%%%%%%%%%%% BODY OF PAPER %%%%%%%%%%%%%%%%%%

\section{Introduction}

It is one of the central predictions of the $\Lambda$ cold dark matter ($\Lambda$CDM) model that there exists a large number of dark matter haloes where the largest objects may have masses of $M \sim 10^{15} \rm{M}_\odot$ and the smallest objects may be as light as a few earth masses or less, depending on the nature of the dark matter particle \citep{Bringmann_20019, Profumo_2006}. Further, cosmological simulations show that haloes may be populated by a large number of smaller haloes \citep[e.g.][]{tormen_1997, Moore_1999, Klypin_1999, Gao_2004, Springel_2005, springel_2008, frenk_white_2012, angulo_2022}\revcom{added Tormen (1997)} -- so called subhaloes -- which orbit inside their host halo and are heavily affected by its tidal fields.  Modelling how tidal fields strip matter from orbiting subhaloes is challenging, but it is crucial for the correct interpretation of many observational probes.

Accurate models of subhaloes are important to constrain the nature of dark matter, e.g. to distinguish cold from warm dark matter through the counts of satellite galaxies \citep{lovell2014, newton_2021}, through the subhaloes' impact on tidal streams \citep{Yoon_2011, Banik_2018}, and through their effects on gravitational lensing \citep{vegetti_2018, ritondale_2019} and on flux-ratio anomalies \citep{gilman_2020, hsueh_2020}.  

Further, properly accounting for the tidal stripping process may be of significant importance for the interpretation of future galaxy surveys and for inferring the correct cosmology from them. It may be that a significant fraction of the galaxies that will be detected in such surveys lie within subhaloes that are orbiting in large clusters of galaxies. Predictions rely on abundance matching techniques or on the semi-analytic modelling of galaxies which in turn rely on the subhalo populations that are inferred from dark matter simulations \citep{guo_2011, moster_2013, moster_2018, contreras_2021}. Therefore, the accuracy of these models is affected by the degree of artificial subhalo disruption. In order to alleviate this, various models adopt the so-called "orphan" galaxies which represent a population of galaxies not hosted by any dark matter subhalo \citep{guo_2014, delfino_2022}. Although this improves the numerical convergence of the results, additional assumptions e.g. about dynamical friction are required which inevitably impose a degree of uncertainty in the predictions.

Moreover, predictions of possibly measurable dark matter self-annihilation signals in our Milky Way require accounting for the annihilation signal that is caused by the Milky Way's subhalo population. It has been suggested in earlier work that subhaloes may even dominate the overall annihilation luminosity \citep{calcaneo_2000, berezinsky_2003, springel_2008b, gao_2012}, while more recent studies have argued that the annihilation luminosity of the Milky Way's main halo should dominate any detectable annihilation radiation \citep{sahnchezconde_2014, Monline_2017, grand_white_2021}.
Modelling the Milky Way's subhalo population is quite difficult in practice, because of several reasons. The hardest challenge here is the resolution limit of today's simulations. Even state-of-the-art hydrodynamical simulations cannot resolve subhaloes that are less massive than $\sim 10^6 \mathrm{M}_\odot$ \citep[e.g.][]{grand_white_2021}. However, an accurate prediction of the annihilation signal in a WIMP dark matter scenario would require resolving subhaloes with masses as small as earth masses. Therefore, such predictions cannot be obtained from simulations alone, but require the extrapolation of results from the numerical accessible range to scales that are many orders of magnitude smaller \citep[e.g.][]{springel_2008b,Monline_2017, grand_white_2021}.

The next challenge is that baryonic effects have a very strong impact onto the strength of tidal disruption \citep{sawala_2017, garrison_2017, richings_2020}. The difference in the tidal fields in the vicinity of the galactic disk between cases that include and that neglect the baryonic component of the Milky Way can easily be a factor of ten. Such a difference can change the predicted mass loss and annihilation luminosities significantly. Cosmological simulations that include baryonic effects have only recently been able to resolve part of the satellite populations of Milky Way-like galaxies \citep[e.g.][]{sawala_2017, garrison_2017, richings_2020, grand_2021, grand_white_2021}.

Finally, it is even quite difficult to estimate the annihilation luminosity of dark matter subhalos that can be resolved in simulations. Annihilation luminosities depend on the square of the density field and therefore the strongest contributions come from the very centres of haloes. However, these centres are also the most difficult parts to resolve and most sensible to numerical noise. Estimating the annihilation luminosity of simulated subhaloes typically requires fits and other simplifying assumptions \citep[e.g.][]{grand_white_2021}. 

The large space of uncertainty in the cosmological context has lead several authors to attempt understanding the tidal stripping problem in highly simplified and controlled numerical experiments \citep[e.g.][to name a few]{vandenbosch_2018, ogiya_2019, Errani2020}. Such simulations have revealed some uncertainties in the realism of cosmological simulations. \citet{vandenbosch_2018, vandenbosch_2018b} have argued that the often found complete disruption of subhaloes in cosmological simulations must be a numerical artefact. If numerical parameters are carefully controlled, a subhalo should always leave behind a small orbiting remnant. An extreme resilience of dark matter subhaloes to tidal effects has been reported by several other studies \citep{Kazantzidis_2004, Penarrubia_2008, Errani2020, errani_2021, amorisco_2021}. Idealized simulations have further been used to discover interesting phenomenological relations. For example, it has been found that the radius and the velocity at which the circular velocity profile reaches its maximum evolve along a one dimensional relation known as a `tidal track' \citep{Penarrubia_2008, penarrubia_2010, Errani2020}. Further, \citet{errani_2021} have found with idealized simulations of the long-term limit of subhaloes orbiting in an isothermal sphere host potential that such subhaloes will eventually follow a simple orbital frequency relation. Beyond such phenomenological relations, other studies have also tried to predict the fate of subhaloes through machine learning techniques and have, for example, found that the orbital pericentre distance may be the most relevant parameter \citep{Nadler_2018, Petulante_2021}.

\begin{figure*}
    \centering
    \includegraphics[width=\textwidth]{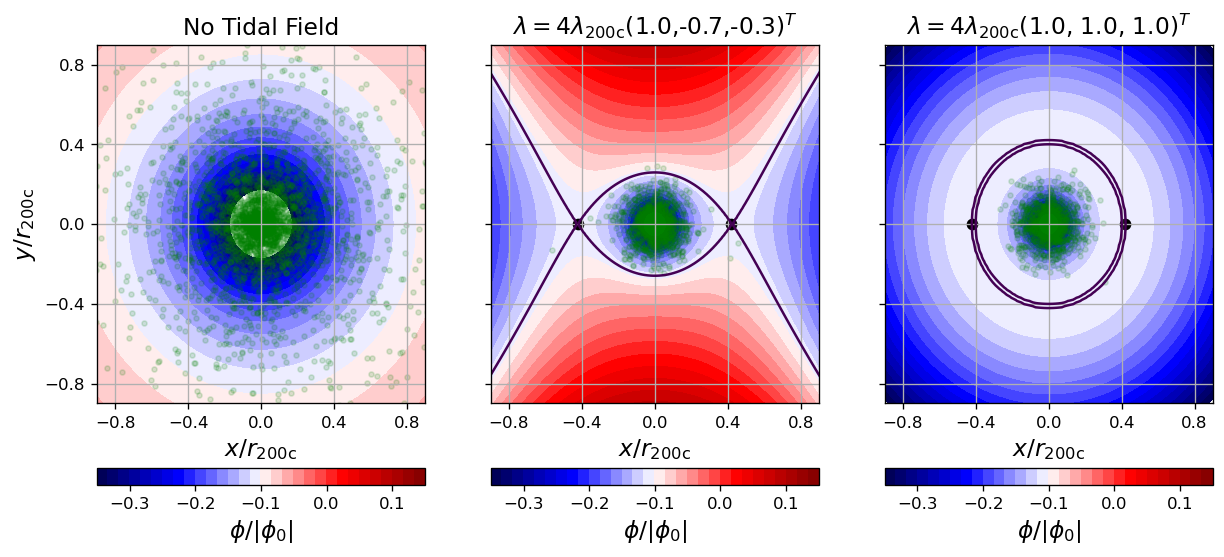}
    \caption{\revcom{Added saddle-point markers}
    Different potential landscapes (contours) and bound particles (green). Left:  for an NFW without any tidal field. Centre: A halo that has been exposed to a trace-free tidal field \rev{with eigenvalues $\myvec{\lambda}$} causing many particles to escape.  Right: A halo that has been exposed to a spherically symmetric tidal field with the same largest eigenvalue. At first sight the potential field seems radically different than for the trace-free tidal field. However, the energy levels of the saddle-point (black lines) are identical and the group of particles that remains bound is quite similar. \rev{The black dots in the right panel indicate the saddle-point locations of the anisotropic case (in the middle panel) to facilitate comparison.} In energy space both cases are almost equivalent. \rev{This reflects also in similar bound mass fractions which are $14.7\%$ for the anisotropic and $12.2\%$ for the isotropic case.} The units and the simulations of this figure are explained in detail in Sections \ref{sec:structuretide} and \ref{sec:atides_sim}.
    }
    \label{fig:tidalexperiment}
\end{figure*}

While there is a lot of literature on numerical experiments, only very few truly analytical models have been proposed. Several semi-analytical models exist which are based on a set of heuristic assumptions of mass loss rates e.g. in relation to the instantaneous tidal radius \citep[e.g.][]{Taylor_2001, penarrubia_2005, vandenbosch_2005, Zentner_2005, Kampakoglou_2007, Pullen_2014, ogiya_2019, Errani2020, jiang_2011}. However, these models usually leave free several parameters which are then calibrated through simulations to match the simulated outputs. Such models are useful as simplified parameterizations of simulation results, but not for obtaining an analytic understanding nor for extrapolating beyond known results. On the other hand, \citet{drakos_2017, drakos_2020} have proposed energy truncation models as a simple approach to gain insights. Such models operate on the idea that the tidal stripping process peels orbiting subhaloes from the outside-in in energy space \citep{choi_2009, stuecker_2021_bp}, first removing particles that have the highest energy levels and then subsequently moving to smaller and smaller energies.  Therefore, the particles that remain in the subhalo may be approximately inferred by a sharp truncation in initial energy space. \citet{amorisco_2021} has extended this idea by additionally following the revirialization process of the truncated remnant through an N-body simulation and has shown that this simple model can already recover the measured tidal tracks. However, there is no first principle way of matching the energy truncation models to the behaviour of individual orbiting subhaloes, since it is unknown which subhaloes should exhibit which degree of energy truncation.

It would be desirable to have a simple analytic model of tidal mass loss that can be derived from first principles. Here, we introduce the \textsc{adiabatic-tides} model to understand the tidal  stripping process in the asymptotic limit. The basic idea behind this model is to first understand how a halo reacts to a tidal field in the adiabatic limit -- which is arguably the simplest possible case of tidal mass loss -- and then see how this can be applied to the more complicated tidal stripping process of orbiting subhaloes. 

The adiabatic limit of tidal stripping can be inferred through the following experiment: We start with an equilibrium Navarro, Frenk and White \citep{nfw1996} halo (later: NFW) in complete isolation as can be seen in the left panel of Figure~\ref{fig:tidalexperiment}. Then we slowly add a tidal field to the potential landscape \rev{(parameterized through the three eigenvalues $\myvec{\lambda}$ of the tidal tensor)} and see how the halo reacts. The tidal field lowers the effective escape energy of the halo to the saddle-point energy level which is significantly lower than the escape energy in absence of a tidal field. This is shown in the central panel of Figure~\ref{fig:tidalexperiment}. This allows particles that have high enough energy to escape. Further, the halo will go through an energy redistribution process which in turn may cause further particles to escape. If the tidal field is increased infinitely slowly, this process approaches the adiabatic limit. In this case the solution is unique and does not have any secondary dependencies.

In this paper we present an analytic model for the adiabatic limit of tidal stripping. We could not find a way to directly calculate the adiabatic limit for the case of anisotropic tidal fields. However, if all three eigenvalues of the tidal tensor are the same then the problem has spherical symmetry and can be solved (right panel of Figure~\ref{fig:tidalexperiment}). While the potential looks quite different between the two cases, it turns out the anisotropic case behaves quantitatively very similar to the isotropic case with the same largest eigenvalue. This is so, since the two cases are almost identical when considered from an energy-space perspective. We will show this quantitatively later in this article, but an intuitive impression can already be gained in Figure~\ref{fig:tidalexperiment}. We make our implementation of the \textsc{adiabatic-tides} model publicly available alongside this article, and we hope that it will be used to improve future predictions.

The article is organized as follows: In Section \ref{sec:adlim} we explain the \textsc{adiabatic-tides} model and show how it can be calculated by using the adiabatic invariance of action variables. In Section  \ref{sec:simulations} we validate the model with numerical simulations of the adiabatic limit and we test in how far it can be used to infer the asymptotic limit of subhaloes. In Section \ref{sec:predictions} we present predictions about the mass loss, luminosities and scaling relations of subhaloes. In Section \ref{sec:literature} we compare the model with other analytic models in the literature and we use the model to test whether the most recent extrapolations of subhalo annihilation luminosities are sound. Finally, in Section \ref{sec:conclusions} we summarize our findings and discuss possible future applications and extensions of the model.

\section{The Adiabatic Limit} \label{sec:adlim}
In this section we present a procedure to predict the behaviour of an NFW system that is exposed to a slowly increasing spherically symmetric tidal field. We will show later in Section \ref{sec:simulations} that such models are also a good approximation for more common highly anisotropic tidal fields and even for orbiting subhaloes.
Under the assumption that the mass loss is proceeding slowly (since the tidal field is applied slowly) we can treat this problem in the adiabatic limit where the actions -- also known as adiabatic invariants -- are conserved \citep[e.g.][]{BinneyTremaine2008}.

Therefore, the phase space distribution as a function of the actions remains conserved and can be used to calculate the final profile. Such an algorithm has first been presented by \citet{young_1980} to adiabatically grow a  central black hole in a stellar cluster and it has later been used by \citet{wilson_2004} and \citet{sellwood_2005} to calculate the reaction of NFW haloes \citep{nfw1996} to a slowly growing baryonic component. Note, that there exists a widely used approximate solution to the adiabatic contraction problem of NFW haloes \citep{blumenthal_1986, gnedin_2004} which is, however, not an exact reconstruction of the adiabatic limit \citep{sellwood_2005}.

Here, we largely follow the procedure as described in \citet{sellwood_2005}, but using the external perturbation from a tidal field instead of a baryonic component.  We argue that this is the simplest possible way to model the mass loss of an NFW halo due to tidal fields in a self-consistent way with full self-gravity. We publish an efficient  \textsc{python} implementation of this algorithm alongside this article in the \textsc{adiabatic-tides} repository\footnote{\label{repository}\url{https://github.com/jstuecker/adiabatic-tides}}.

\subsection{The NFW profile}
The NFW density profile is given by
\begin{align}
    \rho(r) &= \frac{\rho_{\rm{c}}}{\left(1 + \frac{r}{r_{\rm{s}}} \right)^2 \frac{r}{r_{\rm{s}}}}
\end{align}
where $\rho_{\rm{c}}$ and the scale radius $r_{\rm{s}}$ are parameters that may be different for each halo \citep{nfw1996}. The corresponding potential is given by
 \begin{align}
    \phi_{\rm{NFW}}(r) &= \phi_0 \frac{r_{\rm{s}}}{r} \log \left( 1 + \frac{r}{r_{\rm{s}}} \right)\\
    \phi_0 &= - 4 \pi G \rho_{\rm{c}} r_{\rm{s}}^2 . \label{eqn:nfwphi0}
 \end{align}
Under the assumption of an isotropic velocity dispersion, the phase space distribution function of a spherically symmetric NFW profile depends only on energy. It can be evaluated numerically through Eddington inversion \citep{Eddington1916}
\begin{align}
    f_0(E) &= \frac{1}{\sqrt{8} \pi^2} \int_E^0 \frac{d^2 \nu}{d^2 \Phi} (\Phi - E)^{-1/2} d \Phi \label{eqn:nfw_f}
\end{align}
where $\nu(\phi)$ is the density as a function of the potential \citep[see also][]{Errani2020} \rev{and where we have already omitted a term that vanishes if both $\Phi \rightarrow 0$ and $d\nu / d\Phi \rightarrow 0$ for $r \rightarrow \infty$ (which is the case for the NFW potential).}

It is sometimes convenient to express the NFW profile through a mass parameter and a concentration parameter $c$. For this we use $\mvir$ as the mass inside the radius $\rvir$ inside which the mean density of the halo is 200 times the critical density of the universe and the concentration
\begin{align}
    c = \frac{\rvir}{r_{\rm{s}}} \text{.}
\end{align}
We will sometimes loosely refer to $\rvir$ as the virial radius. Note that many numerical studies assume a sharp truncation of their NFW haloes beyond $\rvir$. Such an initial truncation radius should in principle be treated as a third parameter. However, in our model we will assume an initially infinitely extended NFW profile that is only truncated through the tidal field.
\subsection{A uniform tidal field}
It is useful to expand the gravitational potential landscape around the location of a halo or subhalo through a Taylor expansion up to second order
\begin{align}
    \phi(\myvec{x}) &= \phi_{\rm{s}} (\myvec{x}) - \frac{1}{2} \myvec{x} \Tid \myvec{x} \label{eqn:anatidpot}
\end{align}
where $\phi_{\rm{s}}$ is the self-potential of the halo that we consider and $\Tid$ is a symmetric 3x3 matrix and where we have neglected zeroth and first order terms, since these are irrelevant to the internal dynamics of a system \citep[see e.g.][]{renaud_2011, stuecker_2021_bp}. This expansion up to second order is also often called the ``distant tide'' approximation which is fairly accurate as long as third and higher order terms are small across the extent of an object. Whether this is the case depends on the details of the potential landscape of the host halo, the radius of the subhalo and the orbit. Approximately, it holds for most configurations with a mass ratio $M / M_{\rm{h}} \lesssim \SI{e-3}{}$. Additionally, for an orbiting subhalo the tidal field will not be constant, but have a strong time-dependence. However, in the \textsc{adiabatic-tides} model we will consider the idealized case of a uniform tidal field with negligible time-dependence and we will later check how this applies to these complicated more general situations.

The alignment of the tidal field is generally not of interest so that out of the six components of the tidal tensor only the three eigenvalues $\lambda_1 \geq \lambda_2 \geq \lambda_3$ matter. However, we argue that for estimating the mass bound to the adiabatic remnant, primarily just the largest eigenvalue $\lambda_1$ matters, whereas the two smaller eigenvalues might only introduce minor corrections. The idea behind this is illustrated in Figure~\ref{fig:tidalexperiment} where we show the three different potential fields of (1) an unperturbed NFW halo (2) the remnant of an NFW halo after applying a highly anisotropic tidal field with eigenvalues $(\lambda_1, -0.7\lambda_1, -0.3\lambda_1)$ and (3) the remnant of an NFW halo after applying an isotropic tidal field with the same largest eigenvalue $(\lambda_1, \lambda_1, \lambda_1)$. 

At first sight, the two cases with tidal field seem radically different: the anisotropic tidal field seems more realistic, since it has a truly external, trace-free tidal field, whereas the isotropic tidal field is equivalent to adding a uniform negative density everywhere. The potential landscape of the anisotropic tidal field has a saddle-point in the $x$-direction (the direction of the largest eigenvalue) whereas it steeply increases in the $y$-direction. In the isotropic case the potential landscape has an extremum in every direction at the same level. However, if we look at the particles which remain bound to the system (in green), they seem almost equivalent in the two cases. Further, the depth of the potential valley -- that is, the difference between the energy value at the saddle-point and at the minimum -- is identical. As a matter of fact, when considered from energy space, these two cases seem almost identical: In both cases the tidal field introduces an energy-level beyond which particles can easily escape the potential well and this energy level is the same in both cases, since it depends only on the largest eigenvalue of the tidal tensor. See also \citet{stuecker_2021_bp} for a more in depth discussion of the energy-space perspective.

In this section we will simply assume a spherically symmetric tidal field, but we will show in Section \ref{sec:atides_sim} through numerical experiments that these calculations in spherical symmetry are also good approximations for highly anisotropic tidal fields. A spherical tidal field can be described through a single eigenvalue $\lambda$:
\begin{align}
    \phi(r) &= \phi_{\rm{s}} (r) - \frac{1}{2} \lambda r^2 \label{eqn:spheretidpot}
\end{align}
We will often measure tidal fields in units of the tidal field that is necessary to create a saddle point in an NFW potential at the virial radius $\rvir$ at redshift $z = 0$, where $\rvir$ is the radius at which the mean enclosed density is 200 times the critical density of the universe:
\begin{align}
    \lvir &= \frac{\partial_r \phi_{{\rm{NFW}}}(\rvir)}{\rvir} \\
                   &= 100 H_0^2
\end{align}
where $H_0$ is the hubble parameter at redshift $z=0$. Note that due to the definition of the virial radius, this `virial tidal field' is a constant that is independent of halo parameters.

\subsection{Bound and unbound orbits} \label{sec:boundorbits}
\begin{figure}
    \centering
    \includegraphics[width=1\columnwidth]{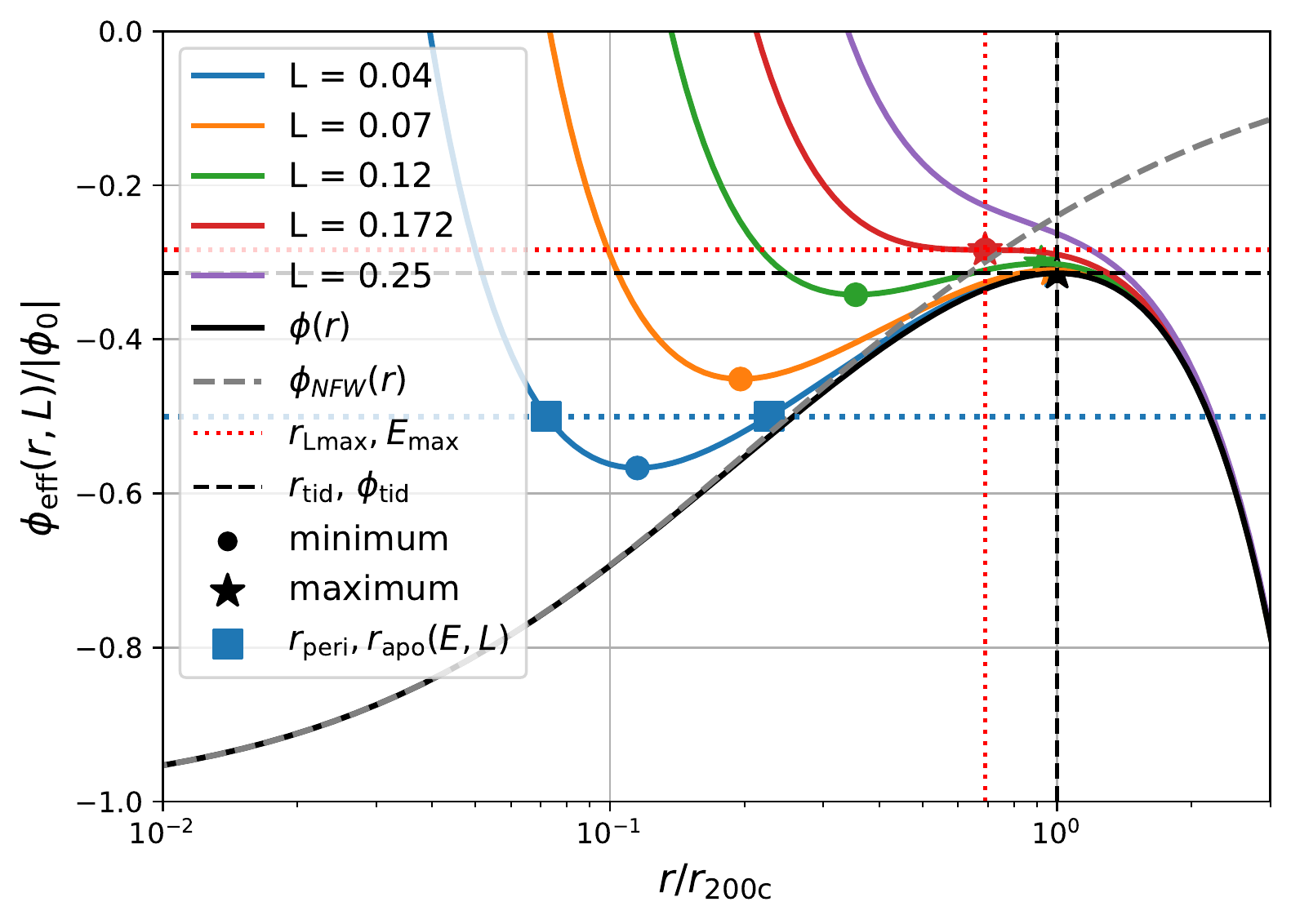}
    \caption{The effective potential of an NFW potential with tidal field. The grey dashed line shows the usual NFW potential, the solid black line the NFW potential with tidal field and the colored solid lines the effective potential for different angular momenta. For bound orbits the intersection $E = \phi_{\rm{eff}}(r, L)$ defines the peri-/apocentre (blue squares show one example). There is a maximum energy $E_{\rm{max}}$ above which no bound orbits are possible.}
    \label{fig:tidal_boundaries}
\end{figure}
\rev{We need to distinguish the populations of ``bound'' particles that are trapped in the potential well of our halo versus ``unbound'' particles that can escape due to the tidal field. In general setups with time-dependent tidal fields this is a very complicated problem that may depend on the definitions used} \citep[e.g.][]{penarrubia_2022}\rev{. However, in our simplified setup with a static, spherically symmetric tidal field, it is possible to distinguish clearly between bound and unbound populations.}

A tidal field introduces the tidal radius $r_{\rm{tid}}$ at which the potential is maximal:
\begin{align}
    \partial_r \phi(r_{\rm{tid}}) = 0 .
\end{align}
Beyond the tidal radius the acceleration points away from the halo centre so that particles will eventually escape arbitrarily far away from the halo. Further, we define the tidal energy level $\phi_{\rm{tid}} = \phi(r_{\rm{tid}})$. A reasonable approximation to which particles can be bound to the potential is given by which particles have an energy-level $E < \phi_{\rm{tid}}$ \citep{stuecker_2021_bp}. However, this is not exact. For a more precise evaluation of which orbits can be bound, we have to consider the effective potential
\begin{align}
    \phi_{\rm{eff}}(r,L) = \phi(r) + \frac{L^2}{2 r^2}
\end{align}
where $L = \lVert\myvec{x} \times \myvec{v}\rVert$ is the angular momentum of a particle within the subhalo. With the effective potential the motion of a particle can be understood effectively as a one dimensional problem, where the radius changes at the rate $\dot{r} = - \partial_r \phi_{\rm{eff}}(r,L)$. A particle is on a bound orbit if it is trapped inside a valley of the effective potential. Given its energy $E$ and angular momentum $L$, its pericentre radius $r_{\rm{p}}$ and apocentre radius $r_{\rm{a}}$ are defined implicitly through the equation 
\begin{align}
    E = \phi_{\rm{eff}} \left(r_{\rm{p/a}},L \right) . \label{eqn:rperiapo}
\end{align}
We illustrate the effective potential for the case of an NFW with a tidal field $\lambda = \lvir$ in Figure~\ref{fig:tidal_boundaries} for different angular momenta. We have also indicated, for one example pair ($E$, $L$), how the peri- and apocentres of a particle are determined (blue dotted line and squares). Note that in a monotonously increasing potential equation \eqref{eqn:rperiapo} has generally two roots, whereas in our case with tidal field the potential goes to $\phi \rightarrow - \infty$ at $r \rightarrow \infty$ it can have one or three roots. If it has three roots, then the first two roots correspond two the peri and apocentre of a bound orbit and the third root lies outside the tidal radius. This indicates that it is also possible to be on an unbound orbit outside the tidal radius with the same energy and angular momentum level. If equation \eqref{eqn:rperiapo} has only one root, then it is not possible to have a bound orbit with $(E,L)$.

It turns out that it is possible to have bound orbits with slightly higher energy than $\phi_{\rm{tid}}$, since some part of the energy can be stored in angular momentum, as indicated by the effective potential. We discuss this in more detail in Appendix \ref{app:boundaries}, where we also show that this maximal possible energy lies at the radius of maximum circular angular momentum. We label this radius $r_{\rm{Lmax}}$, the corresponding circular angular momentum $L_{\rm{max}}$ and the corresponding highest possible energy $E_{\rm{max}}$. This means that there will not be an exact sharp cut in the energy distribution at $\phi_{\rm{tid}}$, but there can exist a small population of particles with $\phi_{\rm{tid}} < E < E_{\rm{max}}$ with bound orbits in some angular momentum range $L_{\rm{min}}(E) < L < L_{\rm{max}}(E)$ where these boundaries are defined and explained in Appendix \ref{app:boundaries}. This explains why some particles with $E > \phi_{\rm{tid}}$ survive tidal stripping as was observed in \citet{stuecker_2021_bp}.

\subsection{Adiabatic invariants}
If a potential is perturbed adiabatically -- that is through a slowly growing perturbation -- the action variables of orbiting particles are conserved. Therefore, the actions are often called \emph{adiabatic invariants}. In a spherically symmetric potential two of the actions can be taken to be the absolute value of the angular momentum $L$ and one of its components $L_z$. 

The third action variable is given by the radial action $J_r$ which can be evaluated numerically through the integral
\begin{align}
    J_r(E,L) &= \frac{1}{\pi} \int_{r_{\rm{p}}}^{r_{\rm{a}}} \sqrt{2(E - \phi_{\rm{eff}}(E, L))} \rm{d}r  . \label{eqn:action}
\end{align}
The conservation of the actions implies also that the phase space distribution function $f$ remains invariant under adiabatic state changes when written as a function of the actions \citep{BinneyTremaine2008}:
\begin{align}
    f(J_r, L)(t) = f(J_r, L)(t=0)
\end{align}
Here we have dropped the $L_z$ adiabatic invariant, since the distribution function is independent of it for spherically symmetric systems. 

Note that applying a tidal field is not a truly reversible adiabatic state change, since some particles that were bound in the initial system, end up on unbound orbits in the final system without well defined actions. If the tidal field would be turned off again (infinitely slowly) those particles would not return to their former orbits and the initial state of the system could not be recovered, hence the process is irreversible. Therefore, we only assume that the actions are conserved for orbits which remain bound and for unbound orbits we assume that they have zero contribution to the final system. 

\subsection{Young's method} \label{sec:youngsmethod}
We follow \citet{sellwood_2005}'s description of Young's method \citep{young_1980} for the iterative construction of the adiabatically modified system. 

Given a phase space distribution function $f$ of an isotropic system one can evaluate the density profile by integration of $f$ over the three dimensional velocity space. Due to the spherical symmetry this can be simplified to a two dimensional integral
\begin{align}
    \rho(r) &= \int \int \int f(\myvec{x},\myvec{v}) \rm{d}^3 v\\ 
            &= 4 \pi \int_{\phi(r)}^{E_{\rm{max}}(r)} \int_{L_{\rm{min}}(r,E)}^{L_{\rm{max}}(r,E)} \frac{L f(E, L)}{r^2 u(r,E,L)} \rm{d}L \rm{d}E  \label{eqn:f_to_dens}
\end{align}
where $u$ is the radial velocity
\begin{align}
    u(r,E,L) &= \sqrt{2 (E - \phi_{\rm{eff}(r, L)})}
\end{align}
and the integration boundaries are chosen so that the integrals go over all possible bound orbits that can contribute at radius $r$. For an infinitely extended NFW halo these would be $E_{\rm{max}} = 0$, $L_{\rm{min}} = 0$ and $L_{\rm{max}} = r \sqrt{2 (E - \phi(r))}$. However, for a non-monotonic profile with tidal field the integration boundaries take a more complicated shape, where $E_{\rm{max}}$ and $L_{\rm{min}}$ can be non-zero. This is discussed in detail in Appendix \ref{app:boundaries}.

Now, evaluating equation \eqref{eqn:f_to_dens} requires knowledge of the phase space distribution as a function of energy and angular momentum. If we assume that our profile is the adiabatic image of a profile with initial phase space distribution $f_0(E, L)$, then we can express the final distribution function through the initial one:
\begin{align}
    f(E,L) = f_0(E_0(J_r(E,L), L), L) \label{eqn:fadiabatic}
\end{align}
Here $E_0(J_r,L)$ 
is a function that estimates the energy of an orbit with the actions $(J_r,L)$ in the initial profile, and $J_r(E,L)$ is the action as a function of energy and angular momentum in the final profile. The function $E_0(J_r,L)$ only requires knowledge of the initial profile and can be expressed e.g. through mesh-free interpolators. However, $J_r(E,L)$ depends implicitly on the density profile $\rho(r)$, since evaluation of the action requires knowledge of the potential \eqref{eqn:action}, which is related to the density through Poisson's equation
\begin{align}
    \nabla^2 \phi_{\rm{s}} &= \frac{1}{r^2} \partial_r \left(r^2 \frac{\partial \phi_{\rm{s}}}{\partial r} \right) = 4 \pi G \rho \label{eqn:poisson}
\end{align}
Therefore equation \eqref{eqn:f_to_dens} is an implicit equation for the density profile when combined with equations \eqref{eqn:fadiabatic}, \eqref{eqn:action} and \eqref{eqn:poisson}. It is not possible to solve this system of equations fully analytically, but it can be solved numerically through an iterative procedure \citep{sellwood_2005, BinneyTremaine2008}. For this we start with a guess of the density profile $\rho_0(r) = \rho_{\rm{NFW}}(r)$. Then for each step of the iteration we construct $\phi_n(r)$ through integration of Poisson's equation \eqref{eqn:poisson}, $J_n(E,L)$ through equation \eqref{eqn:action}, the distribution function $f_n(E,L)$ through equation \eqref{eqn:fadiabatic} and finally a revised estimate of the density profile $\rho_{n+1}(r)$ through \eqref{eqn:f_to_dens}.

This procedure requires solving several integrals and setting up interpolation tables in each iteration for $\rho_n(r)$, $\phi_n(r)$ and $J_n(E,L)$ (and once for $f_0(E)$ and $E_0(J_r,L)$). We have implemented a \textsc{python} code, named \textsc{adiabatic-tides}, which does this procedure and it is publicly available.
We describe more of the numerical details in Appendix \ref{app:numerics}. We have checked that our numerical choices guarantee that the NFW profile is reconstructed to better than $1\%$ relative accuracy on the interval $r \in [10^{-10}, 10^{4}] r_{\rm{s}}$ for a case without tidal field ($\lambda = 0$). We expect a similar accuracy in the case of non-zero $\lambda$ and we validate our implementation through independent tests from N-body simulations in Section \ref{sec:simulations}. If about 100 radial bins are used, each iteration step takes about one second which allows to calculate all information about a tidally truncated halo, even for cases with extremely strong tidal fields, in about a minute. 

\begin{figure}
    \centering
    \includegraphics[width=\columnwidth]{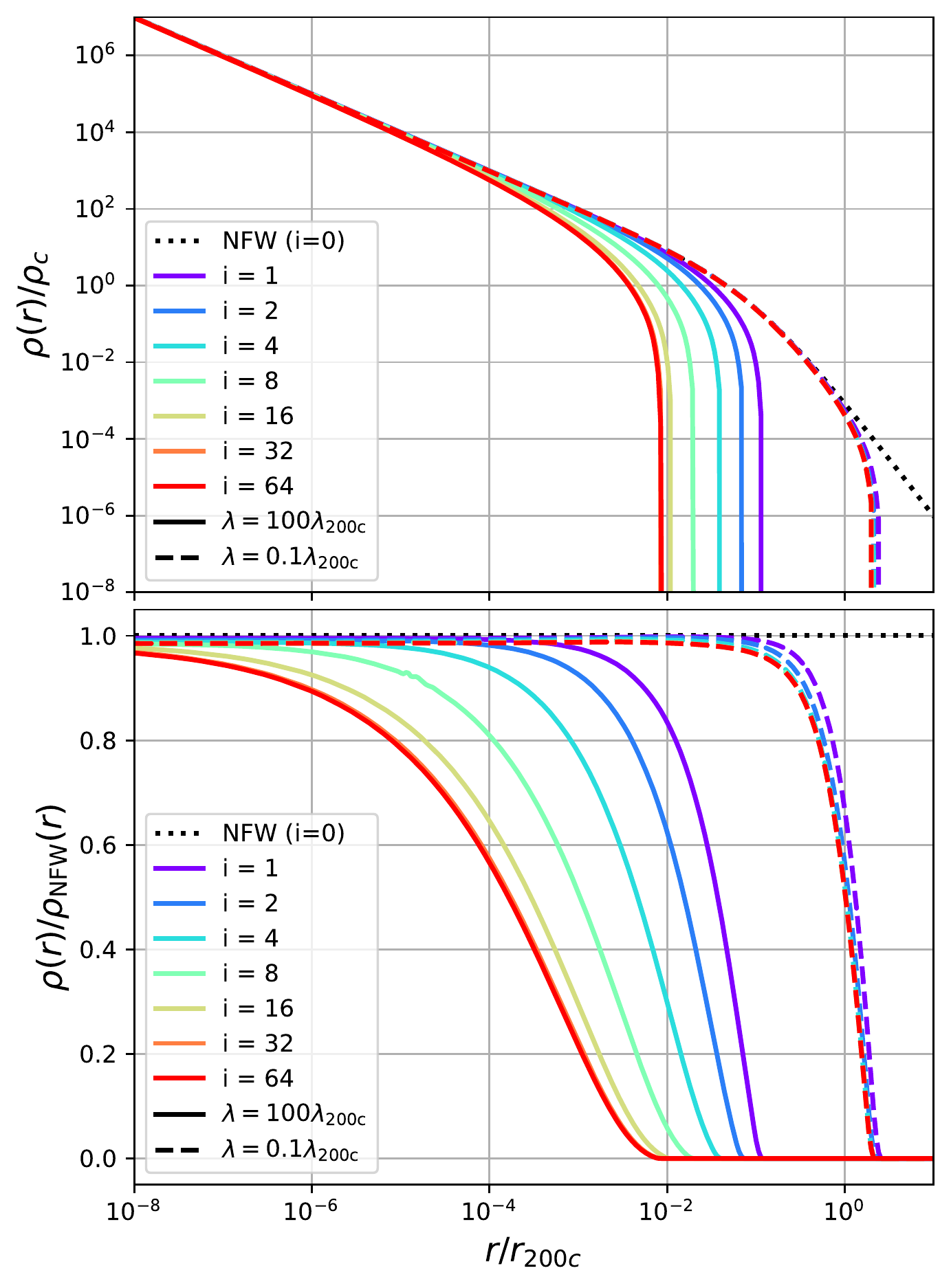}
    \caption{Convergence versus number of iterations for two different tidal fields. The top panel shows the density profile and the bottom panel the ratio between the density profile and the initial NFW profile. If the profile is cut in the $r^{-3}$ part (dashed lines), convergence is quick (4-8 iterations are enough) and the cut-off is reasonably steep. If it is cut in the $r^{-1}$ regime (solid lines), convergence is slow (requires about 32 iterations) and the cut-off is quite flat, spanning about 6 orders of magnitude in radial scale.}
    \label{fig:iteration_convergence}
\end{figure}
We show two examples of the iterative procedure in Figure~\ref{fig:iteration_convergence} for an NFW profile with concentration $c=10$. The first example uses a very weak tidal field with $\lambda = 0.1 \lvir$ which truncates the profile well beyond the scale radius $r_{\rm{s}}$ where the density profile approaches an $r^{-3}$ dependence. In this case the the iterative procedure converges quickly in about $4-8$ iterations. The second example shows a much stronger tidal field $\lambda = 100 \lvir$ which truncates the profile inside the scale radius $r_{\rm{s}}$. In this case it takes about $32$ iterations until convergence is reached and a huge amount of mass is lost. Note that the cut-off goes over six orders of magnitude in spatial scale, which means that such remnants will be hard to simulate while resolving the whole truncation. For all further plots we use $100$ iterations so that we are always in the very well converged regime.

\subsection{The structure - tide degeneracy} \label{sec:structuretide}

At first it might seem that the considered problem has three independent relevant parameters. Two parameters are necessary to describe the initial NFW profile (either $\rho_{\rm{c}}$ and $r_{\rm{s}}$, or $\mvir$ and $c$) and one parameter $\lambda$ to describe the tidal field. 

However, in the adiabatic limit, as well as in the case of orbiting subhaloes, some scales can be eliminated due to the invariances of the Vlasov-Poisson system. The Vlasov-Poisson equations are invariant to a spatial scaling and a time-rescaling
\begin{align}
  \myvec{r}^* &\rightarrow \alpha \myvec{r}   \\
  t^* &\rightarrow \beta t
\end{align}
if velocities are rescaled as $\myvec{v}^* \rightarrow \alpha \beta^{-1} \myvec{v}$, masses as $m^* \rightarrow m \alpha^3 \beta^{-2}$ and any explicit external potential scales as $\phi_{\rm{ext}}(\myvec{r}^*, t^*) = \alpha^2 \beta^{-2} \phi_{\rm{ext}}(\myvec{r}, t)$. If we consider our subhalo as a Vlasov-Poisson system and treat it in the distant tide approximation \rev{(which holds if its mass $M$ is much smaller than the host halo's mass $M_{\rm{h}}$)}, then we model the hosts influence through an explicitly time-dependent potential
\begin{align}
    \phi_{\rm{ext}}(t) &= - \frac{1}{2} \myvec{r} \Tid \myvec{r}
\end{align}
which follows the rescaling relations if the tidal tensor rescales as $\Tid^*(t^*) \rightarrow \beta^{-2} \Tid(t)$. 

This has important implications for the mass dependence and the concentration dependence of tidal stripping: If we compare two initial NFW subhaloes with different scale radii $r_{\rm{s},1}$ and $r_{\rm{s},2}$ (or equivalently: different masses) then we can interpret them as rescaled versions of each other with $\alpha = r_{\rm{s},1} / r_{\rm{s},2}$ and $\beta = 1$. If those two haloes follow the same orbit, then they are exposed to the same tidal fields $\Tid_1(t) = \Tid_2(t)$ at all times and they are exactly rescaled versions of each other (e.g. $M_1(r/r_{\rm{s},1}, t) = \alpha^3 M_2(r/r_{\rm{s},2}, t)$). Therefore, the relative mass-loss of subhaloes is independent of their scale radii or their masses when compared at a fixed orbit \citep[e.g.][]{aguirre_2023}. This degeneracy has already been widely appreciated and used, and deviations occur in realistic scenarios only \rev{for relatively massive subhaloes ($M / M_{\rm{h}} \gtrsim 10^{-3}$}) due to effects of the host halo that are not properly accounted for through a mere distant-tide approximation such as dynamical friction \citep[e.g.][]{ogiya_2021} \rev{or higher order terms of the multipole expansion of the potential} \citep[e.g.][]{aguirre_2023}. 

However, less attention has been paid to the time-rescaling invariance of the system. Two haloes with different characteristic densities $\rho_{c1} / \rho_{c2} = \beta^{-2}$ should also be exactly time-rescaled versions of each other if they are exposed to different tidal fields so that $\Tid_1(\beta t) = \beta^{-2} \Tid_2(t)$. This implies that the subhalo-disruption problem exhibits a degeneracy between the characteristic density of a subhalo (equivalently: its concentration) and the amplitude of the tidal field. Increasing the characteristic density of an initial subhalo by some factor has an \emph{exactly} equivalent effect to decreasing the tidal field it is exposed to by the same factor. We call this degeneracy the `structure-tide' degeneracy, and we will explore some important implications of this in this paper. 

Instead of keeping track explicitly of the rescaling factors $\alpha$ and $\beta$, it is convenient to express all quantities in reduced units, so that all results are independent of $\alpha$ and $\beta$. We formulate all radii in units of the initial scale radius $r_{\rm{s}}$, times in units of the initial circular orbit time at the scale radius $t_{\rm{s}}$, densities in units of the characteristic density $\rho_{\rm{c}}$, masses in units of the mass initially enclosed inside the scale radius $M_{\rm{s}}$, potentials in units of the initial central potential $\phi_0$ and most importantly, tidal fields in units of the scale-tide $\lambda_{\rm{s}}$: 
\begin{align}
  M_{\rm{s}} &= \frac{\ln(2) - 1/2}{\ln(1+c) - c/(1+c)} \mvir \\
  t_{\rm{s}} &= 2 \pi \sqrt{\frac{r_{\rm{s}}^3}{GM_{\rm{s}}}} \\
  \rho_{\rm{c}} &= \frac{200 \rho_{\rm{crit}} c^3}{3 [\ln(1+c) - c/(1+c)]} \\
  \lambda_{\rm{s}} &= \frac{\ln(2) - 1/2}{\ln(1+c) - c/(1+c)} c^3 \lvir \\ 
            &= 4 \pi G \rho_{\rm{c}} (\ln(2) - 1/2)
\end{align}
where we defined $\lambda_{\rm{s}}$ so that it is the tidal field necessary to unbind an NFW profile at the scale radius $\lambda_{\rm{s}} = \partial_r \phi_{\rm{NFW}}(r_{\rm{s}})/r_{\rm{s}}$. Note that $\lambda_{\rm{s}}$ is proportional to the characteristic density of the halo. The conditions for two orbiting subhaloes to be rescaled versions of each other in these units is
\begin{align}
    \frac{\Tid_1(t/t_{\rm{s},1})}{\lambda_{\rm{s},1}} &= \frac{\Tid_2(t/t_{\rm{s},2})}{\lambda_{\rm{s},2}}
\end{align}
where $t_{\rm{s}}$ is the circular orbit time at the scale radius.

It may be difficult to find practical cases where the tidal fields on two different orbits relate exactly in this way. \citet{Delos_2019} has pointed out that such a degeneracy can be used to reduce the effective parameter space for subhaloes that are orbiting very far inside the scale radius of an NFW-host. However, for more general setups we may expect that also an approximate correspondence of tidal fields (e.g. by only matching their largest eigenvalue) may produce subhaloes that disrupt to a very similar degree. To give an example: A subhalo with concentration $c=10$ which is on a circular orbit in a Milky Way-like NFW halo ($\mvir = 10^{12}\rm{M}_\odot$, $c=6$) at a distance of $0.5$ the host's virial radius (with the largest eigenvalue of the tidal tensor $\lambda = 5.2 \lvir = 0.04 \lambda_{\rm{s}}$) might keep a similar amount of mass in units of its scale mass $M_{\rm{s}}$ to a higher concentrated $c=20$ halo at a smaller radius of $0.155$ times the host radius (corresponding to a larger tidal field of $\lambda = 29.7 \lvir = 0.04 \lambda_{\rm{s}}$). We'd expect that such cases match at first order and that other aspects only introduce some smaller secondary dependencies, e.g. on the orbital frequency (and the associated Coriolis and Centrifugal force), on the value of the other two eigenvalues of the tidal tensor and on dynamical friction processes. We  will investigate numerically whether such an approximate matching is possible in Section \ref{sec:structide_sim}.

However, in the adiabatic limit the dependence on time-scales disappears completely (since $t \rightarrow \infty$) and for our isotropic model we only consider a single eigenvalue for the tidal tensor. Therefore, in our model all cases that have the same value of $\lambda / \lambda_{\rm{s}}$ are exactly rescaled versions of each other. Therefore, we can summarize the three parameters $\lambda, r_{\rm{s}}$ and $\rho_{\rm{c}}$ into one effective parameter $\lambda / \lambda_{\rm{s}}$ -- which we will often refer to as the \emph{effective tide}. The following is an intuitive way to think about this reduction to a single parameter: it does not matter what the precise amplitude of the tidal field, the density scale and the spatial scale of the initial NFW are, it only matters at what fraction of the scale radius the halo gets tidally truncated.

We use the structure-tide degeneracy to reduce the number of adiabatic models that have to be evaluated to a one dimensional grid with different values of $\lambda / \lambda_{\rm{s}}$ that can be interpolated later to evaluate for different combinations of structures and tidal fields. Further, we think that the structure-tide degeneracy is also quite relevant outside the adiabatic context in this paper. Therefore, we show in Section \ref{sec:structide_sim} that the structure-tide degeneracy can be recovered in numerical simulations and we show in Section \ref{sec:predictions} how it can be used to simplify the understanding of the tidal disruption parameter space. Finally, in \citet{aguirre_2023}  \rev{we show} applications of this degeneracy under less idealized circumstances than what is considered in this paper.

\subsection{Monte Carlo realisations}
In principle, any interesting quantities of the tidal remnant besides the density profile -- like for example the energy and angular momentum distribution -- can be evaluated by projecting the phase space distribution to the corresponding axes, similar to equation \ref{eqn:f_to_dens}. However, a less tedious approach is to create a Monte-Carlo realisation of the tidal remnant and infer such distributions via histograms. We found that such a realization $(\myvec{x_i}, \myvec{v_i},m_i)$ can be generated efficiently by first creating a Monte-Carlo particle realization of the initial NFW profile $(\myvec{x_{i,0}}, \myvec{v_{i,0}}, m_{i,0})$ up to the tidal radius $r_{\rm{tid}}$ of the final profile. We follow the procedure described in \citet{Errani2020} for this. Then, to get a realization of the final profile we can use the same positions and velocities, but re-weight the masses by the ratio of the distribution functions
\begin{align}
    m_i = m_{i,0} \frac{f(\myvec{x}_i, \myvec{v}_i)}{f_{0} (\myvec{x}_i, \myvec{v}_i)}
\end{align}
\revcom{slightly modified this equation}
\rev{where $f$ is given for bound particles through the adiabatic mapping}
\begin{align}
    f(\myvec{x}_i,\myvec{v}_i) = \begin{cases}
      f_0(J_r(E_i,L_i),L_i)  & \text{for bound orbits} \\
      0         & \text{for unbound orbits}
    \end{cases}
\end{align}
\revcom{new equation}
Distributions that are presented in this article, that do not directly follow from the density profile, were all created through such a weighted Monte-Carlo sampling with a sufficiently high number of samples.

We note that this procedure can also be used to set up initial conditions for N-body simulations that start with a tidal remnant\footnote{If needed, particle masses can be made uniform through an additional rejection sampling step.}. This way one could follow such a remnant at much higher resolution than what would be possible if using an N-body simulation that starts with the initial NFW profile. For example, this could be used to test whether the tidal remnants cannot disrupt, even when the tidal field has a strong time dependence. Such experiments are beyond the scope of this paper, but we note that all necessary tools are implemented in the \textsc{adiabatic-tides} repository for the benefit of potential future studies.

\section{Simulation Validation} \label{sec:simulations}
In this section we test our model of adiabatic tidal remnants against simulations. We do this in four steps: In Section \ref{sec:atides_sim} we use numerical simulations to infer the adiabatic limit of tidal mass loss to validate that our implementation of the adiabatic limit is correct and to show that the calculations in spherical symmetry are even useful for highly anisotropic tidal fields. In Section \ref{sec:simcircular} we check in how far the \textsc{adiabatic-tides} model predicts the asymptotic behavior for orbiting subhaloes on circular orbits when considered from the corotating frame. In Section \ref{sec:orbiting_sim} we check in how far the adiabatic limit corresponds to the asymptotic limit of more generic non-circular orbiting subhaloes. Finally, in Section \ref{sec:structide_sim} we show at the hand of a few examples that the structure-tide degeneracy is a powerful tool -- even in the case of only partially disrupted subhaloes.

\subsection{Adiabatic tides simulation} \label{sec:atides_sim}
\begin{table}
    \centering
    \begin{tabular}{c|c|c|c|c|c|c}
         Type & $\lambda / \lvir$ & $N$ & $\tau / \tvir$ & $F_{\rm{cut}}$ & $M/\mvir$ & $M_{\rm{at}}/\mvir$  \\
         \hline
         iso & 1 & $2^{21}$ & 2.5 & 4 & 0.398 & 0.399 \\
         aniso & 1 & $2^{21}$ & 20 & 2 & 0.404 & 0.399 \\
         iso & 4 & $2^{20}$ & 2.5 & 2 & 0.122 & 0.127 \\
         aniso & 4 & $2^{21}$ & 10 & 2 & 0.147 & 0.127 \\
         iso & 16 & $2^{21}$ & 1 & 2 & 0.0051 & 0.0054 \\
         aniso & 16 & $2^{22}$ & 20 & 2 & 0.0046 & 0.0054
    \end{tabular}
    \caption{Simulation parameters that were used for inferring the adiabatic limit numerically. The first column indicates whether an isotropic or anisotropic tidal tensor was used, $\lambda$ indicates the strength of the tidal field, $N$ the particle number, $F_{\rm{cut}}$ the initial truncation radius in units of $\rvir$, $M$ the simulated final mass and $M_{\rm{at}}$ the remnant mass that is predicted by the \textsc{adiabatic-tides} model.}
    \label{tab:adiabaticsim}
\end{table}

\begin{figure}
    \centering
    \includegraphics[width=\columnwidth]{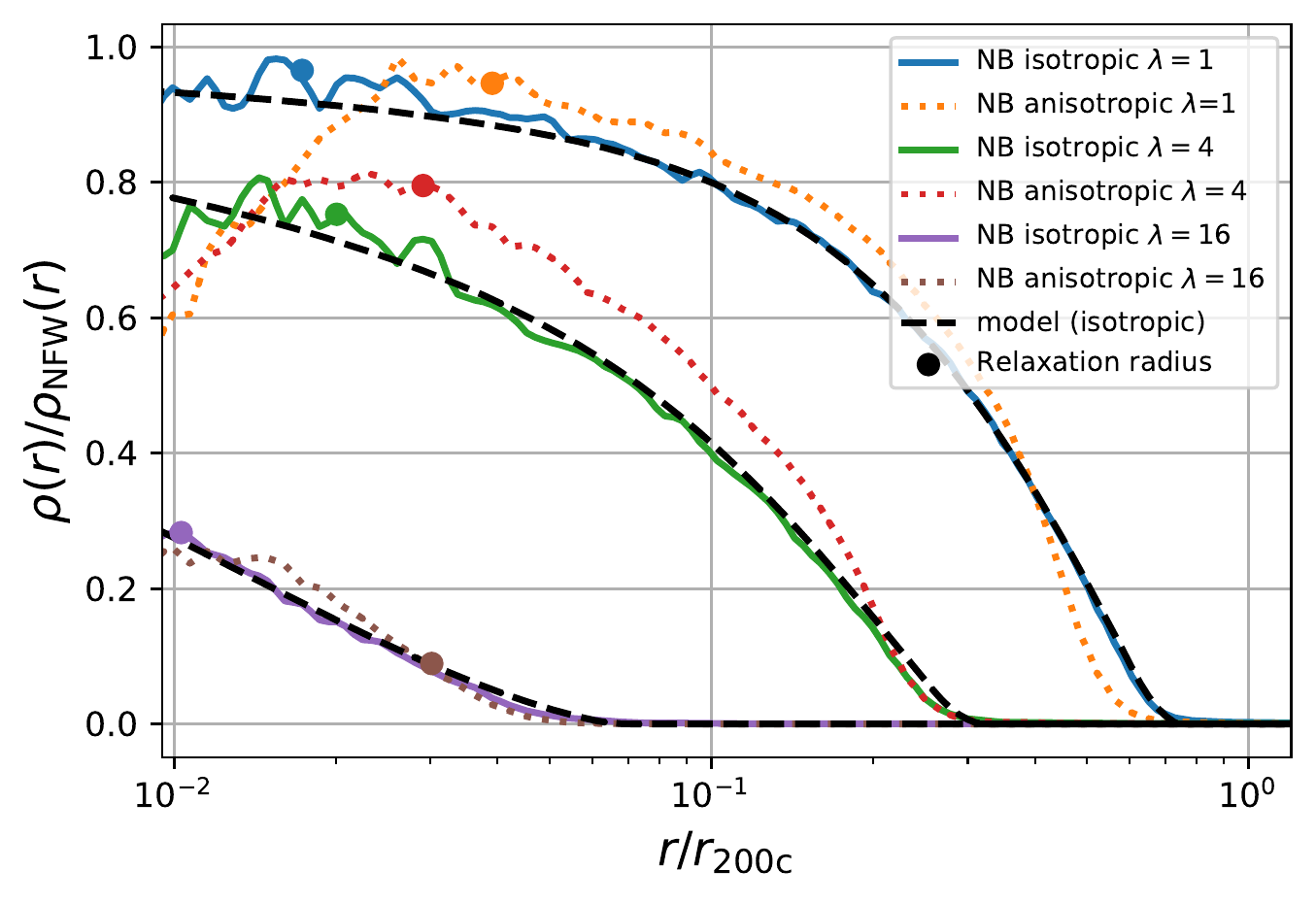}
    \caption{\revcom{Added relaxation radius to this Figure.} Transfer function of the density profiles $\rho/\rho_{\rm{NFW}}$ of adiabatically tidally truncated NFW haloes. Different colours correspond to different simulations, whereas the black dashed lines correspond to our \textsc{adiabatic-tides} model. \rev{The markers indicate the} \citet{Power_2003} \rev{convergence radius below which relaxation effects make the simulation unreliable.} Our model (dashed lines) exactly predicts the adiabatic limit for isotropic cases (solid lines) and provides a reasonable approximation to the anisotropic cases (dotted lines) within $10\%-15\%$ accuracy. The drop at small radii in some simulations is due to relaxation effects. The anisotropic $\lambda=16$ case may be affected by relaxation effects and should not be fully trusted.}
    \label{fig:simcomp_density}
\end{figure}

We run a set of N-body simulations in the adiabatic limit. For this we initialize a Monte-Carlo realization of an NFW halo with concentration $c=10$ up to a truncation radius of $F_{\rm{cut}}$ times the virial radius. We then evolve the particle distribution under their self-gravity $\phi_{\rm{s}}$ plus an additional time-dependent tidal field
\begin{align}
    \phi(\myvec{x}) &= \phi_{\rm{s}} (\myvec{x}) - \frac{1}{2} \myvec{x} \Tid(t) \myvec{x} \label{eqn:tidpot}
\end{align}
where we slowly increase the tidal field over the time-scale $\tau$ and then keep it fixed up to the end of the simulation which we set to be at $t = 2\tau$
\begin{align}
\Tid(t)=
\begin{cases}
\frac{t}{\tau} \Tid_0 &\text{if } 0 < t \leq \tau\\
\Tid_0 &\text{if } t > \tau
\end{cases}
\end{align}
Note that for arbitrarily large $\tau$ it would not be necessary to proceed the simulation after $t = \tau$. However, for finite $\tau$ it is helpful to continue the simulation after the maximum tidal field has been reached, so that all particles that can escape have time to do so. We run these simulations with the N-body tree \citep{barnes_hut_1986} code developed by \citet{Ogiya_2013}. This code has been optimized for utilizing graphics processing unit (GPU) clusters and has been used to create the \textsc{DASH}-library of dynamical subhalo evolution \citep{ogiya_2019}. 
A few problem-specific modifications to the code were necessary and they are discussed in Appendix \ref{app:adsimnumerics}. Further we present convergence tests to the adiabatic limit in Appendix \ref{app:adsimconvergence} to show that simulations converge to the unique adiabatic solution for $\tau \rightarrow \infty$. Here, we present only the results of runs that are already converged  to the adiabatic limit $\tau \rightarrow \infty$ and with the particle number $N$ large enough that relaxation effects are irrelevant. For the final tidal fields $\mathbfss{T}_0$ we choose two different scenarios for isotropic cases and anisotropic ones:
\begin{align}
    \mathbfss{T}_{0,\rm{iso}} &= \rm{diag} (\lambda, \lambda, \lambda) \\
    \mathbfss{T}_{0,\rm{aniso}} &= \rm{diag} (\lambda, -0.7 \lambda, -0.3 \lambda)
\end{align}
where diag(...) stands for a 3x3 diagonal matrix with the given diagonal elements. 
We list the parameters of our converged simulations in Table \ref{tab:adiabaticsim}. The time-scales are measured in units of the time $\tvir$ that is needed for a circular orbit at the virial radius 
\begin{align}
    \tvir &= 2 \pi \sqrt{ \frac{\rvir^3}{G\mvir}} .
\end{align}
Note that the anisotropic simulations need quite a bit longer than the isotropic ones to reach the adiabatic limit, since the spatial extent of the escape route is much smaller. Particles that are slightly above the escape-energy can escape in the isotropic case in every direction, whereas in the anisotropic case they can only escape along a narrow gap along the x-axis which would permit their energy-level. 
Further, to get exactly converged results, it is important to set up the NFW profile to beyond the virial radius so that most mass is removed adiabatically. (A truncation of the profile at the virial radius would correspond to an instantaneous removal of mass beyond this radius.)

% Version of this Figure before revision
% \begin{figure*}
%     \centering
%     \includegraphics[width=0.32\textwidth]{img/simulation/eini_nb_versus_model.pdf}
%     \includegraphics[width=0.32\textwidth]{img/simulation/eself_nb_versus_model.pdf}
%     \includegraphics[width=0.32\textwidth]{img/simulation/etot_nb_versus_model.pdf}
%     \caption{Histograms of initial, final self-energy and final total energy in comparison between model and simulations. The \textsc{adiabatic-tides} model exactly recovers the isotropic adiabatic simulations and is a very good approximation to anisotropic ones. An adiabatically increased tidal field produces a quite sharp truncation in energy space. The anisotropic $\lambda=16$ case cannot be fully trusted, since it is affected by relaxation effects.}
%     \label{fig:simcomp_energy}
% \end{figure*}

\begin{figure*}
    \centering
    \includegraphics[width=0.48\textwidth]{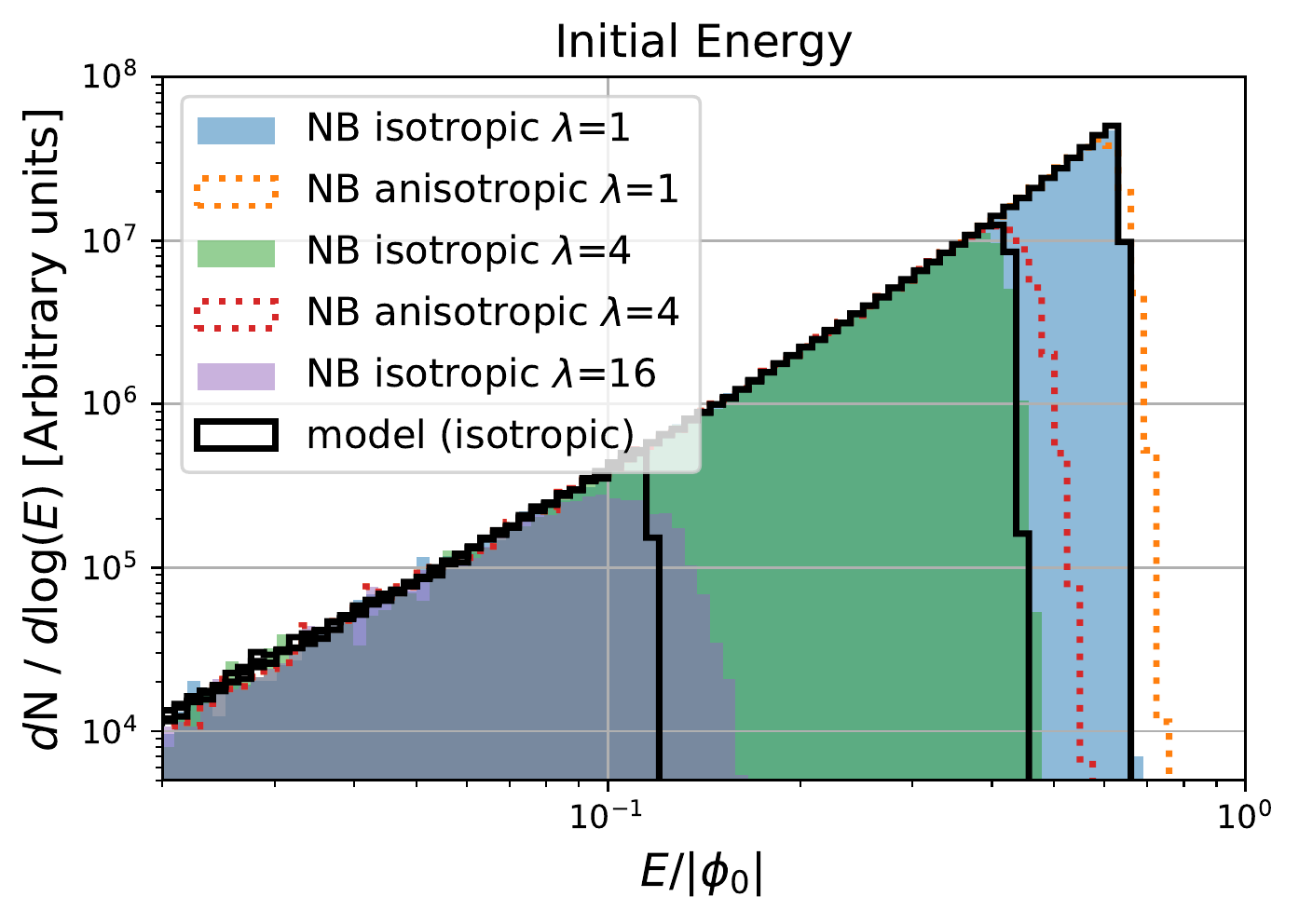}
    \includegraphics[width=0.48\textwidth]{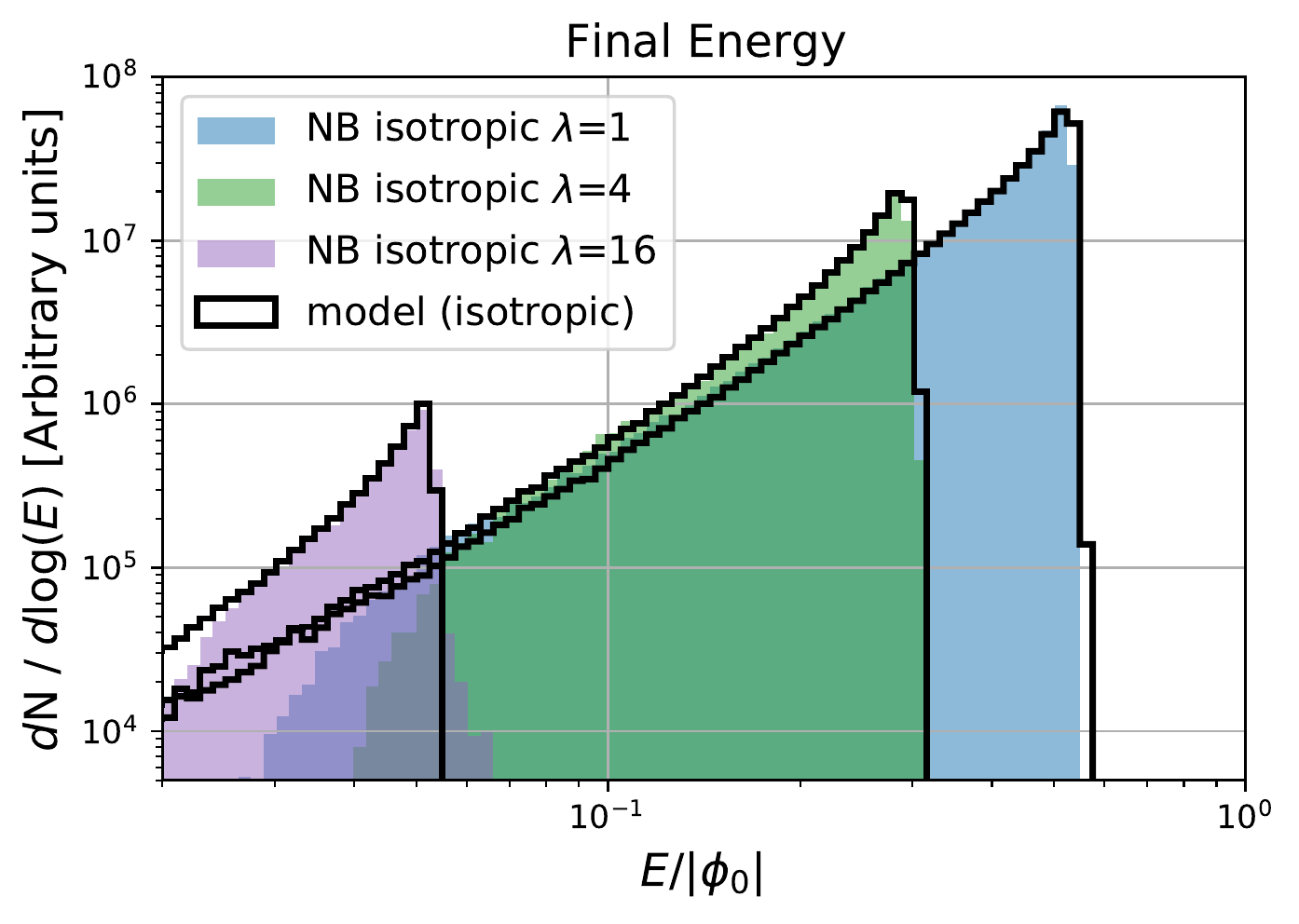}
    \caption{Histograms of initial and final total energy in comparison between model and simulations. The \textsc{adiabatic-tides} model exactly recovers the isotropic adiabatic simulations and is a good approximation to anisotropic ones. An adiabatically increased tidal field produces a quite sharp truncation in energy space. Deviations occur due to numerical relaxation effects, which smooth the boundary in initial energy space and evacuate small energies in final energy space. \revcom{Changed Figure~to log-log-space, removed one panel, increased font size. Also adapted caption}}
    \label{fig:simcomp_energy}
\end{figure*}

We have already shown an example visualization of the simulations with $\lambda=4 \lvir$ in Figure~\ref{fig:tidalexperiment}. We show the density transfer functions
\begin{align}
    T(r) = \frac{\rho(r)}{\rho_{\rm{NFW}} (r)}  
\end{align}
of the simulated profiles versus the one predicted by our adiabatic model in Figure~\ref{fig:simcomp_density}. We note that our adiabatic model \emph{exactly} predicts the density profile of the adiabatic limit of the isotropic simulations. This is expected and shows that our implementation is correct and that the iterative procedure indeed converges to the correct solution. Further, we note that the isotropic and anisotropic cases have slightly different transfer functions, but they still agree within approximately 10\% at every radius and the relative difference between transfer functions does not get worse than about $25\%$ -- except very close to the tidal radius.  This is so despite the huge difference in the potential landscape (compare Figure~\ref{fig:tidalexperiment}). Further, we list in Table \ref{tab:adiabaticsim} the fraction of mass that remains bound in comparison to the \textsc{adiabatic-tides} model prediction and they also agree very well. We note that at small radii some cases have a downturn (especially the anisotropic $\lambda = 1$ case) due to relaxation effects. This may be most problematic for the anisotropic $\lambda = 16$ case where the regime affected by relaxation may lie reasonably close to the tidal radius and this simulation can therefore not be fully trusted.\footnote{Note that we have chosen a two times higher particle number for this simulation which should still make the relaxation effects a little less significant at those radii than for the $\lambda=1$ case. Unfortunately we could not afford to use even higher resolutions to further suppress relaxation effects.}

Beyond the density profile, the \textsc{adiabatic-tides} model can also be used to predict the energy distribution of particles that remain bound to the halo. There are \rev{two} different ways of visualizing these energy distributions. The first way is by visualizing the initial (NFW) energies of all particles that remain bound to the final remnant. We show this in the left panel of Figure~\ref{fig:simcomp_energy}. \rev{We exclude the anisotropic $\lambda=16$ case here, since it is dominated by relaxation effects at all energy levels.} These histograms are again exactly reproduced for isotropic cases and approximately for anisotropic ones. The tidal field induces a quite sharp cut-off in the initial energy space, which justifies the usage of energy truncation models \citep{drakos_2017,drakos_2020} and it validates the energy space perspective that was proposed in \citet{stuecker_2021_bp}.

\rev{The second way of visualizing the energy histogram is through the final energies of bound particles. Here, we calculate the final potential as the sum of self-potential energy plus the tidal field. Therefore, the right panel shows the energy that would be found by measuring the `boosted potential', instead of the self-potential as explained in} \citet{stuecker_2021_bp}. \rev{However, we have also checked the self-energies and they look very similar here.} We didn't include the corresponding plots for the anisotropic cases, since for these it was not so trivial to calculate the energy in post-processing. We can see that these energy distributions are both accurately reproduced by our model. Further, we note that in the final energy there is a quite sharp cut at $E = \phi_{\rm{tid}}$, but there is indeed a small population of particles with larger energies, which are however limited to the small energy range $\phi_{\rm{tid}} < E < E_{\rm{max}}$ as discussed in Section \ref{sec:boundorbits}.

\subsection{Subhaloes on circular orbits} \label{sec:simcircular}

While the results from the last section suggest that the \textsc{adiabatic-tides} model accurately reproduces the adiabatic limit of tidal mass loss, it is not a-priori clear whether the adiabatic limit is at all relevant for subhaloes orbiting in a host halo potential. The main difference is that for orbiting subhaloes the tidal field is oscillating in orientation and amplitude over time. This implies that energy is not only changed through the internal redistribution of the subhalo (which is included in the \textsc{adiabatic-tides} model), but it may also be redistributed through the time-dependence of the tidal field. Further, depending on the orbit of a subhalo, the tidal field may change on very small time-scales, far from the adiabatic limit. 

However, most of these concerns should be irrelevant for subhaloes that are on circular orbits. For circular orbits we can consider the problem from the corotating frame where the natural energy notion is given by the Jacobi potential
\begin{align}
    \phi_{\rm{J}}(\myvec{x}) &= \phi_{\rm{s}} (\myvec{x} - \myvec{x_{\rm{s}}}) + \phi_{\rm{h}} (\myvec{x}) - \frac{1}{2} (\myvec{\omega} \times \myvec{x})^2
\end{align}
where $\phi_{\rm{s}}$ is the self-potential of the subhalo, $\phi_{\rm{h}}$ the host potential, $\omega$ is the angular frequency vector and the last term describes the centrifugal-effect. The force in the corotating frame is given by the gradient of the Jacobi potential plus the Coriolis force.  If the self-potential of the subhalo does not change over time then the associated Jacobi energy is conserved \citep{BinneyTremaine2008}. Of course the self-potential will change over time due to the subhalo's massloss, but this is type of energy redistribution is included in the \textsc{adiabatic-tides} model. Now, if we treat the Jacobi-potential in the distant-tide-approximation we can expand the host-potential and the centrifugal contribution around the subhalo centre:
\begin{align}
    \phi_{\rm{J}}(\Delta \myvec{x}) &= \phi_{\rm{s}}(\Delta \myvec{x}) - \frac{1}{2} \Delta \myvec{x} \Tid^* \myvec{\Delta \myvec{x}} \\
                    T_{ij}^*      &= - \partial_i \partial_j \left. \left(\phi_{\rm{h}}(\myvec{x})- \frac{1}{2} (\myvec{\omega} \times \myvec{x})^2 \right) \right|_{ \myvec{x}=\myvec{x}_{\rm{s}}}
\end{align}
where $\Delta \myvec{x} = \myvec{x} - \myvec{x}_{\rm{s}}$, where we have neglected an irrelevant absolute offset in the potential and where we have defined an effective tidal tensor $\Tid^*$ that includes the centrifugal effect. If we assume a spherical host-potential, $\Tid^*$ has the eigenvalues
\begin{align}
    \lambda_r^* &= \lambda_r + \omega^2 \label{eqn:lameff} \\
    \lambda_{2,3}^* &= \lambda_{2,3}
\end{align}
where $\lambda_r$ and $\lambda_{2,3}$ are the eigenvalues of the tidal tensor of the host potential associated with the radial and the two tangential directions respectively. Therefore, the centrifugal force effectively increases the radial eigenvalue of the tidal tensor, but leaves the other two eigenvalues unchanged.

Now, if we evaluate the \textsc{adiabatic-tides} prediction with the effective tidal field $\lambda_r^*$, then the estimated remnant mass should be a true lower limit for the amount of mass that remains bound on a circular orbit. This is so, since the circular orbit scenario is very similar to the adiabatic limit calculations when considered from the corotating frame, with only a few subtle differences: (1) The tidal field has not been increased adiabatically to its current amplitude, but instantaneously and then held constant. (2) Our model assumes an isotropic tidal field whereas the actual effective tidal tensor is anisotropic. (3) Particles in the orbiting subhalo experience additionally the effect of the Coriolis force. 

We don't expect that aspect (1) has a very big effect, since the energy that can be injected into the remnant by a single instantaneous change in the tidal field is not that large. (This is discussed in Appendix \ref{app:tidal_heating}.) We have already tested (2) -- the difference between isotropic an anisotropic scenarios -- in the last section. Generally, it seems that the isotropic scenario will lose slightly more mass than the anisotropic scenario, and therefore the isotropic case should still provide a true lower barrier. Probably the most important difference is point (3), that the orbiting subhalo experiences the Coriolis force. Effects of the Coriolis force can be highly unintuitive and difficult to predict. For example, \citet{HansWalter_1989} have found that for the case of dumb-bell galaxies the Coriolis force allows prograde orbits with significantly higher Jacobi-energies than the saddle-point energy level to remain restricted to an object. However, we note that the opposite is not possible, i.e. all particles that are below the saddle-point energy level can not leave the system even when considering the Coriolis force. Therefore, we expect that the Coriolis force can only act to keep additional particles restricted to the subhalo, but it can not reduce the number of bound particles. Thus the \textsc{adiabatic-tides} prediction should be a lower limit for the bound mass of subhaloes on circular orbits that can not be crossed even for arbitrarily long times. However, due to the aforementioned effects it may be that the actually asymptotically bound mass is slightly higher than the \textsc{adiabatic-tides} prediction.

\begin{figure}
    \centering
    \includegraphics[width=\columnwidth]{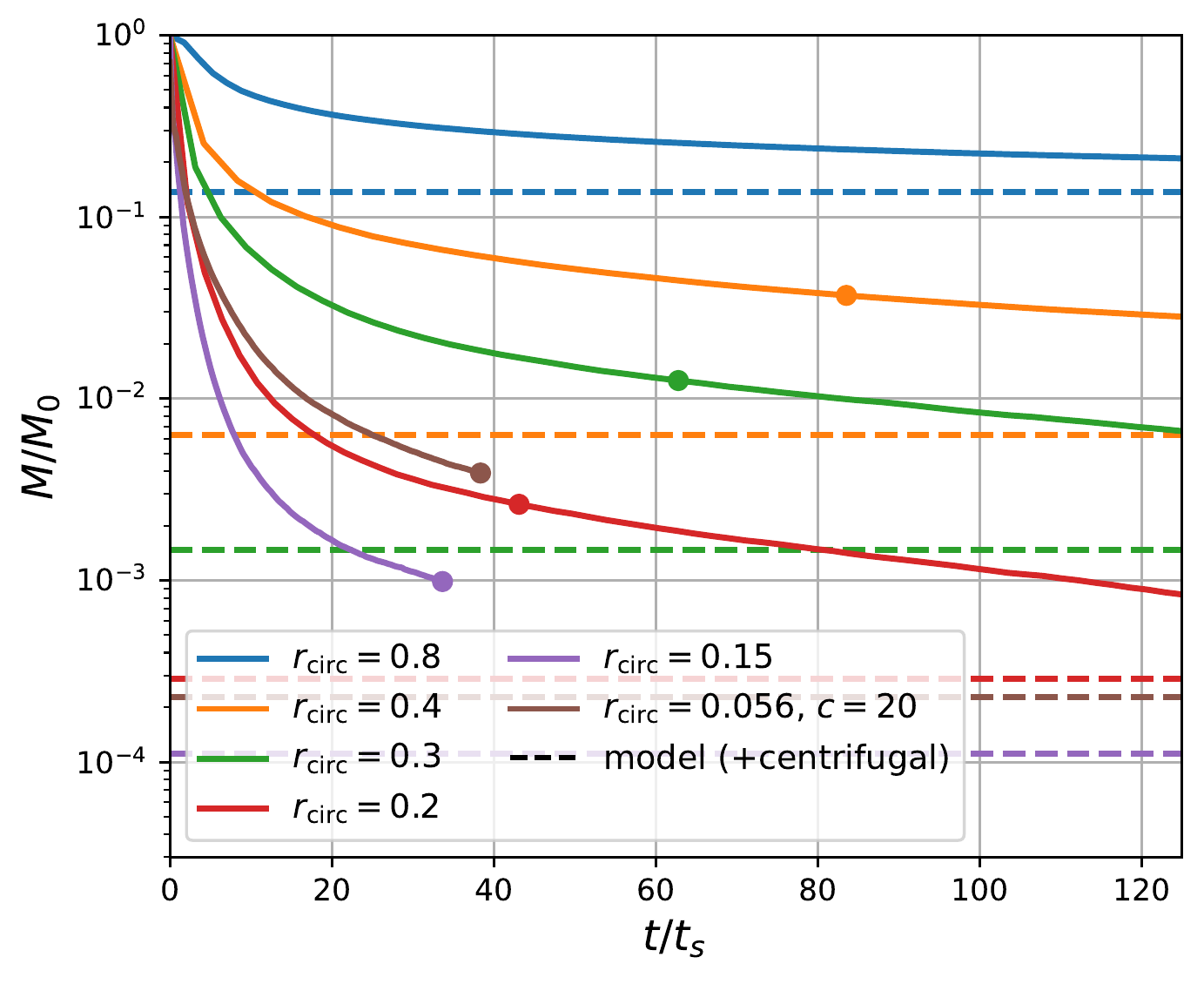}
    \caption{Mass fraction as a function of time in units of the circular orbit timescale at the scale radius of the subhalo. Unlike Figure~\ref{fig:massfraction_versus_orbits}, here the dashed lines indicate the prediction of the \textsc{adiabatic-tides} model when the effect of the centrifugal force is included. The marker indicates the time at which twenty orbits were reached. We have extended some simulations to longer times to check that the adiabatic limit is indeed never crossed.  Circular radii $r_{\rm{circ}}$ are given in units of the hosts virial radius $\rhvir$.}
    \label{fig:massfrac_centrifugal}
\end{figure}

We set up a set of numerical experiments to test whether the \textsc{adiabatic-tides} model indeed describes the asymptotic remnants of tidal stripping. For this we simulate orbiting NFW subhalos with different concentrations at different radii of a Milky Way-like NFW host halo with concentration $c=6$ and mass $\mvir = 10^{12} \rm{M}_\odot$. We state the parameters of these simulations in Table \ref{tab:orbitsims}. The upper half of the table denotes circular orbit simulations that we discuss in this section and the lower half non-circular orbits which we will discuss in the next section. For each of the presented simulations we have checked that they are converged with mass- and force-resolution.

\begin{table*}
    \centering
    \begin{tabular}{c|c|c|c|c|c|c}
         $r_{\rm{p}} / \rhvir$ & $r_{\rm{a}} / \rhvir$ & c & N &  $M_{15}/\mvir$ & $M_{AT} / \mvir$ & $\lambda / \lambda_{\rm{s}}$ \\%& $(\lambda + \lambda_c) / \lambda_{\rm{s}}$ 
         \hline    
0.8 & 0.8 & 10 & $2^{19}$ & 2.1 $\cdot 10^{-1}$ & 2.5 $\cdot 10^{-1}$ & 1.63 $\cdot 10^{-2}$\\ %& 2.91 $\cdot 10^{-2}$\\
0.4 & 0.4 & 10 & $2^{17}$ & 4.3 $\cdot 10^{-2}$ & 4.2 $\cdot 10^{-2}$ & 5.95 $\cdot 10^{-2}$\\ %& 1.17 $\cdot 10^{-1}$ \\
0.3 & 0.3 & 10 & $2^{19}$ & 5.9 $\cdot 10^{-3}$ & 1.2 $\cdot 10^{-2}$ & 9.45 $\cdot 10^{-2}$\\ %& 1.96 $\cdot 10^{-1}$\\
0.2 & 0.2 & 10 & $2^{21}$ & 3.4 $\cdot 10^{-3}$ & 2.3 $\cdot 10^{-3}$ & 1.67 $\cdot 10^{-1}$\\ %& 3.82 $\cdot 10^{-1}$\\
0.15 & 0.15 & 10 & $2^{21}$ & 1.3 $\cdot 10^{-3}$ & 9.3 $\cdot 10^{-4}$ & 2.35 $\cdot 10^{-1}$\\ %& 5.88 $\cdot 10^{-1}$\\
0.056 & 0.056 & 20 & $2^{19}$ & 5.4 $\cdot 10^{-3}$ & 8.8 $\cdot 10^{-3}$ & 0.31 $\cdot 10^{-2}$\\ %& 3.65 $\cdot 10^{-1}$\\
\hline
0.3 & 0.9 & 10 & $2^{19}$ & 2.0 $\cdot 10^{-2}$ & 1.2 $\cdot 10^{-2}$ & 9.45 $\cdot 10^{-2}$\\ %& 1.96 $\cdot 10^{-1}$\\
0.3 & 0.6 & 10 & $2^{19}$ & 9.4 $\cdot 10^{-3}$ & 1.2 $\cdot 10^{-2}$ & 9.45 $\cdot 10^{-2}$\\ %& 1.96 $\cdot 10^{-1}$\\
0.3 & 0.3 & 10 & $2^{19}$ & 5.9 $\cdot 10^{-3}$ & 1.2 $\cdot 10^{-2}$ & 9.45 $\cdot 10^{-2}$\\ %& 1.96 $\cdot 10^{-1}$\\
0.2 & 0.6 & 10 & $2^{19}$ & 6.8 $\cdot 10^{-3}$ & 2.3 $\cdot 10^{-3}$ & 1.67 $\cdot 10^{-1}$\\ %& 3.82 $\cdot 10^{-1}$\\
0.2 & 0.3 & 10 & $2^{19}$ & 4.8 $\cdot 10^{-3}$ & 2.3 $\cdot 10^{-3}$ & 1.67 $\cdot 10^{-1}$\\ %& 3.82 $\cdot 10^{-1}$\\
0.2 & 0.2 & 10 & $2^{21}$ & 3.4 $\cdot 10^{-3}$ & 2.3 $\cdot 10^{-3}$ & 1.67 $\cdot 10^{-1}$ %& 3.82 $\cdot 10^{-1}$\\

%\hline
%1.1 & 1.1 & 5 & $2^{17}$ & 1.4e-01 & 1.3e-01 & x & x\\
%0.5 & 0.5 & 10 & $2^{17}$ & 7.8e-02 & 8.9e-02 & x & x\\
%0.2 & 0.2 & 20 & $2^{17}$ & 4.9e-02 & 6.4e-02 & x & x\\
    \end{tabular}
    \caption{Simulations of orbiting subhaloes that are presented in Figure~\ref{fig:massfrac_centrifugal} and Figure~\ref{fig:massfraction_versus_orbits}. Listed are the pericentre and apocentre radius in units of the hosts virial radius, the subhalo concentration, the number of particles $N$, the mass fraction after 15 orbits, the long-term mass fraction as predicted by the \textsc{adiabatic-tides} model (without centrifugal contribution) and the effective tidal field at pericentre. The horizontal line separates the circular orbit simulations in Figure~\ref{fig:massfrac_centrifugal} from the non-circular orbit simulations in the right panel of Figure~\ref{fig:massfraction_versus_orbits}.}
    \label{tab:orbitsims}
\end{table*}

We show the mass-fractions that remain bound to the subhalo versus time in Figure~\ref{fig:massfrac_centrifugal} for subhaloes which have all the same concentration $c=10$ except for one line with $c=20$. We added the $c=20$ line so that we could test one case at a much smaller radius of $r_{\rm{circ}} = 0.056 \rhvir$ where the $c=10$ simulations had too much mass loss to be resolved. The corresponding estimated asymptotic mass-fractions from the \textsc{adiabatic-tides} model (including the centrifugal effect) are indicated as horizontal dashed lines. Originally we have run all simulations for 20 orbits (indicate by a marker in Figure~\ref{fig:massfrac_centrifugal}). However, we have later extended some simulations to longer times to check that the adiabatic limit is indeed never crossed. We note that it takes an extremely long time until the mass-loss of the circularly orbiting haloes saturates -- this can even be several times longer than the age of the universe. Cases which have less mass loss (e.g. the blue line) preferentially need less time until they reach their asymptotic limit than cases which have a very strong mass loss. This difference is likely due to the qualitatively different behaviour of the tidal truncation in the $r^{-3}$ and the $r^{-1}$ regime of the NFW profile, as is apparent in Figure~\ref{fig:iteration_convergence}. For truncation in the $r^{-3}$ regime only a few iterations were already enough to determine the stable remnant. However, in the $r^{-1}$ regime it required a huge number of iterations. In each of those iterations all mass that can escape the instantaneous potential landscape is lost. The mass-loss in subsequent iterations is a response to the lowering of the escape threshold due to the reduced self-potential and it cannot be lost before the before the previous mass has gone. We expect that mass-loss proceeds in physical setups in a similar manner and therefore the many required iterations indicate that also very long times are needed physically. Note that this behaviour is somewhat counter-intuitive, since the orbital time-scale at the tidal radius is shorter for cases with larger tidal field than for those with smaller tidal field, but the number of crossing times needed to reach the limit increases, still leading to a net increase in the total time needed to reach the adiabatic limit.

We conclude that the \textsc{adiabatic-tides} model with centrifugal correction provides a true lower limit to the asymptotically bound mass for circular orbits. For typical cosmological cases where a subhalo is experiencing a strong tidal field and has only gone through a few orbits, the mass is far from converged yet. We note that even the simulations of the `asymptotic limit' that were considered by \citet{errani_2021} which had about 10-15 orbits did not reach the true asymptotic limit yet. The mass-loss in those cases has not yet saturated, but the subhalo probably goes through a series of quasi-equilibrium states. Reaching the true asymptotic limit where no mass is lost anymore may take much longer than the age of the universe and seems therefore to be rather of academic than of practical interest. Further, circular orbits are quite rare and unnatural for subhaloes in cosmological scenarios.

\subsection{Subhaloes on generic orbits} \label{sec:orbiting_sim}

We have discussed in the last section how the \textsc{adiabatic-tides} model provides a true lower limit for the remnant mass of subhaloes on circular orbits when the centrifugal contribution is included. For circular orbits the tidal field is constant (in the corotating frame) which makes the dynamics quite similar to what is considered in the adiabatic limit calculations. However, it is less clear how this should apply to non-circular orbits where the tidal field is changing over time -- no matter the frame of reference. In this case, additionally, energy can be redistributed through the tidal field. 

Despite these concerns, we think that the \textsc{adiabatic-tides} model should be a good approximation to the asymptotic limit of non-circular orbits if we evaluate the model with the largest tidal field that a subhalo will encounter on its orbit. This will be typically the tidal field at the pericentre (plus a centrifugal correction). Particles that can survive the pericentre tidal field for arbitrary long time-scales can also survive the weaker tidal fields that may be experienced at other points of the orbit. However, if the subhalo goes through many orbits, we might expect that eventually all particles that cannot survive the pericentre tidal field arbitrarily long, will leave the system. 

Further, we argue that the energy redistribution through the tidal field is weak for all particles that remain in the adiabatic remnant. In \citet{stuecker_2021_bp} we have found that particles which leave a subhalo drastically change their energy-levels through the oscillating tidal field. However, particles that remain bound asymptotically change their energies only slightly. This may be because most of these are adiabatically shielded \citep{weinberg_1994a, weinberg_1994b, Spitzer_1987, gnedin_ostriker_1999} and since particles inside the tidal boundary can only experience moderate energy changes through the tidal field. We make a worst-case estimate for the amount of energy that can be injected through a tidal field of a limited amplitude in Appendix \ref{app:tidal_heating} and we find that even in the worst case scenario the remnant should be relatively stable to time-dependent tidal fields, because particles that would react strongly to time-dependent changes of a tidal field of a given order are also removed in the adiabatic limit.
\begin{figure*}
    \centering
    \includegraphics[width=\textwidth]{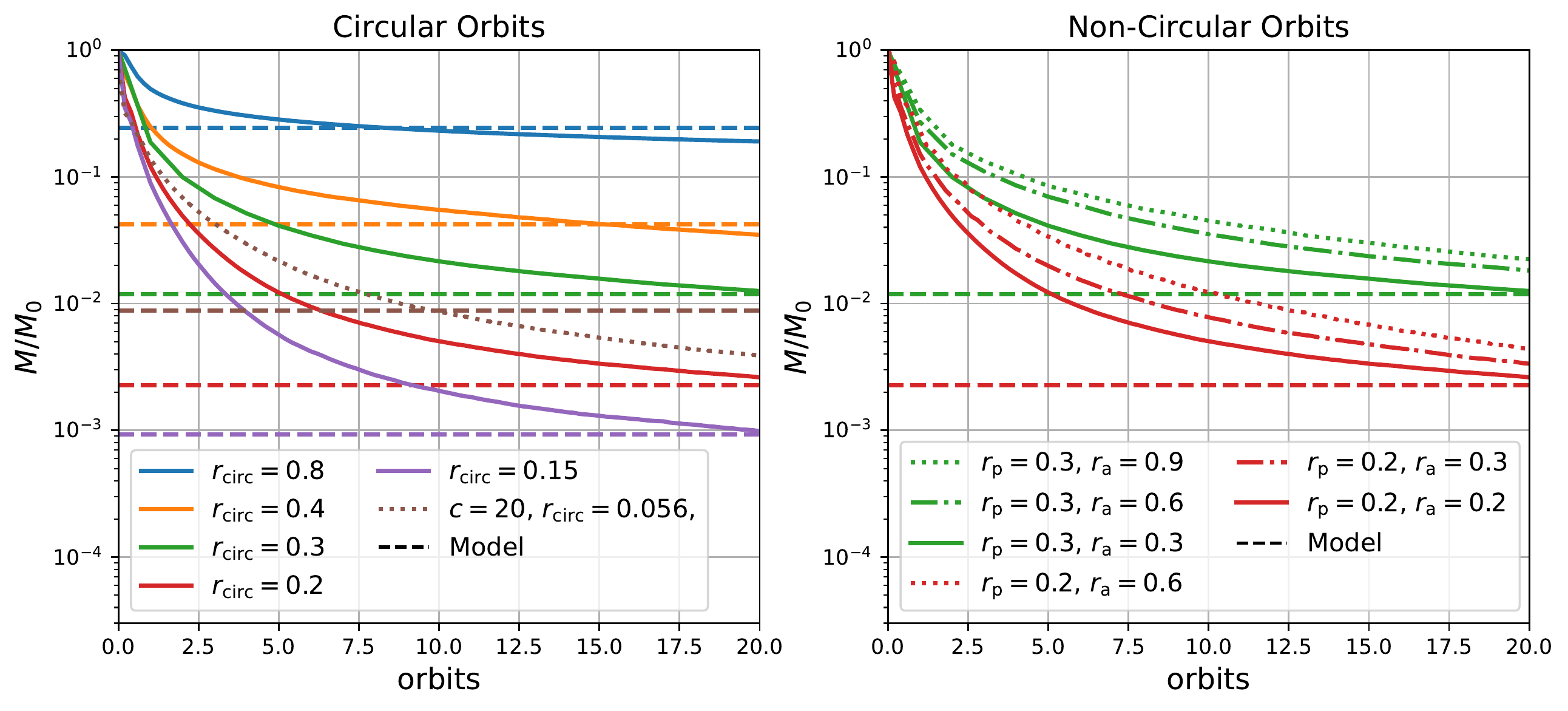}
    \caption{Left: the bound mass fraction versus the number of orbits for circular orbit simulations with different radii $r_{\rm{circ}}$, where $r_{\rm{circ}}$ is given in units of the host's virial radius. Dashed lines indicate the predictions from the \textsc{adiabatic-tides} model. We find that in all cases the \textsc{adiabatic-tides} model gives a reasonable order-of-magnitude prediction for the bound mass after $10-20$ orbits when the evolution slows down drastically. Right: Same, but for non-circular orbits. The predictions still hold approximately if they are matched at the pericentre tidal field.}
    \label{fig:massfraction_versus_orbits}
\end{figure*}

The largest tidal field that a subhalo encounters will be typically the tidal field at pericentre. However, as we have seen for the case of circular orbits in the last section, the effective tidal field that the subhalo `experiences' may be larger than just the gravitational tide due to the effect of the centrifugal force. The centrifugal effect for non-circular orbits varies in general with time \citep[see e.g.][for a detailed discussion]{renaud_2011}. However, we are only interested in its amplitude at pericentre and we show in Appendix \ref{app:centrifugal} that the centrifugal effect should increase the effective tidal field at pericentre as
\begin{align}
    \lambda_r^* &= \lambda_r + \omega_{\rm{circ}}^2 \left(\frac{v_{\rm{circ}}}{v_{\rm{p}}} \right)^2
\end{align}
where $\lambda_r$ is the largest tidal tensor eigenvalue at pericentre, $\omega_{\rm{circ}}$ is the circular orbital frequency at the pericentre radius, $v_{\rm{circ}}$ the circular orbit velocity at pericentre and $v_{\rm{p}}$ is the actual orbital velocity of the subhalo at pericentre. Note that it always holds $v_{\rm{p}} \geq v_{\rm{circ}}$ so that the effective tidal tensor of non-circular orbits is always smaller than that of a circular orbit with the same pericentre. This is so, since the instantaneous curvature radius is larger for non-circular orbits. In Appendix \ref{app:centnoncirc} we show that typical values of $v_{\rm{p}} / v_{\rm{circ}}$ in an NFW profile are by a factor of about 2-3 or larger and that in general the centrifugal term is of reduced importance for typical non-circular orbits. In principle, the centrifugal term implies that haloes with different eccentricities, but identical pericentres, do not have exactly the same asymptotic limit. However, in the approximation that the centrifugal term is irrelevant, they do. 

To simplify the discussion, we neglect the effect of the centrifugal term for the remainder of this paper and approximate $\lambda^* \approx \lambda$. We note that the estimates for circular orbits will be the most affected by this approximation. Therefore, we show the mass-fractions that remain bound to circular orbits again in comparison to the \textsc{adiabatic-tides} prediction in the left panel of Figure~\ref{fig:massfraction_versus_orbits}. The mass-fractions from the \textsc{adiabatic-tides} model (without centrifugal effect) are indicated as horizontal dashed lines.

The \textsc{adiabatic-tides} prediction (without centrifugal contribution) seems to be a good approximation to the mass that remains bound to a subhalo after $10-20$ circular orbits. It also indicates the mass-scale where the mass loss transitions from a rapid mass loss to a slowly progressing asymptotic case. However, we note that the mass-loss trajectories may cross the \textsc{adiabatic-tides} estimate now, which therefore is not a true lower limit under this approximation. We list the mass-fraction that is bound after $15$ orbits in Table \ref{tab:orbitsims} together with the prediction from the \textsc{adiabatic-tides} model. We find that these numbers always lie roughly within a factor of two -- even for haloes which have very low fractions of their initial mass left, e.g. $M/M_0 = 10^{-3}$.

Further, we find that for non-circular orbits, we get a good prediction if we consider the tidal field at pericentre, as shown in the right panel of Figure~\ref{fig:massfraction_versus_orbits}. We see that orbits with the same pericentre, but very different apocentres, reach a similar limit -- varying only within a factor of a few after e.g. 15 orbits. However, we have compared different cases here at the same number of orbits. If compared at the same absolute time, the cases with larger apocentre will need much longer to reach the same mass loss, since they need much longer for each orbit.

We also had a brief look at the density profiles of the presented simulations after fifteen orbits. However, we don't present them here, since the density profiles are not very well converged and require much more computational resources to be brought to convergence than the bound mass fractions. Our first impression is that the actual profiles have a bit sharper truncation than the predicted adiabatic remnants. The simulated profiles seem to be more similar to the steeper cut-offs that are obtained after only a few iterations of the adiabatic procedure (compare Figure~\ref{fig:iteration_convergence}). We leave a rigorous investigation to future studies.

\subsection{The structure-tide degeneracy} \label{sec:structide_sim}

\begin{table}
    \centering
    \begin{tabular}{c|c|c|c|c|c|c}
         $r_{\rm{p}} / \rhvir$ & $r_{\rm{a}} / \rhvir$ & c & N & $F_{\rm{cut}}$ & $\lambda / \lambda_{\rm{s}}$ & $\lambda^* / \lambda_{\rm{s}}$ \\
         \hline    
1.12 & 1.12 & 5 & $2^{17}$ & 2  & 0.0409 & 0.071\\
0.50 & 0.50 & 10 & $2^{17}$ & 1  & 0.0402 & 0.076\\
0.15 & 0.15 & 20 & $2^{17}$ & 0.5 & 0.0399 & 0.099\\
0.07 & 0.07 & 26 & $2^{19}$ & 0.384  & 0.0398 & 0.135\\
\hline
0.41 & 1.22 & 8.1 & $2^{18}$ & 1.23 & 0.0950 & 0.135\\
0.20 & 0.85 & 12.5 & $2^{18}$ & 0.80 & 0.0962 & 0.135\\
0.12 & 0.66 & 15.7 & $2^{18}$ & 0.64 & 0.0963 & 0.136\\
0.05 & 0.49 & 20.2 & $2^{18}$ & 0.49 & 0.0964 & 0.135
    \end{tabular}
    \caption{Parameters of the simulations that are presented in Figure~\ref{fig:massfraction_structure_tide}. Listed are the pericentre and apocentre radius in units of the hosts virial radius, the concentration, the number of particles $N$, the truncation radius of the initial NFW, the effective tide $\lambda / \lambda_{\rm{s}}$ at pericentre and the effective tide $\lambda^* / \lambda_{\rm{s}}$ when including the centrifugal effect as explained in Appendix \ref{app:centrifugal}.}
    \label{tab:stidesims}
\end{table}
\begin{figure}
    \centering
    \includegraphics[width=\columnwidth]{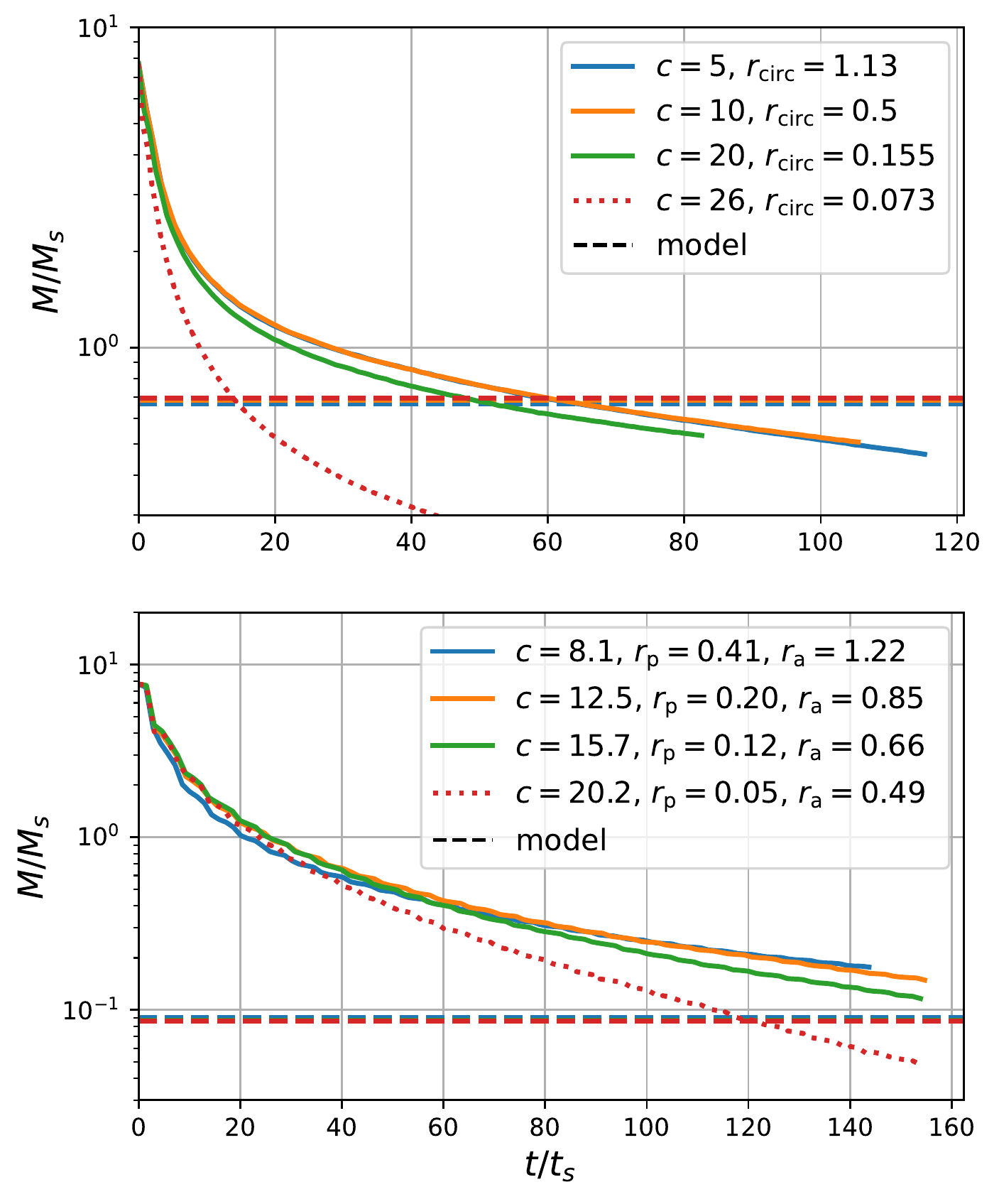}
    \caption{Examples of the structure tide degeneracy. Masses are given in units of the mass contained in the initial scale radius $r_{\rm{s}}$ and time is given in units of the circular orbit time at the initial scale radius $t_{\rm{s}}$. The top panel shows different circular orbit cases which have approximately the same value of the effective tidal field $\lambda / \lambda_{\rm{s}}$. The $c=26$ case (dotted line) deviates strongly since the (here not considered) centrifugal effect is very large. Bottom panel: non-circular orbit cases which have approximately the same value of the effective tide (including and excluding the centrifugal effect) at pericentre. If matched this way, cases with very different physical parameters have almost identical mass loss histories. The horizontal dashed lines represent the \textsc{adiabatic-tides} prediction which is identical for the different cases. The radii in the labels are given in units of $\rhvir.$ }
    \label{fig:massfraction_structure_tide}
\end{figure}

Here, we test whether the structure-tide degeneracy (as explained in Section \ref{sec:structuretide}) can be used to understand dynamical simulations. The main benefit of the structure-tide degeneracy is that it allows to make simple predictions of the concentration dependence of the tidal disruption process. Changing the concentration of a halo increases its characteristic density and makes it more resilient to the effect of a tidal field. If we were to compare the mass loss of two subhaloes that are exposed the same values of the effective tidal field $\lambda / \lambda_{\rm{s}}(c)$ we would expect them to have similar mass-loss histories. As explained in \ref{sec:structide_sim} this correspondence should hold exactly if the full effective tidal tensor matches (in the correct units of time), but we would still expect an approximate correspondence if only the largest eigenvalue matches.

For a first set of simulations we use a circular orbit simulation of a concentration $c=5$ halo at a radius of $r_{\rm{circ}} = 1.13 \rvir$ as a reference case. Then we find the radius in the NFW host halo where the tidal field is stronger by a factor of $\lambda_{\rm{s}}(c) / \lambda_{\rm{s}}(c=5)$ for the concentrations $c=10$, $c=20$ and $c=26$. This way we select cases that have the same value of $\lambda / \lambda_{\rm{s}}$. The corresponding radii are listed alongside other simulation parameters in Table \ref{tab:stidesims}. To further remove any residual dependencies on the initial truncation of the profiles, we have truncated them all at the same value of $r/r_{\rm{s}}$ (leading to different truncation radii in units of the virial radii). 

We show the resulting mass loss histories in the top panel of Figure~\ref{fig:massfraction_structure_tide}. Note that we have plotted the mass in units of the initial mass inside of the scale radius and times in units of the circular orbit time at the scale radius -- which is important to match these simulations. We note that for the cases that orbit at larger radii ($c=5$ and $c=10$) the match between the simulations is almost perfect. The simulation with $c=20$ at $r_{\rm{circ}} = 0.155 \rhvir$ still matches very well, but already shows some slight deviation. Finally, the simulation with $c=26$ and $r_{\rm{circ}} = 0.073 \rhvir$ has a significantly stronger mass loss. This is likely so, since we did not include the centrifugal effect into the tidal field when matching these cases. The centrifugal contribution is dominant for circular orbits at $r \lesssim 0.1 \rhvir$. We have listed the effective tidal field when including the centrifugal contributions $\lambda^* / \lambda_{\rm{s}}$ in Table \ref{tab:stidesims}. We note here that we have also tried experiments on circular orbits where we matched different cases by the centrifugal values of the tidal fields. However, in those cases the mass-loss at the orbit at smaller radii was smaller -- likely due to the effect of the Coriolis force and possibly the larger values of the other components of the tidal tensor.  

We set up a second set of simulations that are on highly non-circular orbits. For these simulations we attempt to match the simulations so that not only their pericentre tidal fields agree, but also so that their tidal fields agree if the centrifugal contribution is accounted for. This is possible for non-circular orbits, since the centrifugal contribution varies with the value of the angular momentum as discussed in Appendix \ref{app:centrifugal}. The considered simulations are listed in the lower half of Table \ref{tab:stidesims}. Note that these simulations all have quite different orbits and also each of the cases spends a different fraction of time at their respective pericentres. Therefore, we cannot expect an exact match.

We show the corresponding mass loss histories in the bottom panel of Figure~\ref{fig:massfraction_structure_tide}. We can see that the mass loss histories match quite well for most of the cases except the one with a very small pericentre $r_{\rm{p}} = 0.05 \rhvir$. This match is quite impressive when considering that the mass loss fraction can easily vary by 2 orders of magnitude if the same orbits would be considered at a fixed concentration (compare Figure~\ref{fig:massfraction_versus_orbits}). We are not entirely sure what causes the $r_{\rm{p}} = 0.05$ to deviate so significantly. A possible reason could be that for this case energy redistribution is more relevant for the progression of mass-loss, since at its pericentre the absolute value of the two non-radial eigenvalues of the tidal tensor are larger relative to the radial one. We discuss energy-redistribution effects in more detail in Appendix \ref{app:tidal_heating}. 

However, we want to emphasize again that the difference between cases with the same value of $\lambda / \lambda_{\rm{s}}$ and different secondary parameters are rather small in comparison to cases which have different value of $\lambda / \lambda_{\rm{s}}$. Therefore, we argue that the pericentre value of $\lambda / \lambda_{\rm{s}}$ should be considered the most important parameter for predictions about the tidal stripping of subhaloes. Other aspects may introduce weaker secondary dependencies.

We conclude that the structure-tide degeneracy may provide a powerful tool for simplifying the parameter-space of tidal stripping -- in particular the dependence on initial concentration. While it is difficult to create exact matches (and therefore predictions) in realistic scenarios because of secondary parameter dependencies, we expect at the very least that strong statistical relations should exist. We show that this is indeed the case in \citet{aguirre_2023}.

\section{Predictions} \label{sec:predictions}
\begin{figure}
    \centering
    \includegraphics[width=\columnwidth]{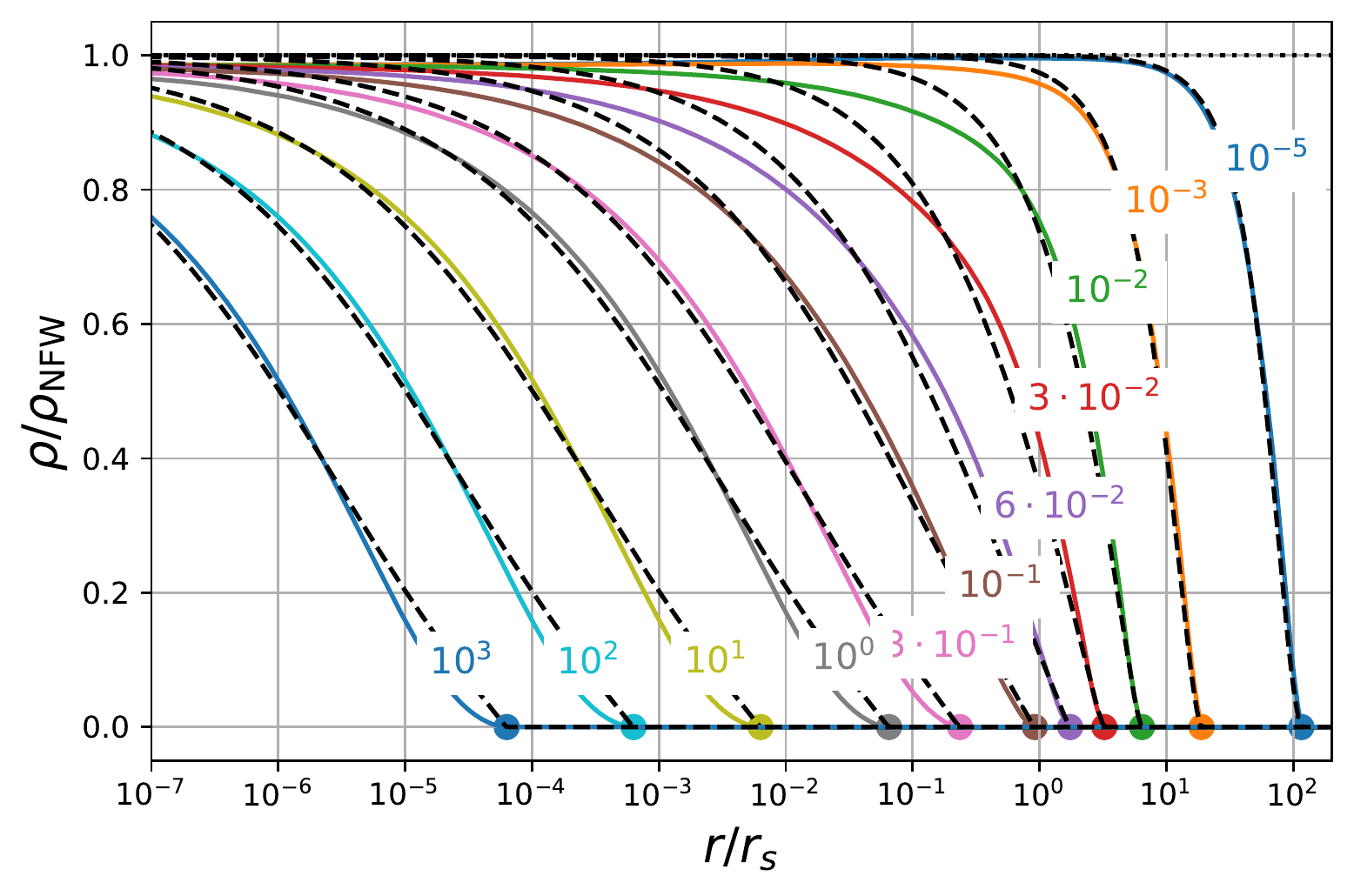}
    \caption{The density transfer function versus the strength of the tidal field. The labels indicate the value of $\lambda/\lambda_{\rm{s}}$. Coloured solid lines correspond to the predictions of our model, and the dashed lines indicate one-parameter fits as in equation \eqref{eqn:transferfit}. The round markers indicate the tidal radius, beyond which densities are exactly zero. Note that the transfer function has two different limiting cases for $\lambda \gg \lambda_{\rm{s}}$ and for $\lambda \ll \lambda_{\rm{s}}$.}
    \label{fig:radial_suppression}
\end{figure}
We have shown in the last section that the \textsc{adiabatic-tides} model is a reasonable approximation for the asymptotic remnants of orbiting NFW haloes if the tidal field at pericentre is considered as the adiabatically applied tidal field. In this section we will use this model to make a variety of predictions. The first set of predictions will be general relations about tidally truncated NFW haloes as a function of the tidal field. The second set of predictions will be for subhaloes that orbit in a Milky Way-like host-potential.

\subsection{Powerlaw profiles}

As a reference case, we have additionally set up a set of powerlaw profiles that have a density profile
\begin{align}
    \rho(r) = \rho_0 \left(\frac{r}{r_0} \right)^\alpha
\end{align}
and calculated their reaction to adiabatically applied tidal fields. The powerlaw profile with $\alpha=-1$ is identical to the central regime of an NFW profile (for $\rho_0 = \rho_{\rm{c}}$ and $r_0 = r_{\rm{s}}$). In plots throughout this section we will often compare with the behaviour of this powerlaw profile and also with powerlaw profiles with different slope (and arbitrary normalizations). Conveniently, the adiabatic limit of a powerlaw profile with a given slope only needs to be calculated once and can easily be rescaled to arbitrary normalizations and tidal fields. All relevant parameters and scaling relations are explained and listed in Appendix \ref{app:powerlaw}.

\subsection{Density transfer functions}
We investigate the density transfer function $T(r) = \rho(r)/\rho_{\rm{NFW}}(r)$ versus the strength of the tidal field. As pointed out previously, the only relevant parameter for this function is the tidal field in units of the scale tide $\lambda/\lambda_{\rm{s}}$. Therefore, we show the transfer functions for different values of $\lambda/\lambda_{\rm{s}}$ in Figure~\ref{fig:radial_suppression}. We additionally have fitted each of the transfer functions through a one parameter fit
\begin{align}
    \rho / \rho_{\rm{NFW}} = \left(1 + \left(\frac{r}{r_{\rm{tid}} - r}\right)^{3 \alpha / 2} \right)^{1/\alpha} \label{eqn:transferfit}
\end{align}
where $r_{\rm{tid}}$ is the tidal radius measured as the saddle-point in the potential (beyond which the density is exactly $0$) and $\alpha$ is the fitted parameter. As can be seen in Figure~\ref{fig:radial_suppression}, these fits are only rough approximations ($\sim 10\%$ accuracy), but they capture the overall behaviour of the transfer function reasonably well. We find that there are two limiting cases, $\alpha \approx 1$ for tidal fields that truncate in the $r^{-3}$ part of the NFW profile and $\alpha \approx \frac{1}{4}$ for tidal fields that truncate the subhalo in the $r^{-1}$ regime.

In comparison to \citet{green_2019}, who focused on measuring the transfer function of tidally stripped subhaloes that were orbiting in a host NFW halo, there are two qualitative differences in our transfer functions. First of all, the adiabatic remnants have exactly zero density outside of the tidal radius. This is so, since mass had arbitrarily large times to escape and therefore no particles in transient states can be found. However, during an ongoing stripping process, the spherical density profile will always pick up particles in the tidal tails, that are about to leave the system. Therefore, our profiles should not be compared in the outskirts to measurements of subhaloes that are still evolving.

The second qualitative difference is that our transfer functions always converge to $T \rightarrow 1$ for $r \rightarrow 0$. \citet{Hayashi_2003} and \citet{green_2019} have proposed transfer functions which approach a limit $T \rightarrow f_{\rm{te}}$ for $r\rightarrow 0$ with values for $f_{\rm{te}}$ that are smaller than one. However, we find that it is not possible -- no matter the strength of the tidal field -- to modify the limiting behaviour of the density for $r \rightarrow 0$ \citep[see also][]{amorisco_2021}. This is so, since the density diverges as $r \rightarrow 0$ and one can always find a central regime which is arbitrary resilient to a finite tidal field. If one checks the measurements of \citet{green_2019} carefully, the data actually seems consistent with a central value of the transfer functions of $1$. The problem is that the cut-off spans many orders of magnitude and the limit for $r \rightarrow 0$ is not clearly measured by \citet{green_2019}. We suggest that the transfer-function descriptions from \citet{green_2019} should be revised to a form that has $\rho/\rho_{\rm{NFW}} \rightarrow 1$ for $r \rightarrow 0$. 

One final point about the transfer functions from \citet{green_2019} is that they explicitly depend on the virial radius $r_{\rm{vir}}$ of the considered object. However, as we have argued in Section \ref{sec:structuretide}, the only spatial scale that matters is the scale radius $r_{\rm{s}}$ and the scale at which the tidal field truncates the profile. Therefore, the dependence on $r_{\rm{vir}}$ may either capture implicitly the effect of the sharp initial truncation of the simulated profiles or it may capture implicitly how the distant tide approximation breaks down when the subhalo's mass gets comparable to the host mass. In either case, we suggest that a simpler, more explicit representation can probably be found to capture such secondary effects. \revcom{Completely rewrote this paragraph.}

%Old version:
%Our final point of criticism about the description from \citet{green_2019} is that it explicitly references the virial radius of the considered object. However, as we have argued in Section \ref{sec:structuretide}, the only spatial scale that matters is the scale radius $r_{\rm{s}}$ and the scale at which the tidal field truncates the profile. Therefore, we speculate that the by \citet{green_2019}  measured $\rvir$ dependence is either an artefact of the sharp truncation of the initial profiles or it means that it should be possible to simplify the description further if the right scales are used.

\subsection{Circular velocity profile and the tidal track}
\begin{figure}
    \centering
    \includegraphics[width=\columnwidth]{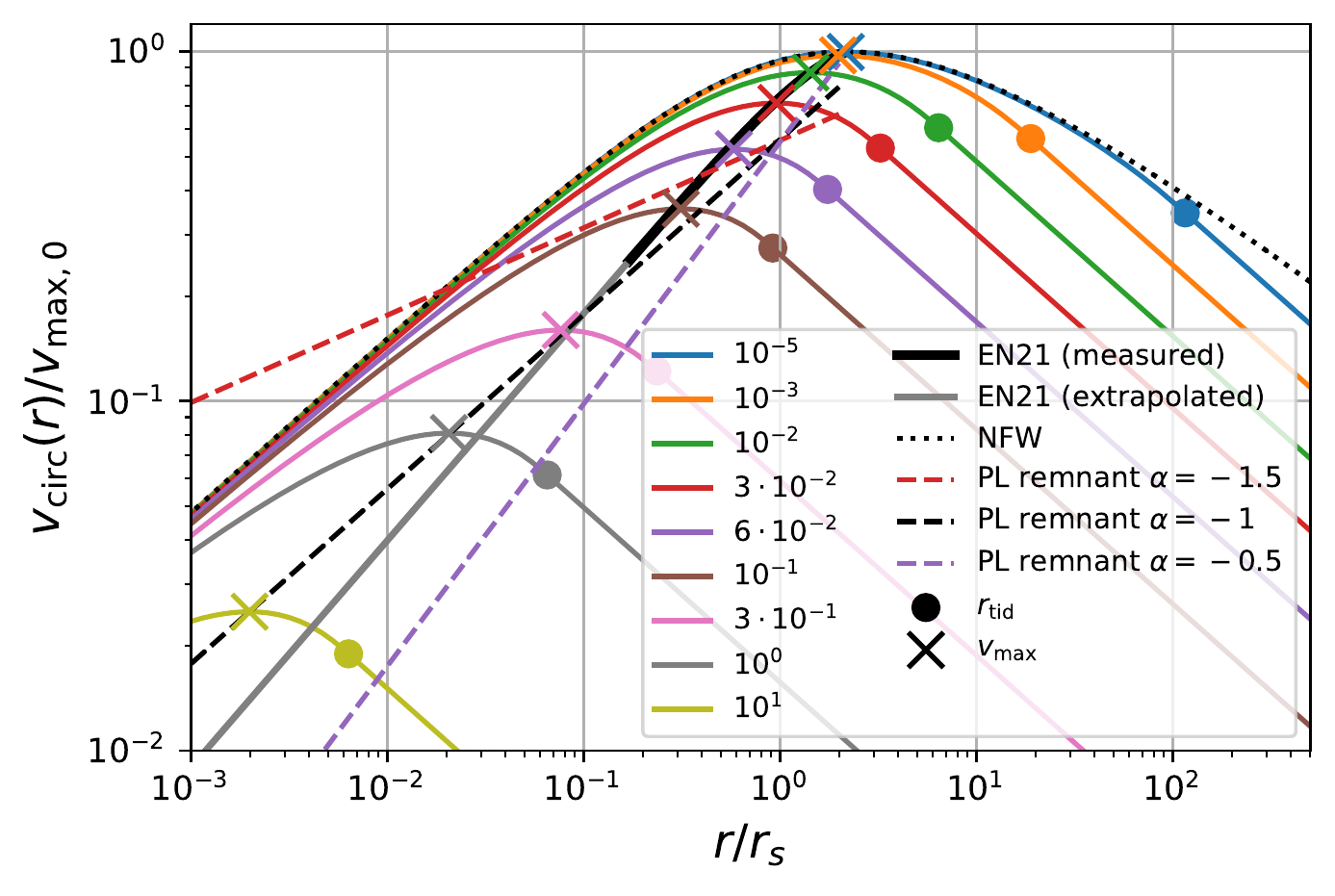}\\
    \includegraphics[width=\columnwidth]{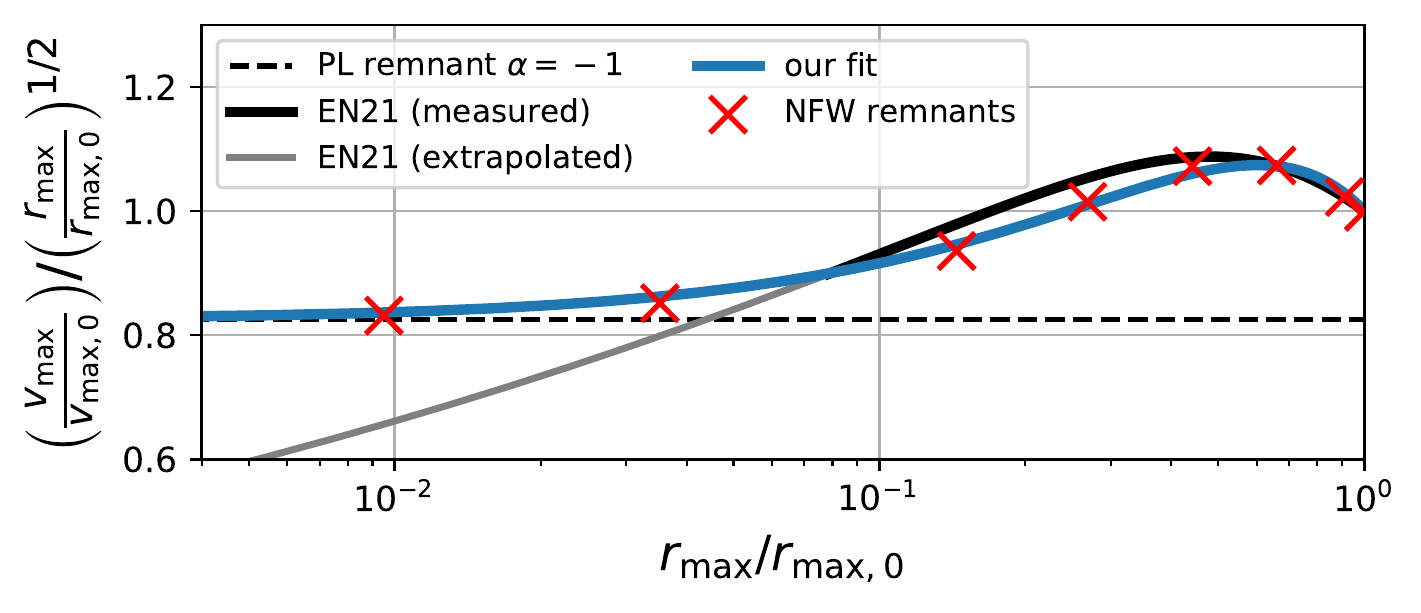}
    \caption{Top: The circular velocity profiles of adiabatically tidally stripped NFW haloes. The crosses mark the maxima of the circular velocity $v_{\rm{max}}$ and the circles mark the tidal radii. For small tidal fields the maximum lies at the same location as for the base NFW, but for very strong tidal fields the final velocity profile can lie drastically below the initial one. The solid black line shows the $v_{\rm{max}}(r_{\rm{max}})$ relation (also known as tidal track) from \citet{errani_2021} where black and grey distinguish between the regime where the relation was measured and where it is an extrapolation. Our adiabatic remnants match the empirical relation very well in the regime where the relation was measured ($v_{\rm{max}} / v_{\rm{max,0}}$ > 0.25), but predict a slightly different asymptotic behaviour in the regime where the relation was extrapolated. Dashed lines indicate remnants of powerlaw profiles with different slopes $\alpha$.
    Bottom: The tidal track and a fit to it in comparison to the relation presented by \citet{errani_2021}. We have divided out the asymptotic slope of 0.5 on the velocity axis here. Our fitting relation appears to be accurate within a few percent and reproduces the correct asymptotic behaviour.}
    \label{fig:vcirc}
\end{figure}

We investigate the circular velocity profile
\begin{align}
    v_{\rm{circ}}(r) &= \sqrt{\frac{G M(r)}{r}} \label{eqn:vcircself}
\end{align}
of adiabaticaly tidally stripped NFW haloes. Note that this definition of the circular velocity uses only the self-gravity of the subhalo, but does not include the potential contribution of the tidal field. In the top panel of Figure~\ref{fig:vcirc} we show the circular velocity profiles for the same examples as from Figure~\ref{fig:radial_suppression}. 

The crosses in Figure~\ref{fig:vcirc} show the radius $r_{\rm{max}}$ and the velocity $v_{\rm{max}}$ of the maximum of the circular velocity profile. As was first found by \citet{Penarrubia_2008} and then later confirmed by other authors \citep{green_2019, errani_2021}, these two follow a scaling relation. This relation has been shown to be relatively independent of the way mass was lost and is therefore independent of the orbital eccentricity, the shape of the host halo etc. A subhalo's position on this `tidal track' is only determined by the fraction of mass that has been lost and a subhalo progressively follows along this line as the tidal stripping progresses.

With our model we can only make clear predictions about the asymptotic limit of orbiting subhaloes, but not about partially stripped subhaloes. However, we can check whether the thus inferred asymptotic remnants are consistent with previous measurements of the tidal track. In Figure~\ref{fig:vcirc} we show as solid black line the parameterization of the tidal track of \citet{errani_2021}:
\begin{align}
    \frac{v_{\rm{max}}}{v_{\rm{max,0}}} = 2^\alpha \left(\frac{r_{\rm{max}}}{r_{\rm{max,0}}} \right)^\beta \left( 1 + \left(\frac{r_{\rm{max}}}{r_{\rm{max,0}}} \right)^2 \right)^{-\alpha} \label{eqn:tidaltrack}
\end{align}
where $v_{\rm{max,0}} = 1.058 \sqrt{G M_s / r_s}$ and $r_{\rm{max,0}} = 2.163 r_s$ and where $\alpha = 0.4$ and $\beta = 0.65$ were inferred by \citet{errani_2021} through a fit which included data down to $v_{\rm{max}}/v_{\rm{max,0}} \gtrsim 0.2$. Note that there are several descriptions of the tidal track in the literature, but this seems to the one that is backed up with the highest resolution simulations up to date \citep{Penarrubia_2008, penarrubia_2010, green_2019}.

We see that the \textsc{adiabatic-tides} model excellently reproduces the empirical tidal track (for $v_{\rm{max}}/v_{\rm{max,0}} \gtrsim 0.2$). This has several interesting implications: This further validates the \textsc{adiabatic-tides} model and shows that much of what is known about tidal stripping can be summarized in this simple picture. Further, this might help explain the existence of the tidal track. As already discussed in Section \ref{sec:structuretide}, the structure-tide degeneracy implies that the parameter space in the adiabatic limit can be reduced to one effective parameter $\lambda/\lambda_{\rm{s}}$. Therefore, also the $v_{\rm{max}}/v_{\rm{max,0}}$ versus $r_{\rm{max}}/r_{\rm{max,0}}$ relation can only be one-dimensional in the adiabatic limit. That also partially disrupted subhaloes follow the same relation has been found empirically in several studies \citep[e.g.][]{Penarrubia_2008,Errani2020}. This might make it possible to apply the \textsc{adiabatic-tides} model even to partially stripped subhaloes at an effective tidal field parameter which would be lower than the pericentre tidal field. This could be done by matching partially evolved subhaloes to the adiabatic model by their bound mass-fraction.

For $v_{\rm{max}}/v_{\rm{max,0}} \ll 0.2$ our models disagree slightly with the tidal track from \citet{errani_2021}. This is outside of the regime where the \citet{errani_2021} tidal track was inferred and we think that equation \eqref{eqn:tidaltrack} gives a poor extrapolation to this regime. Equation \eqref{eqn:tidaltrack} suggests that the asymptotic logarithmic slope for small $r_{\rm{max}}$ is $0.65$, but it is easy to see that this cannot possibly be correct. The scaling relations of the central powerlaw with $\alpha = -1$ (see Appendix \ref{app:powerlaw}) suggest that the limiting slope has to be $0.5$ \citep[see also][]{amorisco_2021}. This is naturally recovered by our model. We also show the predictions for a powerlaw profile with $\alpha = -1$ which is reached for cases with $r_{\rm{max}} < 0.1 r_{\rm{s}}$. Relative to the $\alpha = -1$ tidal track, the NFW tidal track has a slight S-shape behaviour. For reference we show also the (arbitrarily normalized) tidal tracks of powerlaw profiles with $\alpha = -1.5$ and $\alpha = -0.5$. 

We note that our explanation of the tidal track appears to be quite different than the explanation of \citet{benson_2022} who infer it from a tidal heating calculation. However, possibly both calculations are consistent with each other and just approach the problem from different directions. The material that is lost in the adiabatic limit is also very poorly protected from tidal heating, whereas the material that stays in the adiabatic limit is very well protected from heating (see Appendix \ref{app:tidal_heating}). Therefore both approaches may lead to similar results. Our adiabatic limit explanation seems comparatively simpler, since it is a clean prediction without any tunable parameters.

For easy comparison with future studies we fit the adiabatic tidal track through a function of the form
\begin{align}
    \frac{v_{\rm{max}}}{v_{\rm{max,0}}} &= \beta x^{0.5} (1 + \gamma x + (\beta^{-1} - 1 - \gamma) x^{3/2}) \\
    x &= \frac{r_{\rm{max}}}{r_{\rm{max,0}}}
\end{align}
We fix $\beta = 0.8256$ through the requirement of reproducing the behaviour of the powerlaw profile for $x \rightarrow 0$ and find through fitting by eye that $\gamma = 1.5$ describes the tidal track well. We show this function in comparison to the \citet{errani_2021} relation in the bottom panel of Figure~\ref{fig:vcirc}. We find that this functional form describes the tidal track within a few percent accuracy over the whole permitted parameter space $x \in \left[0, 1\right]$.
\subsection{Summary statistics} \label{sec:summarystat}

\begin{figure}
    \centering
    \includegraphics[width=\columnwidth]{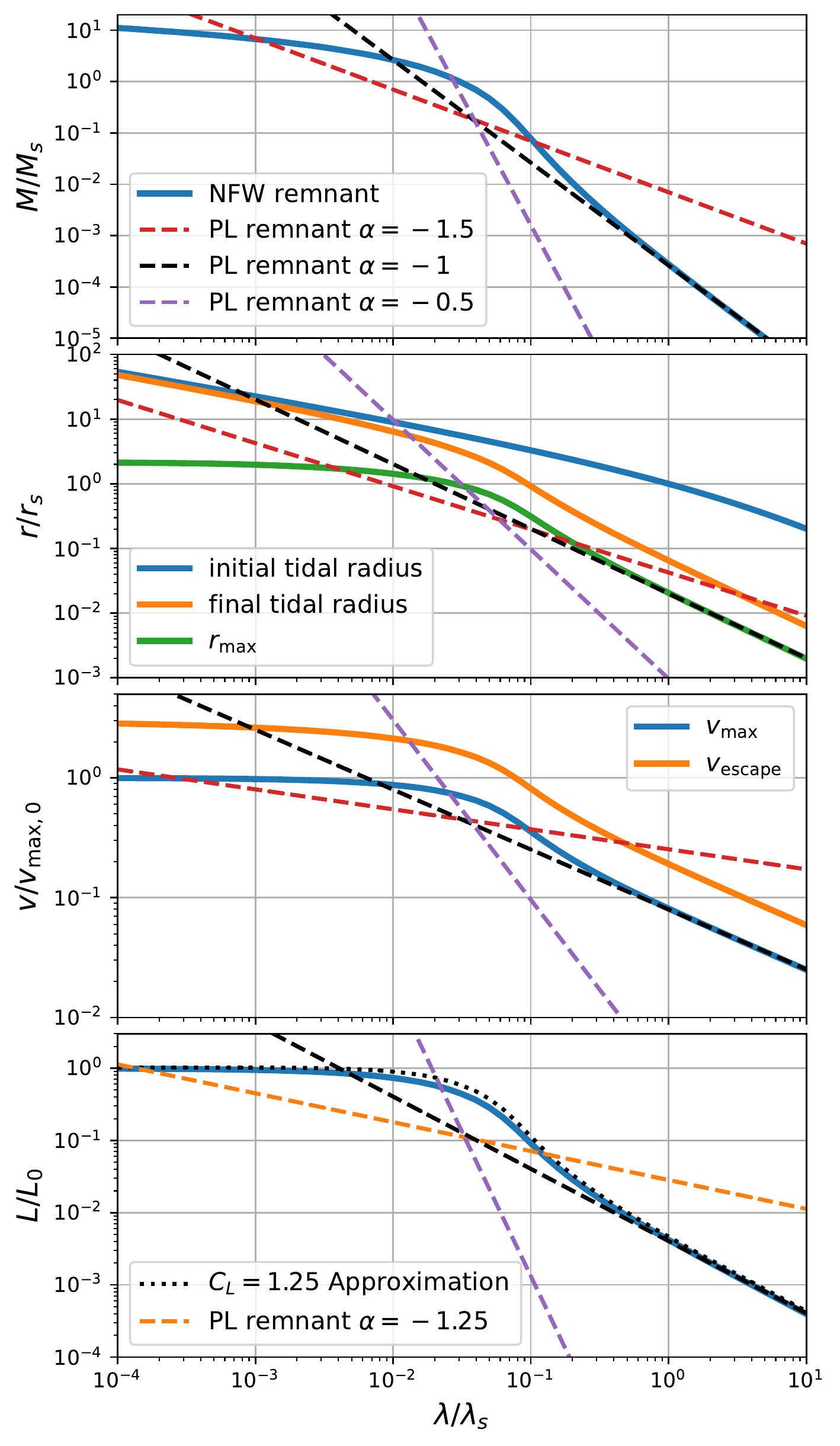}
    \caption{Summary statistics of an NFW halo versus effective strength of the tidal field. Different panels show (1) the mass in units of the scale mass; (2) important radii in units of the scale radius; (3) the escape velocity and the maximum of the maximal circular velocity $v_{\rm{max}}$ in units of the initial value of $v_{\rm{max,0}}$; (4) the annihilation luminosity in units of the initial one. For all statistics the behaviour changes around $\lambda / \lambda_{\rm{s}} \sim 10^{-1}$ when the tidal truncation starts to happen inside of the scale radius. Dashed lines indicated remnants of powerlaw profiles with different slopes $\alpha$. In the second panel we only indicated the powerlaws for $r_{\rm{max}}$ and in the third panel only for $v_{\rm{max}}$. The approximation in the last panel uses equation \eqref{eqn:nfwluminosity} with $C_L = 1.25$.}
    \label{fig:summarystatistics}
\end{figure}

\begin{figure}
    \centering
    \includegraphics[width=\columnwidth]{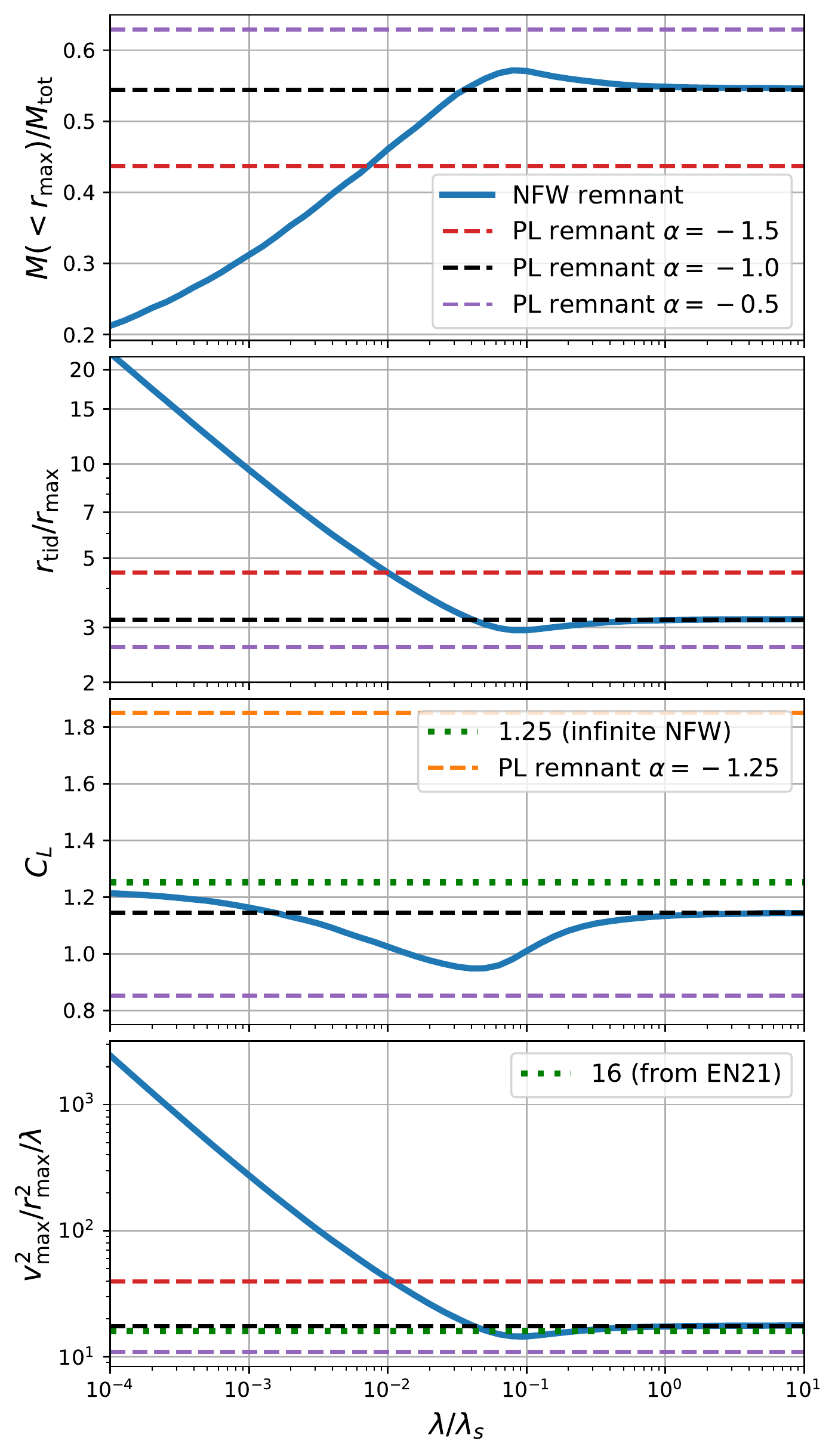}
    \caption{Summary statistics of \textsc{adiabatic-tides} remnants of NFW haloes related to the radius of maximum circular velocity $r_{\rm{max}}$ (blue lines). Dashed lines indicate remnants of powerlaw profiles with different slopes $\alpha$. (1) The mass contained inside the of $r_{\rm{max}}$ as a fraction of the total mass. In the limit of strong tides this approaches $55\%$. (2) The tidal radius in units of $r_{\rm{max}}$ asymptotes to 3.2. (3) The fore-factor of the annihilation luminosity when calculated from scale radius quantities. (4) The tidal ratio as explained in the text. Asymptotically this approaches 17.5 which is close to the value 16 that follows from the measurements of \citet{errani_2021}.}
    \label{fig:summaryvmax}
\end{figure}

In Figure~\ref{fig:summarystatistics} we show several different summary statistics of adiabatically tidally truncated NFW haloes versus the value of the effective tide $\lambda/\lambda_{\rm{s}}$. In each panel we also show powerlaw profiles for comparison. The $\alpha = -1$ profile has an identical density profile to the NFW for $r \rightarrow 0$ and therefore always gives the limit for strong tidal fields. All scaling relations are listed in Appendix \ref{app:powerlaw}.

In the first panel we show the bound mass. As expected, the predicted bound mass reaches for large tidal fields ($\lambda \gtrsim 0.3 \lambda_{\rm{s}}$) the $\alpha=-1$ powerlaw limit. We note that the bound mass might vary drastically for different slopes of the central powerlaw. In the second panel we show different radii of interest. The blue line represents the initial tidal radius, that is the radius at which the saddle-point of the potential can be found when considering the initial unperturbed NFW subhalo plus the tidal field. The orange line represents the final tidal radius corresponding to the saddle-point of the potential after the subhalo as gone through the tidal mass-loss. The green line represents the radius of maximum circular velocity (as in equation \eqref{eqn:vcircself}). If the profile is truncated well outside of the scale radius, the initial and final tidal radius are very close to each other. If it is truncated close to or below the scale radius the final tidal radius can be 1-2 orders of magnitude smaller than the initial one which also reflects itself in the huge amount of mass that is lost as a consequence. In the third panel we show the  maximum of the circular velocity $v_{\rm{max}}$ and the escape velocity $v_{\rm{escape}} = \sqrt{2 \Delta \phi}$ where $\Delta \phi$ is the potential difference between the saddle-point and the minimum of the potential. We indicated the powerlaw cases only for $v_{\rm{max}}$. Both summary  statistics  are indicators of the depth of the potential well and they behave very similar. As a rule of thumb we find $v_{\rm{escape}} \approx 2-3 v_{\rm{max}}$. Both velocities are almost unaffected by weak tidal fields $\lambda \ll 10^{-2} \lambda_{\rm{s}}$, but decrease strongly for $\lambda \gg 10^{-1} \lambda_{\rm{s}}$. 

In the last panel of Figure~\ref{fig:summarystatistics} we estimate the ratio between final and initial dark matter self-annihilation luminosity $L$. Here, we have assumed that the luminosity depends only on the density
\begin{align}
   L &\propto \int_0^\infty 4 \pi r^2 \rho(r)^2 \rm{d}r \label{eqn:luminosity}
\end{align}
and that we can neglect any velocity dependent factors in the self-annihilation calculation. The decrease in annihilation luminosity due to the tidal field is weaker than the decrease in mass, since it is more sensitive to the central part of the halo that is more resilient to the tidal field compared to the outskirts.

For an NFW halo it holds
\begin{align}
    L_{\rm{NFW}} 
      \approx C_{L} \frac{v_{\rm{max}}^4}{G^2 r_{\rm{max}}}. \label{eqn:nfwluminosity}
\end{align}
with $C_{L, \rm{NFW}} \approx 1.25$. We find that equation \eqref{eqn:nfwluminosity} even holds approximately for tidally truncated NFW haloes if the $r_{\rm{max}}$ and the $v_{\rm{max}}$ values are measured in the truncated profile. We show this approximation as a dotted black line in the bottom panel of Figure~\ref{fig:summarystatistics}. We find that this approximation overestimates the annihilation luminosity in the worst case by about 35 $\%$ at around $4 \cdot 10^{-2} \lambda_{\rm{s}}$ and by about $10\%$ in the asymptotic case $\lambda \rightarrow \infty$. We note that this approximation is conceptually the same as the one used by \citet{grand_white_2021}, but for an NFW profile instead of the Einasto profile ($C_{L, \rm{Einasto}} \approx 1.87$). We suggest that equation (1) of \citet{grand_white_2021} or our equation \eqref{eqn:nfwluminosity} are good approaches to approximate the luminosity of tidally stripped subhaloes in cases where the inner profile cannot be resolved well enough for direct integration of the squared density. This holds even if the subhalo profile has been strongly modified through the tidal evolution. We note that the annihilation luminosity depends dramatically on the central slope of the profile. If the profile had for example a central slope of $\alpha = -1.25$, the annihilation luminosity might be much larger overall and would depend much weaker on the effects of tides. For a slope of $\alpha = -1.5$ the luminosity is even divergent which is indeed why we did not plot the luminosity for this case. Precise predictions of the central slope might be crucial for reliable annihilation radiation estimates (consider e.g. \citealt{angulo_2017} versus \citealt{wang_2020}).

In the literature many of these quantities are stated relative to the radius of maximum circular velocity $r_{\rm{max}}$ at a given time. In Figure~\ref{fig:summaryvmax} we present several such relations. The first panel shows the fraction of the total mass of the subhalo that is contained within $r_{\rm{max}}$. This approaches the $\alpha=-1$ powerlaw estimate of $55\%$ in the regime of heavy mass loss. The second panel shows the ratio between tidal radius and $r_{\rm{max}}$. This ratio approaches the powerlaw value of $3.2$ in the strong mass loss limit. This means that for subhaloes in the strong mass loss regime all mass beyond $3.2 r_{\rm{max}}$ can be lost in the long run, and only mass inside $3.2 r_{\rm{max}}$ may be safe from tidal stripping. The third panel shows the dimensionless fore-factor of the annihilation luminosity 
\begin{align}
    C_L = \frac{L G^2 r_{\rm{max}}}{v_{\rm{max}}^4}
\end{align}
(compare equation \eqref{eqn:nfwluminosity}). This factor changes surprisingly little no matter the strength of the tidal field. Understandably the NFW fiducial value of $1.25$ over-predicts the annihilation luminosity, but it does so in the worst case only by about $30\%$. The $\alpha=-1$ powerlaw limit is $1.144$ for this case which is quite close to the NFW value.

Finally we present the tidal ratio which we define as
\begin{align}
    \alpha_{\rm{max}} &= \frac{\partial_r \phi (r_{\rm{max}}) / r_{\rm{max}} }{\lambda} \\
           &= \frac{v_{\rm{max}}^2 / r_{\rm{max}}^2}{\lambda} \label{eqn:tidalratio}
\end{align}
in the last panel of Figure~\ref{fig:summaryvmax}. This ratio compares the attractive force at the $r_{\rm{max}}$ radius to the repulsive tidal field. As can be seen in Figure~\ref{fig:summaryvmax}, this ratio approaches the constant (powerlaw) value of $17.5$ in the heavy mass loss regime. Note that this never gets smaller than $14.5$ -- therefore any stable remnant needs to have $\alpha_{\rm{max}} \geq 14.5$. If a subhalo was exposed to a strong tidal field causing $\alpha_{\rm{max}} \ll 15$ then it would lose mass at least until $\alpha \sim 15$ is established (or an even higher value, when in the weak mass-loss regime).

This prediction is directly supported by the measurements of \citet{errani_2021} and can be seen as a generalization of the orbital time relation that they found. They found that in an isothermal sphere host potential a subhalo becomes stable when the circular orbit time $T_{\rm{max}}$ at $r_{\rm{max}}$ is approximately $T_{\rm{max}} = 0.25 T_{\rm{p}}$, where $T_{\rm{p}}$ is the circular orbit time at the pericentre of the subhalo's orbit. However, it seems rather surprising that there should exist such a precise relation between the circular orbit frequencies, especially for the case of non-circular orbits. A subhalo on a non-circular orbit never experiences directly the value of the pericentre circular orbit frequency\footnote{The value of this frequency is only experienced approximately and indirectly, as it correlates with the time-scales on which the tidal field can vary.}. However, we note that in general circular orbital frequencies are quite correlated with the amplitude of tidal fields, but the precise relation between the two depends on the precise host potential. The isothermal sphere is a special case where the tidal field is directly proportional to the mean density inside a given radius $\lambda(r) = 4 \pi G \overline{\rho}(r) / 3 $. Since the circular orbit time is $T = \sqrt{3 \pi / G \overline{\rho}(r)}$, for an isothermal sphere the special relation \rev{$\lambda = 4 \pi^2 T^{-2}$} \revcom{corrected from $4 \pi T^{-2}$} holds. Therefore, we think that the relation $T_{\rm{max}} = 0.25 T_{\rm{p}}$ is a coincidence for the isothermal sphere and the correct generalization uses the tidal field. Using $T_{\rm{max}} = 2 \pi r_{\rm{max}} / v_{\rm{max}}$ the relation of \citet{errani_2021} implies equation \eqref{eqn:tidalratio} with an asymptotic value of $\alpha_{\rm{max}} \approx 16$. We indicated this implied limit as a dashed line in the bottom panel of Figure~\ref{fig:summaryvmax}. That this limit is so close to our model prediction can be seen as an additional validation.

We note that all quantities in Figure~\ref{fig:summaryvmax} have a very similar behaviour when expressed in units of $r_{\rm{max}}$ and $v_{\rm{max}}$. This is even so for powerlaws with very different slopes. The only exception here is $C_L$ for powerlaws with $\alpha \leq -1.5$ which is not well defined. Therefore, we confirm that defining stripped subhaloes in terms of those quantities, as is common practice in the cosmological community, is very robust. Further, we note that measurements of the quantities as in Figure~\ref{fig:summaryvmax} might be helpful to determine whether subhaloes are in the regime of weak or strong mass loss and possibly to infer their effective central powerlaw slope.

In principle, Figure~\ref{fig:summarystatistics} includes all the relevant information about the disruption of subhaloes in the adiabatic limit. Considerations of different subhalo concentrations, host halo concentrations, baryonic potentials and orbit configurations  all can be summarized into the value of $\lambda / \lambda_{\rm{s}}$. However, to provide some intuition to the reader, how this works in practice, we will show in the following subsections how these simple relations get warped when they get mapped on more commonly used parameter spaces. 

\subsection{Subhalo mass loss}

\begin{figure}
    \centering
    \includegraphics[width=\columnwidth]{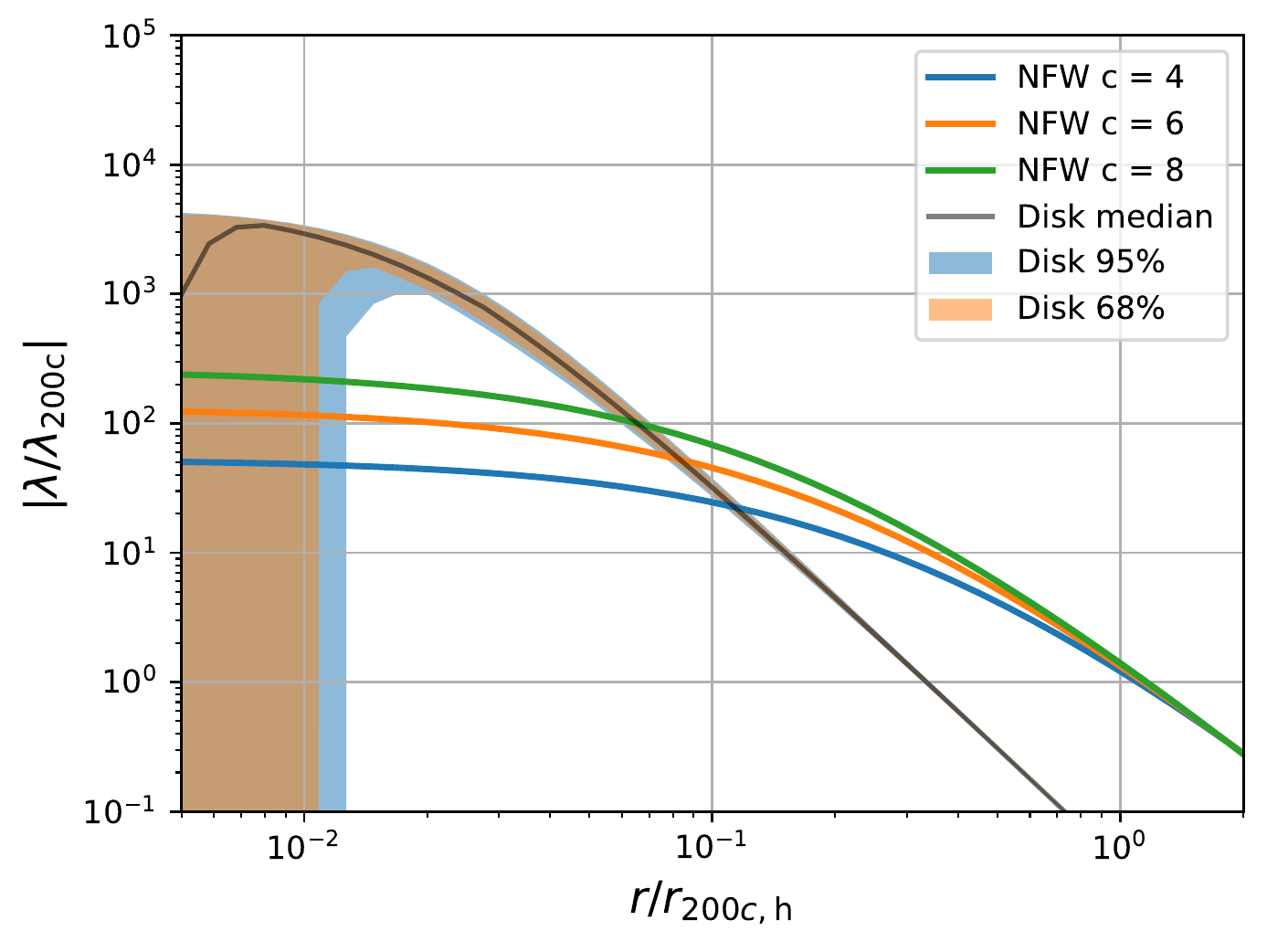}
    \caption{Tidal fields of a Milky Way-like host halo. The blue, orange and green lines show the tidal field of an NFW halo with different concentrations. The shaded regions and the solid line show the percentiles of the distribution of the tidal field that is sourced by a galactic disk potential. In the inner parts of the halo $r < 0.1 \rvir$ the disk dominates the tidal field.}
    \label{fig:host_tides}
\end{figure}

We want to calculate the asymptotic orbit-concentration-massloss relation for a subhalo that is orbiting in the potential of a Milky Way-like host. The relation will depend on the potential that is assumed for the host halo. For this we consider two different cases: The first case is a pure NFW halo with mass $M_{\rm{200c,h}} = 10^{12} \rm{M}_{\odot}$ and concentration $c_{\rm{h}}=6$. For reference we show the tidal field as a function of radius for three different concentrations $c_{\rm{h}} = 4, c_{\rm{h}}=6$ and $c_{\rm{h}} = 8$ in Figure~\ref{fig:host_tides}. Note that we only show the largest eigenvalue of the tidal field here.

However, it has frequently been argued that a baryonic component strongly modifies the tidal field close to the centre of a halo and strongly enhances tidal mass loss \citep[e.g.][]{sawala_2017, garrison_2017, kelley_2019, richings_2020}. Therefore, we additionally consider the effect of a disk component, where the disk is modeled through a Miyamoto Nagai potential \citep{Miyamoto_1975}
\begin{align}
    \phi(R,z) = \frac{-G M_{\rm{d}} }{\sqrt{R^2 + \left(a + \sqrt{z^2 + b^2} \right)^2}}
\end{align}
with parameters $M_{\rm{d}} = 2 \cdot 10^{10} \rm{M}_{\odot}$, $b=\SI{300}{\parsec}$ and $a=\SI{3}{\kilo\parsec}$. Here $z$ is the $z$-coordinate and $R = \sqrt{x^2 + y^2}$ is the distance from the symmetry-axis ($z$-axis). Since at each radius $r$ there can be different values of the tidal field, depending on the direction, we consider the distribution $p_r(\lambda)$ when randomly sampling directions on the unit sphere while fixing the radius $r$. We show the median and the percentiles of this distribution of the largest eigenvalue of the tidal field in Figure~\ref{fig:host_tides}. We can see that for $r \ll 0.1 \rvir \approx \SI{20}{\kilo \parsec}$ the tidal field of the disk is significantly larger than that of the dark matter halo. Therefore, we expect that the mass loss may be dramatically different when considering or not considering the disk in the potential for orbits with $r_{\rm{p}} < 0.1 \rhvir$. 

\begin{figure*}
    \centering
    \includegraphics[width=\columnwidth]{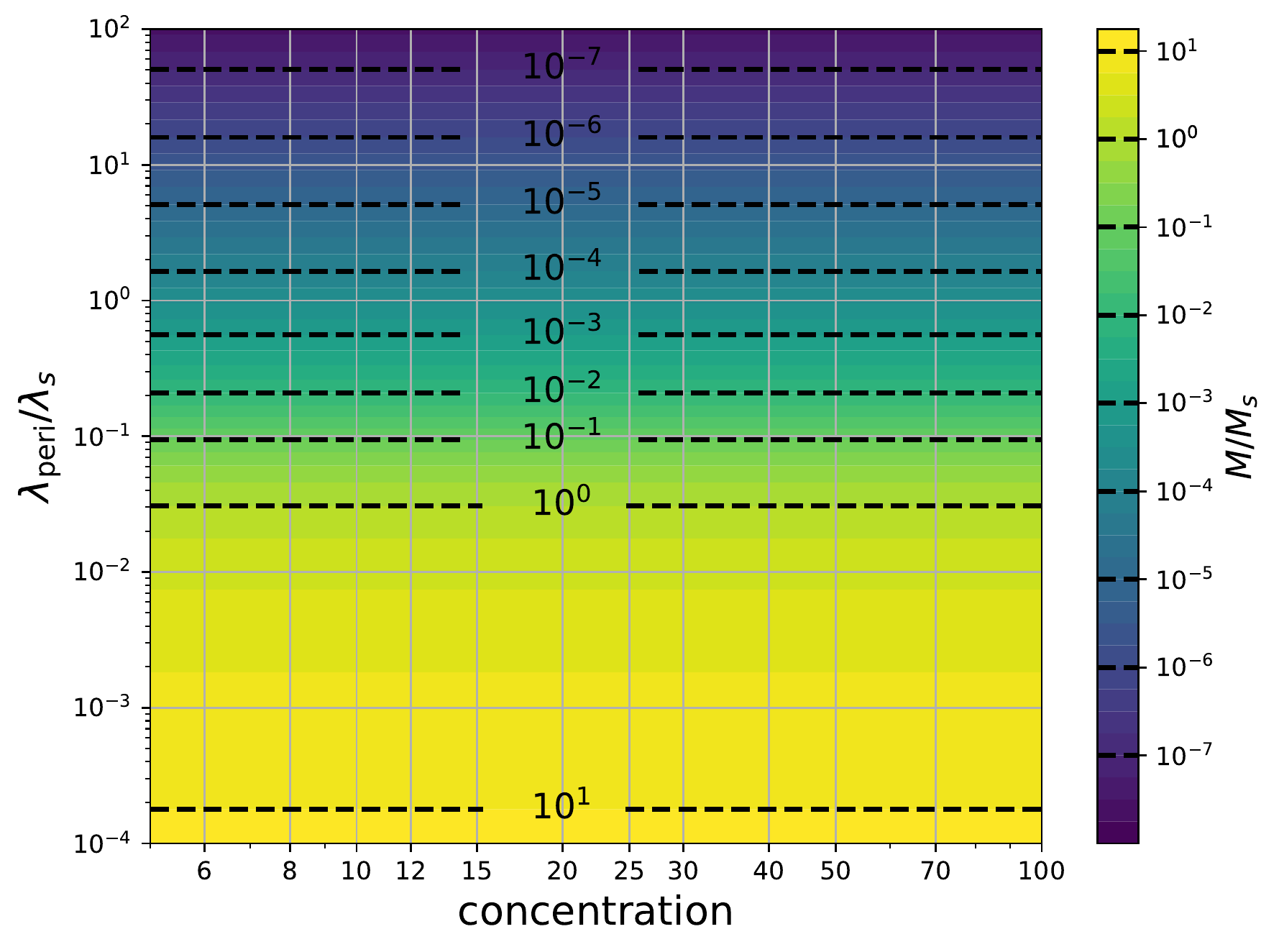}
    \includegraphics[width=\columnwidth]{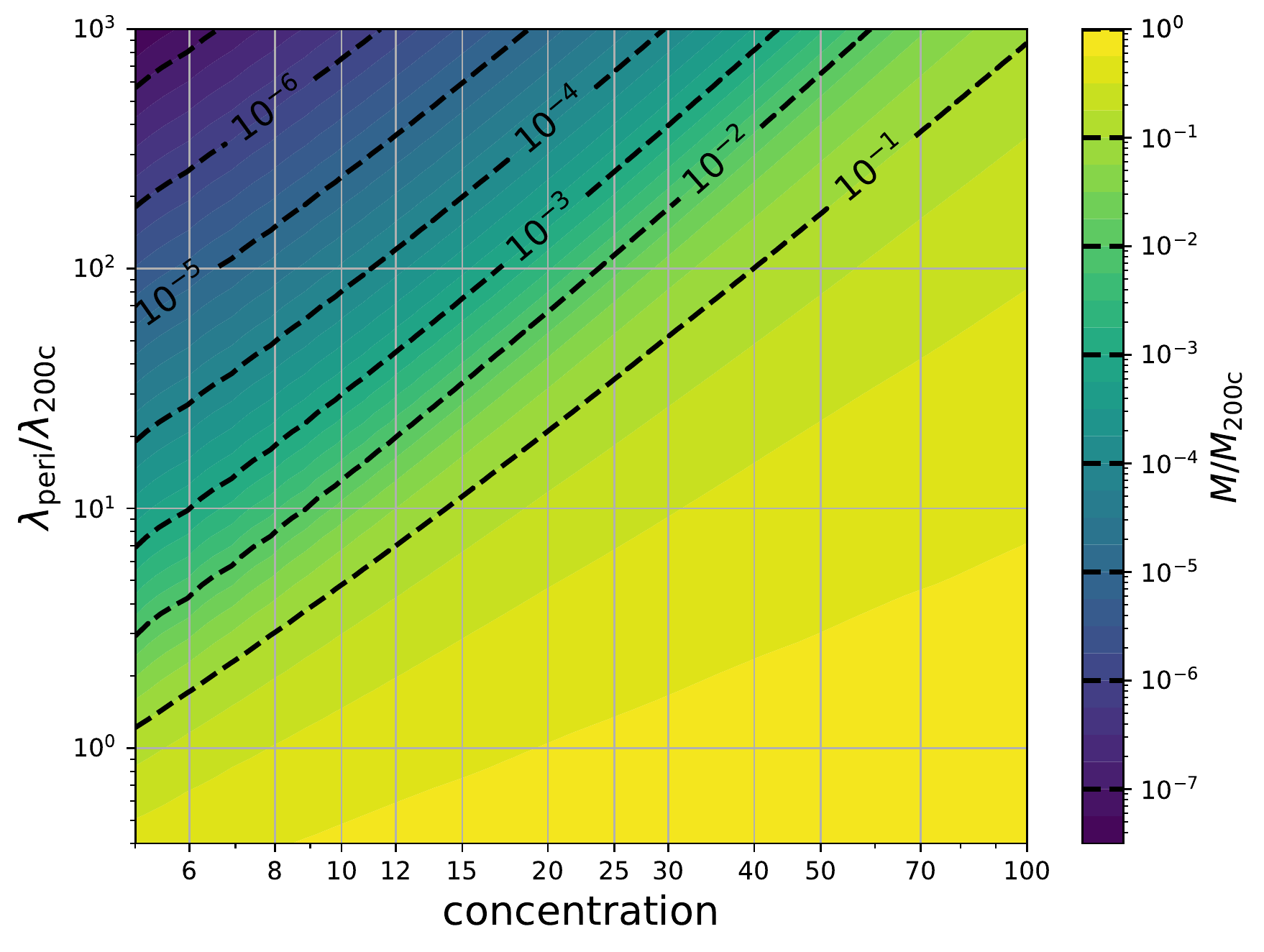}\\
    \includegraphics[width=\columnwidth]{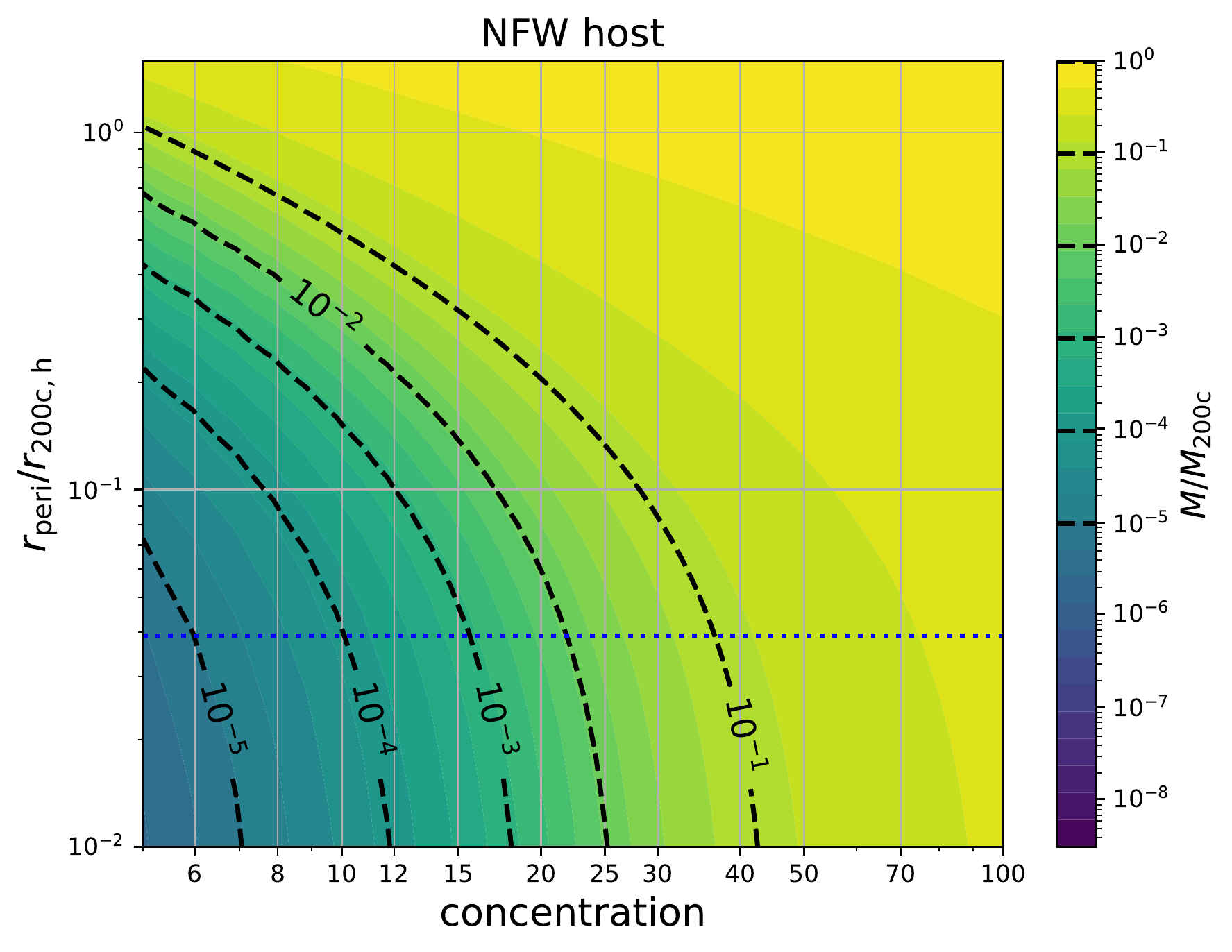}
    \includegraphics[width=\columnwidth]{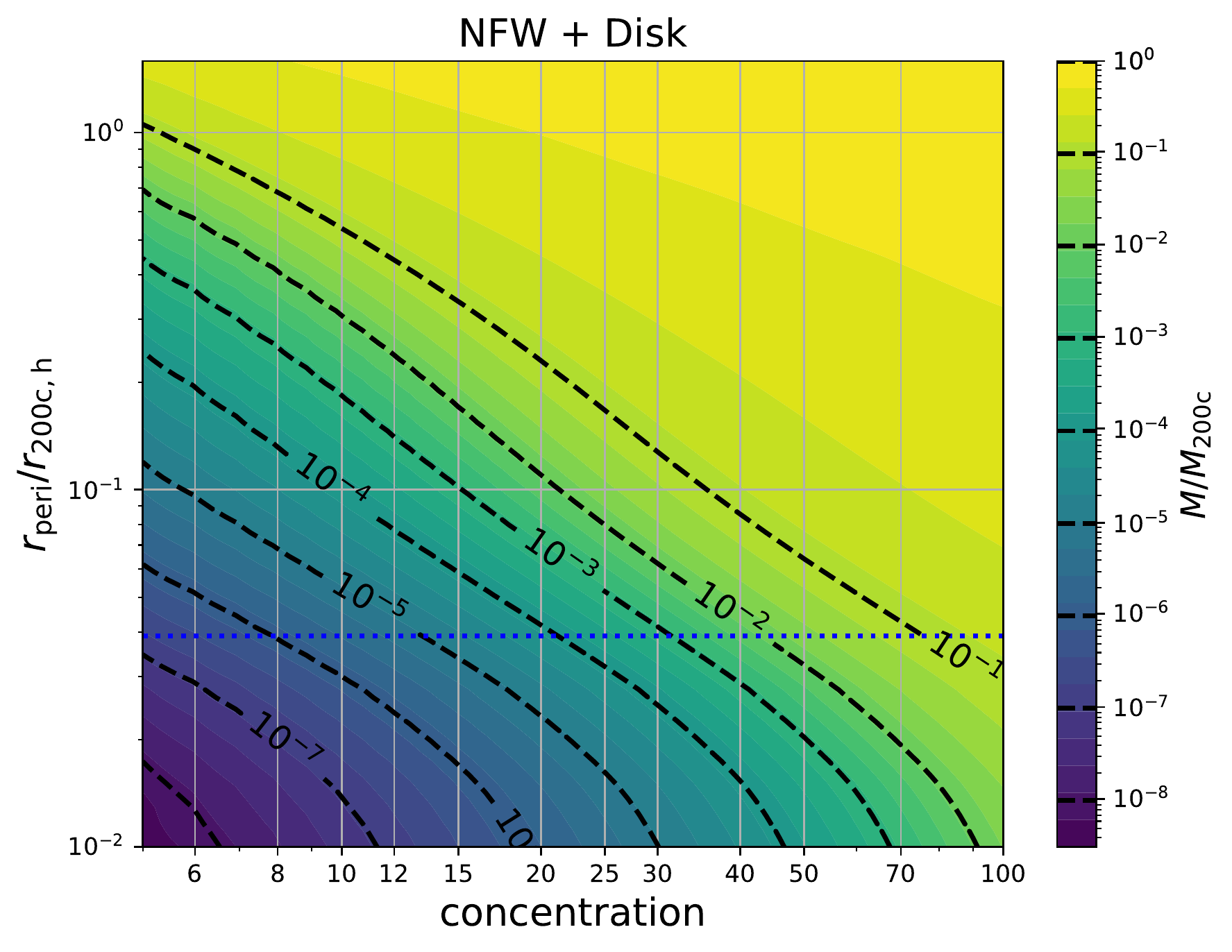}
    \caption{Different ways of visualizing the concentration-tidal field dependence of subhalo mass loss. Top-left: using the mass and the tidal field in units of the scale radius quantities. Presented like this, the concentration dependence disappears due to the structure-tide degeneracy. Top-right: The same plot, but using units of the virial mass and the virial tide. Bottom left: showing on the y-axis the radius in a Milky Way NFW potential needed to reach the tidal field, bottom right: same, but including a baryonic disk component. The blue dotted line indicates the solar radius in the Milky Way. The bottom two panels can be understood as the mass-fraction that is expected to be asymptotically bound for subhaloes orbiting with a given pericentre. We can see dramatic differences between the case with and without disk component for $r < 0.1 \rvir$ and we find overall a very strong concentration dependence.}
    \label{fig:halomassloss}
\end{figure*}

In Figure~\ref{fig:halomassloss} we visualize the asymptotic tidal field -- concentration -- mass loss relation in four different ways: In the top-left panel we show the mass loss in units of the scale mass versus the concentration and the effective tidal field. As discussed previously, if visualized in this manner, the concentration dependence disappears completely in the adiabatic limit. We would also expect that much of the concentration dependence might disappear if results from numerical studies would be presented in this way. We test this proposition \rev{in} \citet{aguirre_2023}.

In the top-right panel we show the same relation, but measuring the mass in units the virial mass and measuring the tidal field in the same units of $\lvir$ for every halo. In this case we can already see a significant concentration dependence which arises from higher concentration haloes responding weaker to the same value of the tidal field and having more mass in units of the scale mass.

Next we show the relation in the bottom left panel of Figure~\ref{fig:halomassloss} when drawing on the y-axis instead of the value of the tidal field, the radius that is needed in the Milky Way NFW potential to reach the corresponding value of the tidal field, as can be read off in Figure~\ref{fig:host_tides}. Finally, we show in the bottom right panel of Figure~\ref{fig:halomassloss} the same relation, but when using the median tidal field at a given radius of the sum of NFW and disk potential. Note that we estimate in Appendix \ref{app:tidal_heating} that the energy which can be injected by 'disk shocking' \citep[e.g.][]{spitzer_1958} to the particles inside the \textsc{adiabatic-tides} remnant is limited and we expect the estimates here to still be a lower limit in practice, even though energy redistribution may enhance the maximum possible amount of mass loss.

As we have argued in Section \ref{sec:orbiting_sim}, the predictions of the \textsc{adiabatic-tides} model can be understood as the asymptotic limit of orbiting subhaloes after $10-20$ orbits, if the pericentre radius is considered for the tidal field. Therefore, the bottom two panels can be understood as such asymptotic predictions for subhaloes that have orbits with a given pericentre. We note that in realistic scenarios, most haloes will not have orbited for so many periods. For those cases the adiabatic limit is a lower limit of their mass $M \geq M_{\rm{ad}}$ \citep[see][]{aguirre_2023}.

When comparing the predictions for the cases with and without disk we note that for $r < 0.1 \rvir$ they are drastically different. For example, a subhalo with concentration $c=10$ and pericentre $r=0.03 \rvir$ should keep about $10^{-4}$ of its initial virial mass, whereas it will go down to only $10^{-6}$ of its initial mass in the case with a disk. It is probably very difficult to resolve such a halo numerically. We suggest that most authors that found complete disruption of dark matter haloes have measured such a very steep limit. However, it is questionable whether it actually matters whether a subhalo completely disrupts or survives with $10^{-6}$ of its initial mass.

\section{Evaluation of Literature Results} \label{sec:literature}
We have seen that the \textsc{adiabatic-tides} model is a physically well motivated model that is consistent with known empirical relations about the tidal stripping of subhaloes. Since all of these results were pure predictions of the model and none of them were fitted, we think the \textsc{adiabatic-tides} model might be the most reliable way of extrapolating resolved simulation results to regimes that are numerically difficult to resolve.

In this section we compare the \textsc{adiabatic-tides} model to other models in the literature and check whether previously made extrapolations appear reasonable.

\subsection{Annihilation luminosities of Milky Way satellites}
One of the most promising ways of detecting dark matter is through its self-annihilation signal. If dark matter has a significant self-annihilation cross-section, we might expect a self-annihilation signal from two different components -- that is from the Milky Way's main dark matter halo and from Milky Way satellites (which might be dark). There have been many studies trying to evaluate which of the components is most relevant and whether the largest flux contributions come from the smallest ($M \sim 10^{-5} \rm{M}_\odot$) or the largest satellites. We will discuss only the measurements and extrapolations of \citet{grand_white_2021} here, since these are based on state-of-the-art baryonic simulations and use the information from most recent measurements of extremely low-mass haloes \citep{wang_2020} for the extrapolation of their results. 

Here, we can test two of the key assumptions that \citet{grand_white_2021} used for extrapolation of their simulation results from resolved subhaloes ($v_{\rm{max}} \geq \SI{10}{\kilo\metre\per\second}$) to unresolved ones. The first assumption is that at any mass, concentration and degree of tidal stripping equation \eqref{eqn:nfwluminosity} (with $C_L = 1.87$ for Einasto profiles in their case) is a good approximation to the annihilation luminosity of a subhalo. We have already shown in Section \ref{sec:summarystat} that this is indeed a robust approximation. For the NFW profile the approximation overpredicts the annihilation by about $35\%$ in the worst case and we'd expect that the behaviour will be similar for the Einasto profile.

The second assumption that we can test regards the extrapolations that \citet{grand_white_2021} assume for the Milky Way subhaloes. Their extrapolation consists in two parts, the first one being that the ratio between the number densities of subhaloes $n_{\rm{sub}} (v_{\rm{max}})$ and field haloes $n_{\rm{field}} (v_{\rm{max}})$ at a given value of $v_{\rm{max}}$ is approximately the same in the resolved regime as in the regime of much smaller unresolved haloes
\begin{align}
   \frac{n_{\rm{sub}} (v_{\rm{max,1}})}{n_{\rm{field}} (v_{\rm{max,1}})} &= \frac{n_{\rm{sub}} (v_{\rm{max,2}})}{n_{\rm{field}} (v_{\rm{max,2}})}
\end{align}
where \citet{grand_white_2021} use the description of $n_{\rm{field}}$ from \citet{angulo_2012}. The second part of the extrapolation assumes that the $v_{\rm{max}}$ -- $r_{\rm{max}}$ relation for subhaloes can be approximated by taking the $v_{\rm{max}}$ -- $r_{\rm{max}}$ relation for field haloes from \citet{wang_2020} and additionally shifting it logarithmically to match the low mass resolved subhaloes of the Auriga simulations. 

Both of these extrapolation assumptions subsume that field haloes and subhaloes relate in a similar way in the resolved regime as in the unresolved regime at much smaller masses. \citet{grand_white_2021} argue that typical concentrations change only weakly by about a factor 1.5 from the resolved regime $M \sim 10^{8} \rm{M}_\odot$ ($c \sim 18$) to the much lower mass unresolved regime (peaking at $M \sim 1 \rm{M}_\odot$, $c \sim 27$), and therefore argue that the effect of tidal stripping should be similar in the resolved and the unresolved regime.

With the \textsc{adiabatic-tides} model we can test this assumption. Instead of directly mimicking the effects of this concentration dependence onto the $n_{\rm{sub}}(v_{\rm{max}}) / n_{\rm{field}}(v_{\rm{max}})$ ratio and the $v_{\rm{max}}$ -- $r_{\rm{max}}$ relation, we try to make a simpler estimate of the effect of the concentration dependence of tidal strippping onto subhalo luminosities. We estimate by how much the ratio between the luminosity of a typical stripped subhalo $L$ and its initial luminosity at infall $L_0$ would change when the initial concentration (at infall) is increased by a factor $1.5$. An increase in concentration will on the one hand increase the initial luminosity $L_0$ -- which is accounted for in the calculations of \citet{grand_white_2021} -- but it will also additionally increase the ratio $L/L_0$ by making the subhalo more resilient to tidal stripping. However, the assumptions of \citet{grand_white_2021} should approximately imply $L / L_0 \sim \rm{constant}$, since neither the subhalo to halo ratio or the constant shift in the $v_{\rm{max}} - r_{\rm{max}}$ relation can account for the concentration dependence of the stripping process. 

As we have argued in Section \ref{sec:structuretide}, a change in concentration can also be interpreted as a change in the effective tide $\lambda/\lambda_{\rm{s}}$ where approximately $\lambda_{\rm{s}} \propto c^3$. Therefore a factor 1.5 in concentration reduces the effective tide $\lambda / \lambda_{\rm{s}}$ by about a factor $3.4$. From Figure~4 in \citet{grand_white_2021} we infer that typical subhaloes in the baryonic case have gone down by about a factor 1.5 - 2 in $v_{\rm{max}}$ (compared to field haloes). Subhaloes that are close to the solar radius have likely gone down by a larger factor. Therefore, it seems reasonable to assume that the relevant baryonic subhaloes from \citet{grand_white_2021} are at least in the regime of intermediate mass loss $\lambda / \lambda_{\rm{s}} \sim 0.05 - 0.1$ (compare Figure~\ref{fig:summarystatistics}). 

We find that a decrease in effective tide by a factor of $3.4$ would cause an increase in $L / L_0$ by a factor of 5 at $\lambda / \lambda_{\rm{s}} = 10^{-1}$. For different starting values of the effective tide $\lambda / \lambda_{\rm{s}}$ of $0.5 \cdot 10^{-1}$ and $2 \cdot 10^{-1}$ we find increases in luminosity by factors of $2.3$ and $7.9$ respectively. Therefore, we argue that the extrapolation used by \citet{grand_white_2021} underestimates the luminosities of typical subhaloes with mass $M \sim \rm{M}_{\odot}$ by a factor between $2-5$ and may underestimate the luminosities of the most stripped subhaloes around the solar radius by up to a factor of $8$. However, we note that the factor that we have estimated here holds only for the differential contribution of the highest concentration haloes at $M \sim \rm{M}_{\odot}$. Haloes with smaller or larger mass have lower concentrations if dark matter is a weakly interacting massive particle \citep[see][]{wang_2020}. The correction to the integrated contribution of all sub-resolution subhaloes is likely lower. In principle it should be possible to improve the quantitative extrapolations of \citet{grand_white_2021} by modelling sub-resolution haloes as shifted versions of resolved haloes where the shift can be estimated from the \textsc{adiabatic-tides} model. However, even if we assume that the total luminosity of the subhalo component in \citet{grand_white_2021} would increase by about a factor of $4$, the qualitative conclusions of \citet{grand_white_2021} would still be correct -- i.e. the smooth halo component would still be by far dominant over the annihilation luminosity of all subhaloes combined and the annihilation flux from the galactic centre should still exceed the brightest subhalo by several orders of magnitude.

\subsection{Comparison with energy truncation models}
Several authors have argued that tidal stripping can be understood as a successive peeling in energy space \citep[e.g.][]{drakos_2017, drakos_2020, stuecker_2021_bp, amorisco_2021}. Based on this idea \citet{amorisco_2021} (later \citetalias{amorisco_2021}) has proposed a model for tidal remnants that fits a very similar purpose as the \textsc{adiabatic-tides} model. We want to briefly compare the two models here in (1) their assumptions and (2) their predictions. Note that the \citetalias{amorisco_2021} model is not the only energy-truncation model \citep[see also][]{drakos_2017, drakos_2020}, but it is the most advanced one and therefore we limit ourselves to a comparison with the \citetalias{amorisco_2021} model. 

\citetalias{amorisco_2021} proposes a model where all particles of an NFW halo with energy beyond some truncation energy level are removed. Since such a truncated NFW is not in dynamical equilibrium \citetalias{amorisco_2021}, runs an N-body simulation of the truncated remnant to model the revirialization process of the remnant. \citetalias{amorisco_2021} argues that such an energy-truncated + revirialized remnant poses a good model for a tidally stripped subhalo. \rev{This allows} \citetalias{amorisco_2021} \rev{to infer the tidal track as well as additional scaling relations, for example between the circular velocity at the initial truncation radius and the final value of $v_{\rm{max}}$.} 

We note a few differences between the \citetalias{amorisco_2021} model and the \textsc{adiabatic-tides} model here: (1) \citetalias{amorisco_2021} assumes a sharp energy truncation whereas the \textsc{adiabatic-tides} model assumes only a tidal field and finds sharp energy truncation as a predicted result (compare Section \ref{sec:atides_sim}). (2) The \citetalias{amorisco_2021} model requires running a simulation to model the revirialization process whereas energy-redistribution is a built-in feature of the \textsc{adiabatic-tides} model. \rev{(3) Due to the necessary discretization to particles the} \citetalias{amorisco_2021}  \rev{model exhibits moderate discreteness effects (e.g. limiting the radially viable profile range) whereas such effects are practically negligible in our case\footnote{We show in extra material in the online repository that $\Delta \rho / \rho < 1\%$ over more than 10 orders of magnitude in spatial scale.}.} (4) The \citetalias{amorisco_2021} model follows the revirialization process in the absence of a tidal field whereas revirialization happens in the presence of a tidal field in the \textsc{adiabatic-tides} model. (5) The \textsc{adiabatic-tides} model makes clear predictions at a given value of the (pericentre-) tidal field.  The \citetalias{amorisco_2021} model requires additional assumptions to match the truncation energy parameter to the tidal field present in a realistic scenario. We think that each of these aspects show that the \textsc{adiabatic-tides} model is simpler and more predictive from a theoretical perspective and additionally easier to apply from a practical perspective. 

When comparing the predictions of the \citetalias{amorisco_2021} and the \textsc{adiabatic-tides} model, we find that most qualitative predictions of \citetalias{amorisco_2021} are consistent with the ones we find from our model. These include the following: (1) NFW subhaloes cannot be completely disrupted. (2) They approach the initial density profile in the centre $\rho / \rho_{\rm{NFW}} \rightarrow 1$ for $r \rightarrow 0$. (3) Asymptotic remnants follow the tidal track consistent with the one measured by \citet{errani_2021} for $v_{\rm{max}} / v_{\rm{max,0}} > 0.2$. (4) The tidal track approaches a slope of $v_{\rm{max}} / v_{\rm{max,0}} \propto (r_{\rm{max}} / r_{\rm{max,0}})^{0.5}$ in the limit of extreme disruption. One important side-node here is that while we agree with the prediction of \citetalias{amorisco_2021} that NFW subhaloes cannot be disrupted, we think that \citetalias{amorisco_2021} presents an incomplete proof. \citetalias{amorisco_2021}'s argumentation includes the assumption that tidal fields will always truncate the subhalo at a finite energy level that is larger than the central potential. In principle, it could have been that the tidal stripping process is a runaway process, i.e. that whenever a remnant is exposed to the same tidal field again it will lose more mass again, never reaching a stable result. It is easy to show that a single truncating iteration as in Figure~\ref{fig:iteration_convergence} always leaves a finite remnant, but it is a non-trivial result that the whole iterative procedure converges to a finite remnant.

Since the \textsc{adiabatic-tides} model also allowed us to make several additional predictions which would not be possible with energy truncation models, we think that it can be considered as an improvement over energy-truncation approaches.

\section{Conclusions} \label{sec:conclusions}
In this article we have introduced the \textsc{adiabatic-tides} model which describes the remnant of a halo that was exposed to a slowly increasing isotropic tidal field. We argue that this is the simplest self-consistent model of tidal mass loss that accounts for the tidal stripping and the internal energy redistribution of the halo. This model effectively depends only on the value of a single parameter -- the effective tidal field -- and all other parameters, such as the halo's concentration, mass and the value of the actual tidal field can be summarized into this effective parameter. We have published our implementation of the \textsc{adiabatic-tides} model here\footnote{\url{https://github.com/jstuecker/adiabatic-tides}}.

We have verified with numerical simulations that our implementation correctly predicts the adiabatic limit and further, that the modeled scenario with isotropic tidal field also provides a good approximation to the effect of anisotropic tidal fields. Further, we have shown that the \textsc{adiabatic-tides} model provides meaningful predictions for realistic scenarios of subhaloes orbiting in the potential of another halo if the largest eigenvalue of the tidal field at pericentre is used as the input parameter. If centrifugal corrections are taken into account, the model predicts a lower limit for the mass of a remnant that cannot be crossed even after arbitrary long times. If centrifugal corrections are neglected, the model gives reasonable predictions for the amount of mass that is left inside the subhalo after about 15 orbits. Since most subhaloes in cosmological scenarios will have gone through much fewer than 15 orbits up to the present epoch, the \textsc{adiabatic-tides} predictions (without centrifugal correction) can be understood in practice as lower bound estimates for the halo mass and annihilation luminosity in such scenarios.

Beyond detailed predictions for individual haloes, the \textsc{adiabatic-tides} model also provides a theoretical understanding for several previously found phenomenological relations about tidally stripped subhaloes. The model correctly predicts the existence of the tidal track and it precisely recovers the measured location of it \citep{Penarrubia_2008, green_2019, errani_2021} while it also correctly transitions to the theoretically expected asymptotic behaviour. Further, it explains the orbital frequency relation that was measured by \citet{errani_2021} for the case of an isothermal sphere host potential. Additionally our understanding through the \textsc{adiabatic-tides} model allowed us to propose a more general tidal ratio relation that should also generalize to other host potentials. Beyond this, our model predicts that complete tidal disruption of subhaloes is not possible as long as the amplitude of the tidal field is finite at all times and the subhalo has initially a centrally divergent density profile. We expect that tidal disruption processes -- both in dark matter only scenarios and scenarios including baryons -- should always leave behind a small remnant of the subhalo. This point has recently been argued for by other authors \citep{vandenbosch_2018, Errani2020, amorisco_2021} and we think that the \textsc{adiabatic-tides} model provides a strong support for this. Finally, in the framework of the \textsc{adiabatic-tides} model it is quite simple to understand the effect of baryonic components like a galactic disk. Baryonic components drastically increase the value of the tidal field in the inner parts of a host halo (e.g. by a factor of ten) and therefore, have a strong impact on the expected mass loss. The enhancement in mass loss can easily be predicted by the \textsc{adiabatic-tides} model.

Additionally, we have used the \textsc{adiabatic-tides} model to make novel predictions which have not been empirically verified yet. The most important of these predictions is the structure-tide degeneracy: increasing the tidal field by a given factor has the same effect onto the disruption of an NFW halo as a decrease in characteristic density by the same factor. Since the characteristic density is mostly determined by the concentration of a halo, it should be possible to understand most of the concentration dependence of the tidal stripping process in terms of the structure-tide degeneracy. For example, we expect that most of the concentration dependence would disappear if mass loss is considered in units of the scale mass as a function of the effective pericentre tidal field and the initial concentration $M/M_{\rm{s}} (\lambda_{\rm{p}}/\lambda_{\rm{s}}, c)$.

Further, the \textsc{adiabatic-tides} model predicts that the density profiles of tidal remnants should always approach the initial NFW profile in the very centre \citep[see also][]{amorisco_2021}. Note that this is in contrast to the relations suggested by \citet{green_2019} which approach a reduced central density profile. Finally, we have predicted lower limits for the bound mass fractions of subhaloes that are orbiting in a Milky Way-like host with and without baryonic components as a function of their orbital pericentre and their concentration. The behaviour of subhaloes that reach pericentres within $0.1 \rvir$ varies dramatically between the two cases. Further, we note that the concentration dependence can be very steep, e.g. changes in concentration by $50\%$ may modify the bound mass fraction by up to a factor of 10 and the ratio between initial and final annihilation luminosity by factors between $2$ and $8$. 

It would be interesting to explore the \textsc{adiabatic-tides} model further in future studies. It would be worthwhile to test the model more rigorously in numerical studies and it would be interesting to see to what degree it can also be applied to partially disrupted subhaloes, if the effective tide is allowed to be varied as a fitting parameter. Since the \textsc{adiabatic-tides} model is consistent with the tidal track and other studies have argued that even partially disrupted subhaloes follow the tidal track \citep{penarrubia_2010, errani_2021}, we think that it is very likely that also partially disrupted subhaloes may be described approximately in this matter. If this is confirmed, one could use the \textsc{adiabatic-tides} model to infer improved  extrapolation estimates of the annihilation luminosity of the smallest haloes that may be orbiting in the Milky Way.

Finally, we note that the \textsc{adiabatic-tides} model can easily be extended and used for different scenarios than the fiducial isotropic NFW scenario. For example, we have also implemented powerlaw profiles and presented all relevant summary statistics in a way that they can analytically be rescaled to arbitrary tidal fields. We found that the slope of a powerlaw can drastically change the degree of disruption and even more dramatically the expected annihilation luminosities. Annihilation luminosities may even diverge for a powerlaw with slope $\alpha < -1.5$. Therefore, we expect that a major uncertainty for the robust prediction of annihilation luminosities of small mass (sub)haloes is the uncertainty about the central slopes of such (sub)haloes. The question for the central slope is not yet completely resolved -- e.g. \citet{angulo_2017} have reported central slopes of $\alpha \sim -1.5$ at the neutralino cut-off scale whereas \citet{wang_2020} have found shallower central slopes consistent with the NFW value of $-1$. 

Further, other initial profiles, like for example an Einasto profile, could easily be incorporated in the \textsc{adiabatic-tides} framework. The code might even be applied for very different scenarios like, for example, globular clusters.  It would also be straightforward to investigate the effect of velocity anisotropy by modification of the initial phase space distribution. Finally, it would be interesting to consider more sophisticated adiabatic potential perturbations. For example, one could easily add the effect of a small baryonic component in the centre of the subhalo and see how much this reduces the tidal disruption of a subhalo. It could also be used to explore analytically under what conditions it is possible to generate galaxies lacking dark matter through the effect of tidal fields \citep[see e.g.][]{dokkum_2018, dokkum_2019, ogiya_2022, Moreno_2022}.

We think that there is plenty of reason for further investigations and we hope that other researchers incorporate the \textsc{adiabatic-tides} model into their investigations to enhance the understanding of tidal disruption. 

\section*{Acknowledgements}
\rev{We thank the anonymous referee for the detailed and constructive feedback to the draft.} JS thanks Simon White for helpful comments to the draft and related discussions. Further, JS and RA thank all members of the cosmology group at Donostia International Physics Center for daily discussions and for the motivating research environment. JS and RA  acknowledge  the  support  of the European  Research Council through grant number ERC-StG/716151. GO was supported by the National Key Research and Development Program of China (No. 2022YFA1602903), the Fundamental Research Fund for Chinese Central Universities (Grant No. NZ2020021, No. 226-2022-00216), and the Waterloo Centre for Astrophysics Fellowship. The work of AAS and MASC was supported by the grants PGC2018-095161-B-I00 and CEX2020-001007-S, both funded by MCIN/AEI/10.13039/501100011033 and by ``ERDF A way of making Europe''. MASC was also supported by the Atracci\'on de Talento contract no. 2020-5A/TIC-19725 granted by the Comunidad de Madrid in Spain. The work of AAS was also supported by the Spanish Ministry of Science and Innovation through the grant FPI-UAM 2018.

%%%%%%%%%%%%%%%%%%%%%%%%%%%%%%%%%%%%%%%%%%%%%%%%%%
\section*{Data Availability}
Simulation data used in this article will be provided on reasonable request to the corresponding author. The \textsc{adiabatic-tides} code is publicly available under \url{https://github.com/jstuecker/adiabatic-tides}.

%%%%%%%%%%%%%%%%%%%% REFERENCES %%%%%%%%%%%%%%%%%%

% The best way to enter references is to use BibTeX:

\bibliographystyle{mnras}
\bibliography{example} % if your bibtex file is called example.bib

\begin{thebibliography}{}
\makeatletter
\relax
\def\mn@urlcharsother{\let\do\@makeother \do\$\do\&\do\#\do\^\do\_\do\%\do\~}
\def\mn@doi{\begingroup\mn@urlcharsother \@ifnextchar [ {\mn@doi@}
  {\mn@doi@[]}}
\def\mn@doi@[#1]#2{\def\@tempa{#1}\ifx\@tempa\@empty \href
  {http://dx.doi.org/#2} {doi:#2}\else \href {http://dx.doi.org/#2} {#1}\fi
  \endgroup}
\def\mn@eprint#1#2{\mn@eprint@#1:#2::\@nil}
\def\mn@eprint@arXiv#1{\href {http://arxiv.org/abs/#1} {{\tt arXiv:#1}}}
\def\mn@eprint@dblp#1{\href {http://dblp.uni-trier.de/rec/bibtex/#1.xml}
  {dblp:#1}}
\def\mn@eprint@#1:#2:#3:#4\@nil{\def\@tempa {#1}\def\@tempb {#2}\def\@tempc
  {#3}\ifx \@tempc \@empty \let \@tempc \@tempb \let \@tempb \@tempa \fi \ifx
  \@tempb \@empty \def\@tempb {arXiv}\fi \@ifundefined
  {mn@eprint@\@tempb}{\@tempb:\@tempc}{\expandafter \expandafter \csname
  mn@eprint@\@tempb\endcsname \expandafter{\@tempc}}}

\bibitem[\protect\citeauthoryear{{Aguirre-Santaella}, {S{\'a}nchez-Conde},
  {Ogiya}, {St{\"u}cker}  \& {Angulo}}{{Aguirre-Santaella}
  et~al.}{2023}]{aguirre_2023}
{Aguirre-Santaella} A.,  {S{\'a}nchez-Conde} M.~A.,  {Ogiya} G.,  {St{\"u}cker}
  J.,   {Angulo} R.~E.,  2023, \mn@doi [\mnras] {10.1093/mnras/stac2921}, \href
  {https://ui.adsabs.harvard.edu/abs/2023MNRAS.518...93A} {518, 93}

\bibitem[\protect\citeauthoryear{Amorisco}{Amorisco}{2021}]{amorisco_2021}
Amorisco N.~C.,  2021, Cold dark matter subhaloes at arbitrarily low masses
  (\mn@eprint {arXiv} {2111.01148})

\bibitem[\protect\citeauthoryear{{Angulo} \& {Hahn}}{{Angulo} \&
  {Hahn}}{2022}]{angulo_2022}
{Angulo} R.~E.,  {Hahn} O.,  2022, \mn@doi [Living Reviews in Computational
  Astrophysics] {10.1007/s41115-021-00013-z}, \href
  {https://ui.adsabs.harvard.edu/abs/2022LRCA....8....1A} {8, 1}

\bibitem[\protect\citeauthoryear{Angulo, Springel, White, Jenkins, Baugh  \&
  Frenk}{Angulo et~al.}{2012}]{angulo_2012}
Angulo R.~E.,  Springel V.,  White S. D.~M.,  Jenkins A.,  Baugh C.~M.,   Frenk
  C.~S.,  2012, \mn@doi [Monthly Notices of the Royal Astronomical Society]
  {10.1111/j.1365-2966.2012.21830.x}, 426, 2046–2062

\bibitem[\protect\citeauthoryear{{Angulo}, {Hahn}, {Ludlow}  \&
  {Bonoli}}{{Angulo} et~al.}{2017}]{angulo_2017}
{Angulo} R.~E.,  {Hahn} O.,  {Ludlow} A.~D.,   {Bonoli} S.,  2017, \mn@doi
  [\mnras] {10.1093/mnras/stx1658}, \href
  {https://ui.adsabs.harvard.edu/abs/2017MNRAS.471.4687A} {471, 4687}

\bibitem[\protect\citeauthoryear{Banik, Bertone, Bovy  \& Bozorgnia}{Banik
  et~al.}{2018}]{Banik_2018}
Banik N.,  Bertone G.,  Bovy J.,   Bozorgnia N.,  2018, \mn@doi [Journal of
  Cosmology and Astroparticle Physics] {10.1088/1475-7516/2018/07/061}, 2018,
  061

\bibitem[\protect\citeauthoryear{{Barnes} \& {Hut}}{{Barnes} \&
  {Hut}}{1986}]{barnes_hut_1986}
{Barnes} J.,  {Hut} P.,  1986, \mn@doi [\nat] {10.1038/324446a0}, \href
  {https://ui.adsabs.harvard.edu/abs/1986Natur.324..446B} {324, 446}

\bibitem[\protect\citeauthoryear{{Benson} \& {Du}}{{Benson} \&
  {Du}}{2022}]{benson_2022}
{Benson} A.~J.,  {Du} X.,  2022, \mn@doi [\mnras] {10.1093/mnras/stac2750},
  \href {https://ui.adsabs.harvard.edu/abs/2022MNRAS.517.1398B} {517, 1398}

\bibitem[\protect\citeauthoryear{Berezinsky, Dokuchaev  \&
  Eroshenko}{Berezinsky et~al.}{2003}]{berezinsky_2003}
Berezinsky V.,  Dokuchaev V.,   Eroshenko Y.,  2003, \mn@doi [Phys. Rev. D]
  {10.1103/PhysRevD.68.103003}, 68, 103003

\bibitem[\protect\citeauthoryear{{Binney} \& {Tremaine}}{{Binney} \&
  {Tremaine}}{2008}]{BinneyTremaine2008}
{Binney} J.,  {Tremaine} S.,  2008, {Galactic Dynamics: Second Edition}

\bibitem[\protect\citeauthoryear{{Blumenthal}, {Faber}, {Flores}  \&
  {Primack}}{{Blumenthal} et~al.}{1986}]{blumenthal_1986}
{Blumenthal} G.~R.,  {Faber} S.~M.,  {Flores} R.,   {Primack} J.~R.,  1986,
  \mn@doi [\apj] {10.1086/163867}, \href
  {https://ui.adsabs.harvard.edu/abs/1986ApJ...301...27B} {301, 27}

\bibitem[\protect\citeauthoryear{Bovy}{Bovy}{2015}]{bovy_2015}
Bovy J.,  2015, \mn@doi [The Astrophysical Journal Supplement Series]
  {10.1088/0067-0049/216/2/29}, 216, 29

\bibitem[\protect\citeauthoryear{{Bringmann}}{{Bringmann}}{2009}]{Bringmann_20019}
{Bringmann} T.,  2009, \mn@doi [New Journal of Physics]
  {10.1088/1367-2630/11/10/105027}, \href
  {https://ui.adsabs.harvard.edu/abs/2009NJPh...11j5027B} {11, 105027}

\bibitem[\protect\citeauthoryear{{Calc{\'a}neo-Rold{\'a}n} \&
  {Moore}}{{Calc{\'a}neo-Rold{\'a}n} \& {Moore}}{2000}]{calcaneo_2000}
{Calc{\'a}neo-Rold{\'a}n} C.,  {Moore} B.,  2000, \mn@doi [\prd]
  {10.1103/PhysRevD.62.123005}, \href
  {https://ui.adsabs.harvard.edu/abs/2000PhRvD..62l3005C} {62, 123005}

\bibitem[\protect\citeauthoryear{{Choi}, {Weinberg}  \& {Katz}}{{Choi}
  et~al.}{2009}]{choi_2009}
{Choi} J.-H.,  {Weinberg} M.~D.,   {Katz} N.,  2009, \mn@doi [\mnras]
  {10.1111/j.1365-2966.2009.15556.x}, \href
  {https://ui.adsabs.harvard.edu/abs/2009MNRAS.400.1247C} {400, 1247}

\bibitem[\protect\citeauthoryear{{Contreras}, {Angulo}  \&
  {Zennaro}}{{Contreras} et~al.}{2021}]{contreras_2021}
{Contreras} S.,  {Angulo} R.~E.,   {Zennaro} M.,  2021, \mn@doi [\mnras]
  {10.1093/mnras/stab2560}, \href
  {https://ui.adsabs.harvard.edu/abs/2021MNRAS.508..175C} {508, 175}

\bibitem[\protect\citeauthoryear{{Delfino}, {Sc{\'o}ccola}, {Cora},
  {Vega-Mart{\'\i}nez}  \& {Gargiulo}}{{Delfino} et~al.}{2022}]{delfino_2022}
{Delfino} F.~M.,  {Sc{\'o}ccola} C.~G.,  {Cora} S.~A.,  {Vega-Mart{\'\i}nez}
  C.~A.,   {Gargiulo} I.~D.,  2022, \mn@doi [\mnras] {10.1093/mnras/stab3494},
  \href {https://ui.adsabs.harvard.edu/abs/2022MNRAS.510.2900D} {510, 2900}

\bibitem[\protect\citeauthoryear{Delos}{Delos}{2019}]{Delos_2019}
Delos M.~S.,  2019, \mn@doi [Physical Review D] {10.1103/physrevd.100.063505},
  100

\bibitem[\protect\citeauthoryear{{Drakos}, {Taylor}  \& {Benson}}{{Drakos}
  et~al.}{2017}]{drakos_2017}
{Drakos} N.~E.,  {Taylor} J.~E.,   {Benson} A.~J.,  2017, \mn@doi [\mnras]
  {10.1093/mnras/stx652}, \href
  {https://ui.adsabs.harvard.edu/abs/2017MNRAS.468.2345D} {468, 2345}

\bibitem[\protect\citeauthoryear{{Drakos}, {Taylor}  \& {Benson}}{{Drakos}
  et~al.}{2020}]{drakos_2020}
{Drakos} N.~E.,  {Taylor} J.~E.,   {Benson} A.~J.,  2020, \mn@doi [\mnras]
  {10.1093/mnras/staa760}, \href
  {https://ui.adsabs.harvard.edu/abs/2020MNRAS.494..378D} {494, 378}

\bibitem[\protect\citeauthoryear{{Eddington}}{{Eddington}}{1916}]{Eddington1916}
{Eddington} A.~S.,  1916, \mn@doi [\mnras] {10.1093/mnras/76.7.572}, \href
  {https://ui.adsabs.harvard.edu/abs/1916MNRAS..76..572E} {76, 572}

\bibitem[\protect\citeauthoryear{Errani \& Navarro}{Errani \&
  Navarro}{2021}]{errani_2021}
Errani R.,  Navarro J.~F.,  2021, \mn@doi [Monthly Notices of the Royal
  Astronomical Society] {10.1093/mnras/stab1215}, 505, 18–32

\bibitem[\protect\citeauthoryear{{Errani} \& {Pe{\~n}arrubia}}{{Errani} \&
  {Pe{\~n}arrubia}}{2020}]{Errani2020}
{Errani} R.,  {Pe{\~n}arrubia} J.,  2020, \mn@doi [\mnras]
  {10.1093/mnras/stz3349}, \href
  {https://ui.adsabs.harvard.edu/abs/2020MNRAS.491.4591E} {491, 4591}

\bibitem[\protect\citeauthoryear{{Frenk} \& {White}}{{Frenk} \&
  {White}}{2012}]{frenk_white_2012}
{Frenk} C.~S.,  {White} S.~D.~M.,  2012, \mn@doi [Annalen der Physik]
  {10.1002/andp.201200212}, \href
  {https://ui.adsabs.harvard.edu/abs/2012AnP...524..507F} {524, 507}

\bibitem[\protect\citeauthoryear{{Gao}, {De Lucia}, {White}  \&
  {Jenkins}}{{Gao} et~al.}{2004}]{Gao_2004}
{Gao} L.,  {De Lucia} G.,  {White} S.~D.~M.,   {Jenkins} A.,  2004, \mn@doi
  [\mnras] {10.1111/j.1365-2966.2004.08098.x}, \href
  {https://ui.adsabs.harvard.edu/abs/2004MNRAS.352L...1G} {352, L1}

\bibitem[\protect\citeauthoryear{{Gao}, {Frenk}, {Jenkins}, {Springel}  \&
  {White}}{{Gao} et~al.}{2012}]{gao_2012}
{Gao} L.,  {Frenk} C.~S.,  {Jenkins} A.,  {Springel} V.,   {White} S.~D.~M.,
  2012, \mn@doi [\mnras] {10.1111/j.1365-2966.2011.19836.x}, \href
  {https://ui.adsabs.harvard.edu/abs/2012MNRAS.419.1721G} {419, 1721}

\bibitem[\protect\citeauthoryear{{Garrison-Kimmel} et~al.,}{{Garrison-Kimmel}
  et~al.}{2017}]{garrison_2017}
{Garrison-Kimmel} S.,  et~al., 2017, \mn@doi [\mnras] {10.1093/mnras/stx1710},
  \href {https://ui.adsabs.harvard.edu/abs/2017MNRAS.471.1709G} {471, 1709}

\bibitem[\protect\citeauthoryear{{Gilman}, {Birrer}, {Nierenberg}, {Treu}, {Du}
   \& {Benson}}{{Gilman} et~al.}{2020}]{gilman_2020}
{Gilman} D.,  {Birrer} S.,  {Nierenberg} A.,  {Treu} T.,  {Du} X.,   {Benson}
  A.,  2020, \mn@doi [\mnras] {10.1093/mnras/stz3480}, \href
  {https://ui.adsabs.harvard.edu/abs/2020MNRAS.491.6077G} {491, 6077}

\bibitem[\protect\citeauthoryear{{Gnedin}, {Hernquist}  \& {Ostriker}}{{Gnedin}
  et~al.}{1999}]{gnedin_ostriker_1999}
{Gnedin} O.~Y.,  {Hernquist} L.,   {Ostriker} J.~P.,  1999, \mn@doi [\apj]
  {10.1086/306910}, \href
  {https://ui.adsabs.harvard.edu/abs/1999ApJ...514..109G} {514, 109}

\bibitem[\protect\citeauthoryear{{Gnedin}, {Kravtsov}, {Klypin}  \&
  {Nagai}}{{Gnedin} et~al.}{2004}]{gnedin_2004}
{Gnedin} O.~Y.,  {Kravtsov} A.~V.,  {Klypin} A.~A.,   {Nagai} D.,  2004,
  \mn@doi [\apj] {10.1086/424914}, \href
  {https://ui.adsabs.harvard.edu/abs/2004ApJ...616...16G} {616, 16}

\bibitem[\protect\citeauthoryear{{Grand} \& {White}}{{Grand} \&
  {White}}{2021}]{grand_white_2021}
{Grand} R. J.~J.,  {White} S. D.~M.,  2021, \mn@doi [\mnras]
  {10.1093/mnras/staa3993}, \href
  {https://ui.adsabs.harvard.edu/abs/2021MNRAS.501.3558G} {501, 3558}

\bibitem[\protect\citeauthoryear{{Grand} et~al.,}{{Grand}
  et~al.}{2021}]{grand_2021}
{Grand} R. J.~J.,  et~al., 2021, \mn@doi [\mnras] {10.1093/mnras/stab2492},
  \href {https://ui.adsabs.harvard.edu/abs/2021MNRAS.507.4953G} {507, 4953}

\bibitem[\protect\citeauthoryear{Green \& van~den Bosch}{Green \& van~den
  Bosch}{2019}]{green_2019}
Green S.~B.,  van~den Bosch F.~C.,  2019, \mn@doi [Monthly Notices of the Royal
  Astronomical Society] {10.1093/mnras/stz2767}, 490, 2091–2101

\bibitem[\protect\citeauthoryear{{Guo} \& {White}}{{Guo} \&
  {White}}{2014}]{guo_2014}
{Guo} Q.,  {White} S.,  2014, \mn@doi [\mnras] {10.1093/mnras/stt2116}, \href
  {https://ui.adsabs.harvard.edu/abs/2014MNRAS.437.3228G} {437, 3228}

\bibitem[\protect\citeauthoryear{{Guo} et~al.,}{{Guo} et~al.}{2011}]{guo_2011}
{Guo} Q.,  et~al., 2011, \mn@doi [\mnras] {10.1111/j.1365-2966.2010.18114.x},
  \href {https://ui.adsabs.harvard.edu/abs/2011MNRAS.413..101G} {413, 101}

\bibitem[\protect\citeauthoryear{Hayashi, Navarro, Taylor, Stadel  \&
  Quinn}{Hayashi et~al.}{2003}]{Hayashi_2003}
Hayashi E.,  Navarro J.~F.,  Taylor J.~E.,  Stadel J.,   Quinn T.,  2003,
  \mn@doi [The Astrophysical Journal] {10.1086/345788}, 584, 541

\bibitem[\protect\citeauthoryear{{Hsueh}, {Enzi}, {Vegetti}, {Auger},
  {Fassnacht}, {Despali}, {Koopmans}  \& {McKean}}{{Hsueh}
  et~al.}{2020}]{hsueh_2020}
{Hsueh} J.~W.,  {Enzi} W.,  {Vegetti} S.,  {Auger} M.~W.,  {Fassnacht} C.~D.,
  {Despali} G.,  {Koopmans} L.~V.~E.,   {McKean} J.~P.,  2020, \mn@doi [\mnras]
  {10.1093/mnras/stz3177}, \href
  {https://ui.adsabs.harvard.edu/abs/2020MNRAS.492.3047H} {492, 3047}

\bibitem[\protect\citeauthoryear{{Jiang}, {Cole}, {Sawala}  \& {Frenk}}{{Jiang}
  et~al.}{2015}]{Jiang_2015}
{Jiang} L.,  {Cole} S.,  {Sawala} T.,   {Frenk} C.~S.,  2015, \mn@doi [\mnras]
  {10.1093/mnras/stv053}, \href
  {https://ui.adsabs.harvard.edu/abs/2015MNRAS.448.1674J} {448, 1674}

\bibitem[\protect\citeauthoryear{{Jiang}, {Dekel}, {Freundlich}, {van den
  Bosch}, {Green}, {Hopkins}, {Benson}  \& {Du}}{{Jiang}
  et~al.}{2021}]{jiang_2011}
{Jiang} F.,  {Dekel} A.,  {Freundlich} J.,  {van den Bosch} F.~C.,  {Green}
  S.~B.,  {Hopkins} P.~F.,  {Benson} A.,   {Du} X.,  2021, \mn@doi [\mnras]
  {10.1093/mnras/staa4034}, \href
  {https://ui.adsabs.harvard.edu/abs/2021MNRAS.502..621J} {502, 621}

\bibitem[\protect\citeauthoryear{{Kampakoglou} \& {Benson}}{{Kampakoglou} \&
  {Benson}}{2007}]{Kampakoglou_2007}
{Kampakoglou} M.,  {Benson} A.~J.,  2007, \mn@doi [\mnras]
  {10.1111/j.1365-2966.2006.11223.x}, \href
  {https://ui.adsabs.harvard.edu/abs/2007MNRAS.374..775K} {374, 775}

\bibitem[\protect\citeauthoryear{Kazantzidis, Mayer, Mastropietro, Diemand,
  Stadel  \& Moore}{Kazantzidis et~al.}{2004}]{Kazantzidis_2004}
Kazantzidis S.,  Mayer L.,  Mastropietro C.,  Diemand J.,  Stadel J.,   Moore
  B.,  2004, \mn@doi [The Astrophysical Journal] {10.1086/420840}, 608, 663

\bibitem[\protect\citeauthoryear{{Kelley}, {Bullock}, {Garrison-Kimmel},
  {Boylan-Kolchin}, {Pawlowski}  \& {Graus}}{{Kelley}
  et~al.}{2019}]{kelley_2019}
{Kelley} T.,  {Bullock} J.~S.,  {Garrison-Kimmel} S.,  {Boylan-Kolchin} M.,
  {Pawlowski} M.~S.,   {Graus} A.~S.,  2019, \mn@doi [\mnras]
  {10.1093/mnras/stz1553}, \href
  {https://ui.adsabs.harvard.edu/abs/2019MNRAS.487.4409K} {487, 4409}

\bibitem[\protect\citeauthoryear{Klypin, Kravtsov, Valenzuela  \& Prada}{Klypin
  et~al.}{1999}]{Klypin_1999}
Klypin A.,  Kravtsov A.~V.,  Valenzuela O.,   Prada F.,  1999, \mn@doi [The
  Astrophysical Journal] {10.1086/307643}, 522, 82

\bibitem[\protect\citeauthoryear{{Li}, {Zhao}, {Jing}, {Han}  \& {Dong}}{{Li}
  et~al.}{2020}]{Li_2020}
{Li} Z.-Z.,  {Zhao} D.-H.,  {Jing} Y.~P.,  {Han} J.,   {Dong} F.-Y.,  2020,
  \mn@doi [\apj] {10.3847/1538-4357/abc481}, \href
  {https://ui.adsabs.harvard.edu/abs/2020ApJ...905..177L} {905, 177}

\bibitem[\protect\citeauthoryear{{Lovell}, {Frenk}, {Eke}, {Jenkins}, {Gao}  \&
  {Theuns}}{{Lovell} et~al.}{2014}]{lovell2014}
{Lovell} M.~R.,  {Frenk} C.~S.,  {Eke} V.~R.,  {Jenkins} A.,  {Gao} L.,
  {Theuns} T.,  2014, \mn@doi [\mnras] {10.1093/mnras/stt2431}, \href
  {https://ui.adsabs.harvard.edu/abs/2014MNRAS.439..300L} {439, 300}

\bibitem[\protect\citeauthoryear{{Miyamoto} \& {Nagai}}{{Miyamoto} \&
  {Nagai}}{1975}]{Miyamoto_1975}
{Miyamoto} M.,  {Nagai} R.,  1975, \pasj, \href
  {https://ui.adsabs.harvard.edu/abs/1975PASJ...27..533M} {27, 533}

\bibitem[\protect\citeauthoryear{{Molin{\'e}}, {S{\'a}nchez-Conde},
  {Palomares-Ruiz}  \& {Prada}}{{Molin{\'e}} et~al.}{2017}]{Monline_2017}
{Molin{\'e}} {\'A}.,  {S{\'a}nchez-Conde} M.~A.,  {Palomares-Ruiz} S.,
  {Prada} F.,  2017, \mn@doi [\mnras] {10.1093/mnras/stx026}, \href
  {https://ui.adsabs.harvard.edu/abs/2017MNRAS.466.4974M} {466, 4974}

\bibitem[\protect\citeauthoryear{Moore, Ghigna, Governato, Lake, Quinn, Stadel
  \& Tozzi}{Moore et~al.}{1999}]{Moore_1999}
Moore B.,  Ghigna S.,  Governato F.,  Lake G.,  Quinn T.,  Stadel J.,   Tozzi
  P.,  1999, \mn@doi [The Astrophysical Journal] {10.1086/312287}, 524, L19

\bibitem[\protect\citeauthoryear{{Moreno} et~al.,}{{Moreno}
  et~al.}{2022}]{Moreno_2022}
{Moreno} J.,  et~al., 2022, \mn@doi [Nature Astronomy]
  {10.1038/s41550-021-01598-4}, \href
  {https://ui.adsabs.harvard.edu/abs/2022NatAs.tmp...36M} {}

\bibitem[\protect\citeauthoryear{{Moster}, {Naab}  \& {White}}{{Moster}
  et~al.}{2013}]{moster_2013}
{Moster} B.~P.,  {Naab} T.,   {White} S. D.~M.,  2013, \mn@doi [\mnras]
  {10.1093/mnras/sts261}, \href
  {https://ui.adsabs.harvard.edu/abs/2013MNRAS.428.3121M} {428, 3121}

\bibitem[\protect\citeauthoryear{{Moster}, {Naab}  \& {White}}{{Moster}
  et~al.}{2018}]{moster_2018}
{Moster} B.~P.,  {Naab} T.,   {White} S. D.~M.,  2018, \mn@doi [\mnras]
  {10.1093/mnras/sty655}, \href
  {https://ui.adsabs.harvard.edu/abs/2018MNRAS.477.1822M} {477, 1822}

\bibitem[\protect\citeauthoryear{{Nadler}, {Mao}, {Wechsler}, {Garrison-Kimmel}
   \& {Wetzel}}{{Nadler} et~al.}{2018}]{Nadler_2018}
{Nadler} E.~O.,  {Mao} Y.-Y.,  {Wechsler} R.~H.,  {Garrison-Kimmel} S.,
  {Wetzel} A.,  2018, \mn@doi [\apj] {10.3847/1538-4357/aac266}, \href
  {https://ui.adsabs.harvard.edu/abs/2018ApJ...859..129N} {859, 129}

\bibitem[\protect\citeauthoryear{{Navarro}, {Frenk}  \& {White}}{{Navarro}
  et~al.}{1996}]{nfw1996}
{Navarro} J.~F.,  {Frenk} C.~S.,   {White} S. D.~M.,  1996, \mn@doi [\apj]
  {10.1086/177173}, \href
  {https://ui.adsabs.harvard.edu/abs/1996ApJ...462..563N} {462, 563}

\bibitem[\protect\citeauthoryear{{Newton} et~al.,}{{Newton}
  et~al.}{2021}]{newton_2021}
{Newton} O.,  et~al., 2021, \mn@doi [\jcap] {10.1088/1475-7516/2021/08/062},
  \href {https://ui.adsabs.harvard.edu/abs/2021JCAP...08..062N} {2021, 062}

\bibitem[\protect\citeauthoryear{Ogiya, Mori, Miki, Boku  \& Nakasato}{Ogiya
  et~al.}{2013}]{Ogiya_2013}
Ogiya G.,  Mori M.,  Miki Y.,  Boku T.,   Nakasato N.,  2013, \mn@doi [Journal
  of Physics: Conference Series] {10.1088/1742-6596/454/1/012014}, 454, 012014

\bibitem[\protect\citeauthoryear{Ogiya, van~den Bosch, Hahn, Green, Miller  \&
  Burkert}{Ogiya et~al.}{2019}]{ogiya_2019}
Ogiya G.,  van~den Bosch F.~C.,  Hahn O.,  Green S.~B.,  Miller T.~B.,
  Burkert A.,  2019, \mn@doi [Monthly Notices of the Royal Astronomical
  Society] {10.1093/mnras/stz375}, 485, 189–202

\bibitem[\protect\citeauthoryear{{Ogiya}, {Taylor}  \& {Hudson}}{{Ogiya}
  et~al.}{2021}]{ogiya_2021}
{Ogiya} G.,  {Taylor} J.~E.,   {Hudson} M.~J.,  2021, \mn@doi [\mnras]
  {10.1093/mnras/stab361}, \href
  {https://ui.adsabs.harvard.edu/abs/2021MNRAS.503.1233O} {503, 1233}

\bibitem[\protect\citeauthoryear{{Ogiya}, {van den Bosch}  \&
  {Burkert}}{{Ogiya} et~al.}{2022}]{ogiya_2022}
{Ogiya} G.,  {van den Bosch} F.~C.,   {Burkert} A.,  2022, \mn@doi [\mnras]
  {10.1093/mnras/stab3658}, \href
  {https://ui.adsabs.harvard.edu/abs/2022MNRAS.510.2724O} {510, 2724}

\bibitem[\protect\citeauthoryear{{Pe{\~n}arrubia}}{{Pe{\~n}arrubia}}{2023}]{penarrubia_2022}
{Pe{\~n}arrubia} J.,  2023, \mn@doi [\mnras] {10.1093/mnras/stac3642}, \href
  {https://ui.adsabs.harvard.edu/abs/2023MNRAS.519.1955P} {519, 1955}

\bibitem[\protect\citeauthoryear{{Pe{\~n}arrubia} \& {Benson}}{{Pe{\~n}arrubia}
  \& {Benson}}{2005}]{penarrubia_2005}
{Pe{\~n}arrubia} J.,  {Benson} A.~J.,  2005, \mn@doi [\mnras]
  {10.1111/j.1365-2966.2005.09633.x}, \href
  {https://ui.adsabs.harvard.edu/abs/2005MNRAS.364..977P} {364, 977}

\bibitem[\protect\citeauthoryear{{Pe{\~n}arrubia}, {Benson}, {Walker},
  {Gilmore}, {McConnachie}  \& {Mayer}}{{Pe{\~n}arrubia}
  et~al.}{2010}]{penarrubia_2010}
{Pe{\~n}arrubia} J.,  {Benson} A.~J.,  {Walker} M.~G.,  {Gilmore} G.,
  {McConnachie} A.~W.,   {Mayer} L.,  2010, \mn@doi [\mnras]
  {10.1111/j.1365-2966.2010.16762.x}, \href
  {https://ui.adsabs.harvard.edu/abs/2010MNRAS.406.1290P} {406, 1290}

\bibitem[\protect\citeauthoryear{Pe{\~n}arrubia, Navarro  \&
  McConnachie}{Pe{\~n}arrubia et~al.}{2008}]{Penarrubia_2008}
Pe{\~n}arrubia J.,  Navarro J.~F.,   McConnachie A.~W.,  2008, \mn@doi [The
  Astrophysical Journal] {10.1086/523686}, 673, 226

\bibitem[\protect\citeauthoryear{{Petulante}, {Berlind}, {Holley-Bockelmann}
  \& {Sinha}}{{Petulante} et~al.}{2021}]{Petulante_2021}
{Petulante} A.,  {Berlind} A.~A.,  {Holley-Bockelmann} J.~K.,   {Sinha} M.,
  2021, \mn@doi [\mnras] {10.1093/mnras/stab867}, \href
  {https://ui.adsabs.harvard.edu/abs/2021MNRAS.504..248P} {504, 248}

\bibitem[\protect\citeauthoryear{{Power}, {Navarro}, {Jenkins}, {Frenk},
  {White}, {Springel}, {Stadel}  \& {Quinn}}{{Power} et~al.}{2003}]{Power_2003}
{Power} C.,  {Navarro} J.~F.,  {Jenkins} A.,  {Frenk} C.~S.,  {White} S.~D.~M.,
   {Springel} V.,  {Stadel} J.,   {Quinn} T.,  2003, \mn@doi [\mnras]
  {10.1046/j.1365-8711.2003.05925.x}, \href
  {https://ui.adsabs.harvard.edu/abs/2003MNRAS.338...14P} {338, 14}

\bibitem[\protect\citeauthoryear{{Profumo}, {Sigurdson}  \&
  {Kamionkowski}}{{Profumo} et~al.}{2006}]{Profumo_2006}
{Profumo} S.,  {Sigurdson} K.,   {Kamionkowski} M.,  2006, \mn@doi [\prl]
  {10.1103/PhysRevLett.97.031301}, \href
  {https://ui.adsabs.harvard.edu/abs/2006PhRvL..97c1301P} {97, 031301}

\bibitem[\protect\citeauthoryear{Pullen, Benson  \& Moustakas}{Pullen
  et~al.}{2014}]{Pullen_2014}
Pullen A.~R.,  Benson A.~J.,   Moustakas L.~A.,  2014, \mn@doi [The
  Astrophysical Journal] {10.1088/0004-637x/792/1/24}, 792, 24

\bibitem[\protect\citeauthoryear{{Renaud}, {Gieles}  \& {Boily}}{{Renaud}
  et~al.}{2011}]{renaud_2011}
{Renaud} F.,  {Gieles} M.,   {Boily} C.~M.,  2011, \mn@doi [\mnras]
  {10.1111/j.1365-2966.2011.19531.x}, \href
  {https://ui.adsabs.harvard.edu/abs/2011MNRAS.418..759R} {418, 759}

\bibitem[\protect\citeauthoryear{{Richings} et~al.,}{{Richings}
  et~al.}{2020}]{richings_2020}
{Richings} J.,  et~al., 2020, \mn@doi [\mnras] {10.1093/mnras/stz3448}, \href
  {https://ui.adsabs.harvard.edu/abs/2020MNRAS.492.5780R} {492, 5780}

\bibitem[\protect\citeauthoryear{{Ritondale}, {Vegetti}, {Despali}, {Auger},
  {Koopmans}  \& {McKean}}{{Ritondale} et~al.}{2019}]{ritondale_2019}
{Ritondale} E.,  {Vegetti} S.,  {Despali} G.,  {Auger} M.~W.,  {Koopmans}
  L.~V.~E.,   {McKean} J.~P.,  2019, \mn@doi [\mnras] {10.1093/mnras/stz464},
  \href {https://ui.adsabs.harvard.edu/abs/2019MNRAS.485.2179R} {485, 2179}

\bibitem[\protect\citeauthoryear{{Rix} \& {White}}{{Rix} \&
  {White}}{1989}]{HansWalter_1989}
{Rix} H.-W.~R.,  {White} S. D.~M.,  1989, \mn@doi [\mnras]
  {10.1093/mnras/240.4.941}, \href
  {https://ui.adsabs.harvard.edu/abs/1989MNRAS.240..941R} {240, 941}

\bibitem[\protect\citeauthoryear{{S{\'a}nchez-Conde} \&
  {Prada}}{{S{\'a}nchez-Conde} \& {Prada}}{2014}]{sahnchezconde_2014}
{S{\'a}nchez-Conde} M.~A.,  {Prada} F.,  2014, \mn@doi [\mnras]
  {10.1093/mnras/stu1014}, \href
  {https://ui.adsabs.harvard.edu/abs/2014MNRAS.442.2271S} {442, 2271}

\bibitem[\protect\citeauthoryear{{Sawala}, {Pihajoki}, {Johansson}, {Frenk},
  {Navarro}, {Oman}  \& {White}}{{Sawala} et~al.}{2017}]{sawala_2017}
{Sawala} T.,  {Pihajoki} P.,  {Johansson} P.~H.,  {Frenk} C.~S.,  {Navarro}
  J.~F.,  {Oman} K.~A.,   {White} S. D.~M.,  2017, \mn@doi [\mnras]
  {10.1093/mnras/stx360}, \href
  {https://ui.adsabs.harvard.edu/abs/2017MNRAS.467.4383S} {467, 4383}

\bibitem[\protect\citeauthoryear{{Sellwood} \& {McGaugh}}{{Sellwood} \&
  {McGaugh}}{2005}]{sellwood_2005}
{Sellwood} J.~A.,  {McGaugh} S.~S.,  2005, \mn@doi [\apj] {10.1086/491731},
  \href {https://ui.adsabs.harvard.edu/abs/2005ApJ...634...70S} {634, 70}

\bibitem[\protect\citeauthoryear{{Spitzer}}{{Spitzer}}{1958}]{spitzer_1958}
{Spitzer} Lyman J.,  1958, \mn@doi [\apj] {10.1086/146435}, \href
  {https://ui.adsabs.harvard.edu/abs/1958ApJ...127...17S} {127, 17}

\bibitem[\protect\citeauthoryear{{Spitzer}}{{Spitzer}}{1987}]{Spitzer_1987}
{Spitzer} L.,  1987, {Dynamical evolution of globular clusters}

\bibitem[\protect\citeauthoryear{{Springel} et~al.,}{{Springel}
  et~al.}{2005}]{Springel_2005}
{Springel} V.,  et~al., 2005, \mn@doi [\nat] {10.1038/nature03597}, \href
  {https://ui.adsabs.harvard.edu/abs/2005Natur.435..629S} {435, 629}

\bibitem[\protect\citeauthoryear{Springel et~al.,}{Springel
  et~al.}{2008a}]{springel_2008}
Springel V.,  et~al., 2008a, \mn@doi [Monthly Notices of the Royal Astronomical
  Society] {10.1111/j.1365-2966.2008.14066.x}, 391, 1685–1711

\bibitem[\protect\citeauthoryear{{Springel} et~al.,}{{Springel}
  et~al.}{2008b}]{springel_2008b}
{Springel} V.,  et~al., 2008b, \mn@doi [\nat] {10.1038/nature07411}, \href
  {https://ui.adsabs.harvard.edu/abs/2008Natur.456...73S} {456, 73}

\bibitem[\protect\citeauthoryear{{St{\"u}cker}, {Angulo}  \&
  {Busch}}{{St{\"u}cker} et~al.}{2021}]{stuecker_2021_bp}
{St{\"u}cker} J.,  {Angulo} R.~E.,   {Busch} P.,  2021, \mn@doi [\mnras]
  {10.1093/mnras/stab2913}, \href
  {https://ui.adsabs.harvard.edu/abs/2021MNRAS.508.5196S} {508, 5196}

\bibitem[\protect\citeauthoryear{Taylor \& Babul}{Taylor \&
  Babul}{2001}]{Taylor_2001}
Taylor J.~E.,  Babul A.,  2001, \mn@doi [The Astrophysical Journal]
  {10.1086/322276}, 559, 716

\bibitem[\protect\citeauthoryear{{Tormen}}{{Tormen}}{1997}]{tormen_1997}
{Tormen} G.,  1997, \mn@doi [\mnras] {10.1093/mnras/290.3.411}, \href
  {https://ui.adsabs.harvard.edu/abs/1997MNRAS.290..411T} {290, 411}

\bibitem[\protect\citeauthoryear{{Vegetti}, {Despali}, {Lovell}  \&
  {Enzi}}{{Vegetti} et~al.}{2018}]{vegetti_2018}
{Vegetti} S.,  {Despali} G.,  {Lovell} M.~R.,   {Enzi} W.,  2018, \mn@doi
  [\mnras] {10.1093/mnras/sty2393}, \href
  {https://ui.adsabs.harvard.edu/abs/2018MNRAS.481.3661V} {481, 3661}

\bibitem[\protect\citeauthoryear{{Wang}, {Bose}, {Frenk}, {Gao}, {Jenkins},
  {Springel}  \& {White}}{{Wang} et~al.}{2020}]{wang_2020}
{Wang} J.,  {Bose} S.,  {Frenk} C.~S.,  {Gao} L.,  {Jenkins} A.,  {Springel}
  V.,   {White} S.~D.~M.,  2020, \mn@doi [\nat] {10.1038/s41586-020-2642-9},
  \href {https://ui.adsabs.harvard.edu/abs/2020Natur.585...39W} {585, 39}

\bibitem[\protect\citeauthoryear{{Weinberg}}{{Weinberg}}{1994a}]{weinberg_1994a}
{Weinberg} M.~D.,  1994a, \mn@doi [\aj] {10.1086/117161}, \href
  {https://ui.adsabs.harvard.edu/abs/1994AJ....108.1398W} {108, 1398}

\bibitem[\protect\citeauthoryear{{Weinberg}}{{Weinberg}}{1994b}]{weinberg_1994b}
{Weinberg} M.~D.,  1994b, \mn@doi [\aj] {10.1086/117162}, \href
  {https://ui.adsabs.harvard.edu/abs/1994AJ....108.1403W} {108, 1403}

\bibitem[\protect\citeauthoryear{{Widrow}}{{Widrow}}{2000}]{widrow_2000}
{Widrow} L.~M.,  2000, \mn@doi [\apjs] {10.1086/317367}, \href
  {https://ui.adsabs.harvard.edu/abs/2000ApJS..131...39W} {131, 39}

\bibitem[\protect\citeauthoryear{{Wilson}}{{Wilson}}{2004}]{wilson_2004}
{Wilson} G.,  2004, PhD thesis, Australian National University, Canberra,
  Australia

\bibitem[\protect\citeauthoryear{Yoon, Johnston  \& Hogg}{Yoon
  et~al.}{2011}]{Yoon_2011}
Yoon J.~H.,  Johnston K.~V.,   Hogg D.~W.,  2011, \mn@doi [The Astrophysical
  Journal] {10.1088/0004-637x/731/1/58}, 731, 58

\bibitem[\protect\citeauthoryear{{Young}}{{Young}}{1980}]{young_1980}
{Young} P.,  1980, \mn@doi [\apj] {10.1086/158553}, \href
  {https://ui.adsabs.harvard.edu/abs/1980ApJ...242.1232Y} {242, 1232}

\bibitem[\protect\citeauthoryear{Zentner, Berlind, Bullock, Kravtsov  \&
  Wechsler}{Zentner et~al.}{2005}]{Zentner_2005}
Zentner A.~R.,  Berlind A.~A.,  Bullock J.~S.,  Kravtsov A.~V.,   Wechsler
  R.~H.,  2005, \mn@doi [The Astrophysical Journal] {10.1086/428898}, 624, 505

\bibitem[\protect\citeauthoryear{{van Dokkum} et~al.,}{{van Dokkum}
  et~al.}{2018}]{dokkum_2018}
{van Dokkum} P.,  et~al., 2018, \mn@doi [\nat] {10.1038/nature25767}, \href
  {https://ui.adsabs.harvard.edu/abs/2018Natur.555..629V} {555, 629}

\bibitem[\protect\citeauthoryear{{van Dokkum}, {Danieli}, {Abraham}, {Conroy}
  \& {Romanowsky}}{{van Dokkum} et~al.}{2019}]{dokkum_2019}
{van Dokkum} P.,  {Danieli} S.,  {Abraham} R.,  {Conroy} C.,   {Romanowsky}
  A.~J.,  2019, \mn@doi [\apjl] {10.3847/2041-8213/ab0d92}, \href
  {https://ui.adsabs.harvard.edu/abs/2019ApJ...874L...5V} {874, L5}

\bibitem[\protect\citeauthoryear{{van den Bosch} \& {Ogiya}}{{van den Bosch} \&
  {Ogiya}}{2018}]{vandenbosch_2018b}
{van den Bosch} F.~C.,  {Ogiya} G.,  2018, \mn@doi [\mnras]
  {10.1093/mnras/sty084}, \href
  {https://ui.adsabs.harvard.edu/abs/2018MNRAS.475.4066V} {475, 4066}

\bibitem[\protect\citeauthoryear{{van den Bosch}, {Tormen}  \& {Giocoli}}{{van
  den Bosch} et~al.}{2005}]{vandenbosch_2005}
{van den Bosch} F.~C.,  {Tormen} G.,   {Giocoli} C.,  2005, \mn@doi [\mnras]
  {10.1111/j.1365-2966.2005.08964.x}, \href
  {https://ui.adsabs.harvard.edu/abs/2005MNRAS.359.1029V} {359, 1029}

\bibitem[\protect\citeauthoryear{{van den Bosch}, {Ogiya}, {Hahn}  \&
  {Burkert}}{{van den Bosch} et~al.}{2018}]{vandenbosch_2018}
{van den Bosch} F.~C.,  {Ogiya} G.,  {Hahn} O.,   {Burkert} A.,  2018, \mn@doi
  [\mnras] {10.1093/mnras/stx2956}, \href
  {https://ui.adsabs.harvard.edu/abs/2018MNRAS.474.3043V} {474, 3043}

\makeatother
\end{thebibliography}

% Alternatively you could enter them by hand, like this:
% This method is tedious and prone to error if you have lots of references
%\begin{thebibliography}{99}
%\bibitem[\protect\citeauthoryear{Author}{2012}]{Author2012}
%Author A.~N., 2013, Journal of Improbable Astronomy, 1, 1
%\bibitem[\protect\citeauthoryear{Others}{2013}]{Others2013}
%Others S., 2012, Journal of Interesting Stuff, 17, 198
%\end{thebibliography}

%%%%%%%%%%%%%%%%%%%%%%%%%%%%%%%%%%%%%%%%%%%%%%%%%%

%%%%%%%%%%%%%%%%% APPENDICES %%%%%%%%%%%%%%%%%%%%%

\appendix

\section{Integration Boundaries} \label{app:boundaries}
\begin{figure*}
    \centering
    \includegraphics[width=1\textwidth]{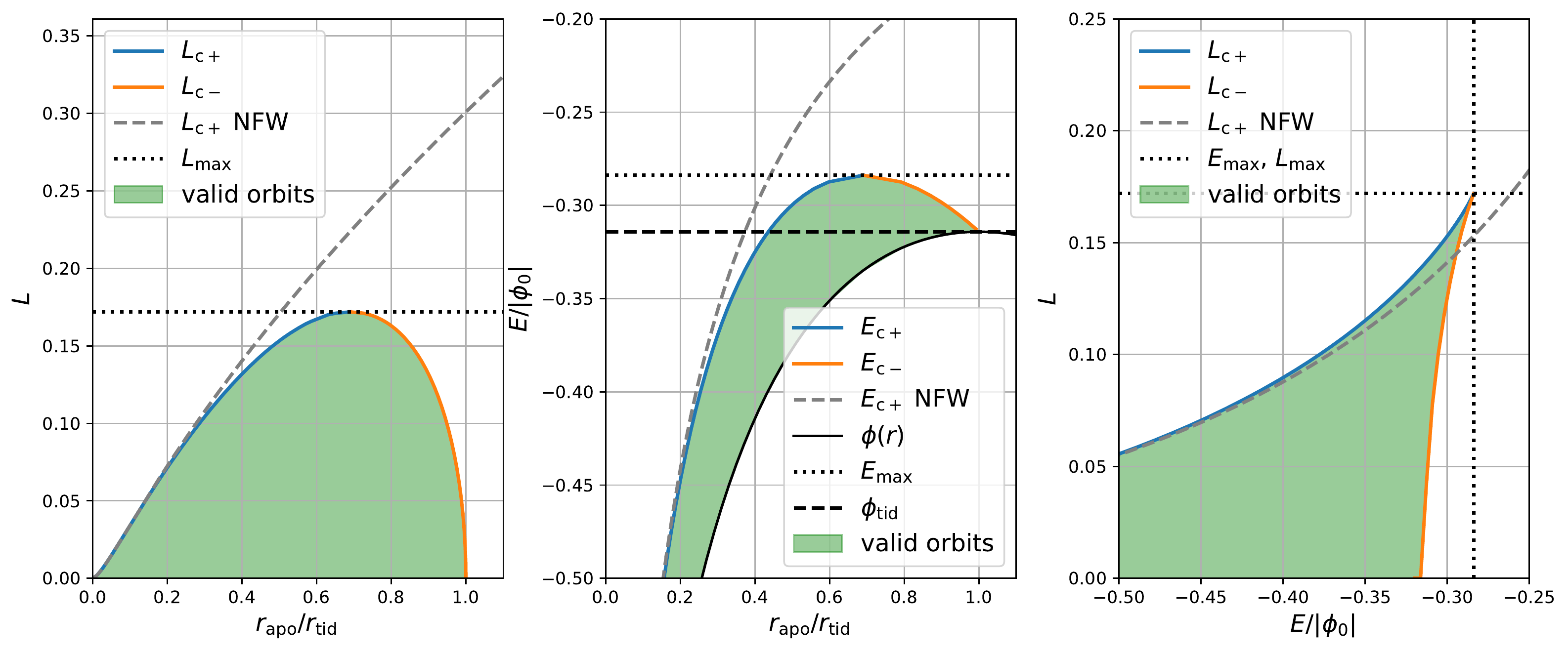}
    \caption{The space of valid orbits for a tidally truncated NFW profile. The circular angular momentum profile (left panel) and the circular energy (central panel) are non montonous which leads to the complicated shape of the space of valid orbits in energy-angular momentum space (right panel).  Note that minimum/maximum means here whether the orbit corresponds to a minimum or maximum of the effective potential. In the first two panels the green contours only indicate the space of orbits that have their apocentre at the given radius.}
    \label{fig:intboundaries}
\end{figure*}
We have briefly discussed in Section \ref{sec:boundorbits} that distinguishing between bound and unbound orbits requires consideration of the effective potential and that the boundaries of the valid orbit space can be more complicated for tidally truncated profiles than for commonly considered monotonous profiles. This is important for choosing the right integration boundaries for equation \eqref{eqn:f_to_dens}. Here, we will explain in more detail how these boundaries can be determined.

An important difference between monotonous profiles and tidally truncated ones is that the circular angular momentum profile
\begin{align}
  L_{\rm{c}}(r) %&= r v_{\rm{c} (r)} \\
                   &= \sqrt{r^3 \partial_r \phi(r)} \label{eqn:lcirc}
\end{align}
is not monotonous. Instead it is increasing up to some radius $r_{\rm{Lmax}}$ to a maximum value $L_{\rm{max}}$ and then decreases to zero at the tidal radius $r_{\rm{tid}}$. We show an example of the circular angular momentum profile for an NFW potential with tidal field $\lambda = \lvir$  in the left panel of Figure~\ref{fig:intboundaries}. That the angular momentum approaches zero at $r_{\rm{tid}}$ means that a particle can stay at the tidal radius if it has zero kinetic energy. 

It is easy to check that a circular orbit corresponds always to an extreme point of the effective potential:
\begin{align}
    \partial_r \phi_{\rm{eff}} (r_{c}, L_{\rm{c}}) &= 0 .
\end{align}
where $r_c$ is the circular orbit radius that was used to infer $L_c$ through \eqref{eqn:lcirc}. However, some of these extreme points can be minima and some of them maxima of the effective potential. It turns out that all extrema for $r_{\rm{c}} < r_{\rm{Lmax}}$ are minima and all extrema for $r_{\rm{c}} > r_{\rm{Lmax}}$ are maxima, and that at $r_{\rm{c}} = r_{\rm{Lmax}}$ we have a saddle-point of the effective potential. In Figure~\ref{fig:intboundaries} we label the minima through an index `c+' and the maxima through an index `c-'. 
The maxima are of special importance, since their associated circular energy
\begin{align}
    E_{\rm{c}} = \phi(r_{\rm{c}}) + \frac{L_{\rm{c}}^2}{2 r_{\rm{c}}^2} \label{eqn:ecirc}
\end{align}
is the highest possible energy that a bound orbit with angular momentum $L = L_{\rm{c-}}$ can have, as can be understood by another look at Figure~\ref{fig:tidal_boundaries}. We visualize the circular energy profile $E_{\rm{c}}$ and the permitted energy space in the central panel of Figure~\ref{fig:intboundaries}. The lowest possible energy at any radius is given by the potential $\phi(r)$ whereas the highest energy is given by $E_{\rm{c}}(r)$. Note, that the location of the highest possible angular momentum and the highest possible circular energy are the same:
\begin{align}
    \partial_r L_{\rm{c}} (r_{\rm{Lmax}}) &= 0 \\
    \Leftrightarrow \partial_r E_{\rm{c}} (r_{\rm{Lmax}}) &= 0
\end{align}

We can infer the the permitted energy-angular momentum space for bound orbits. It lies between $ L_{\rm{min}}(E) \leq  L \leq L_{\rm{max}(E)}$ where
\begin{align}
    L_{\rm{max}}(E) &= L_{c+}(E) \\
    L_{\rm{min}}(E) &= \begin{cases}
      0  &\text{if } E < \phi_{\rm{tid}}\\
      L_{c-}(E) &\text{if } \phi_{\rm{tid}} \leq E < E_{\rm{max}}
    \end{cases}
\end{align}
This is shown in the right panel of Figure~\ref{fig:intboundaries}. Note that orbits with $E > \phi_{\rm{tid}}$ can exist, but they need to have a minimum value of the angular momentum.  Therefore, we expect that there exists a small population of bound particles with $E  > \phi_{\rm{tid}}$, but that there cannot be any bound particles with $E > E_{\rm{max}}$. This explains why energy distributions with a sharp cut at the energy-level $\phi_{\rm{tid}}$ and a small tail to higher energies were observed in \citet{stuecker_2021_bp}.

The considerations in this section were used to define all of the integration boundaries for equation  \eqref{eqn:f_to_dens}.

\section{Numerical Details of the adiabatic-tides Implementation} \label{app:numerics}
There are a large number of numerical issues that had to be handled for the \textsc{adiabatic-tides} implementation. For each necessary step, we will briefly state the numerical method used, the target accuracy and how we verified it. More detailed evaluation scripts and plots can be found inside the \textsc{adiabatic-tides} repository.

\subsection{Initial phase space distribution}
We construct the initial phase space distribution function of the NFW profile as a function of energy $f(E)$ by evaluating it on a grid of energy values through equation \eqref{eqn:nfw_f} and later interpolating it in log-log space through third order interpolation. The limiting cases $E \rightarrow -|\phi_0|$ and $E \rightarrow 0$ of the integral in  equation \eqref{eqn:nfw_f}  are difficult to handle because of singularities at both ends and therefore we use the approximations described in \citet{widrow_2000} for $E < -0.999 |\phi_0|$ and for $E > -0.1 |\phi_0|$. We have benchmarked our implemenation against the one in the \textsc{galpy} library \citep{bovy_2015} and found that they agree to better than $10^{-3}$ relative accuracy in the regime where equation \eqref{eqn:nfw_f} was integrated and to about $5 \cdot 10^{-3}$ relative accuracy in the regime where the approximation from \citet{widrow_2000} is used.

\subsection{Action integration}
To calculate the actions for a given orbit ($r$, $E$, $L$), we have to find its peri- and apocentres and then integrate equation \eqref{eqn:action}. To find the peri- and apocentres, we can check for sign changes in the function
\begin{align}
    h_{E,L}(r) =  E - \frac{L^2}{2 r^2} - \phi(r) .
\end{align}
We do this through a simple binary search. Sample points where the function is definetly negative are given by $r \rightarrow 0$ and by the effective potential maximum $r_{\rm{c-}}(L)$ (where $r_{\rm{c-}} \rightarrow \infty$ for a pure NFW). A location where it is definitely positive is given by the effective potential minimum $r_{\rm{c+}}(L)$. We describe how $r_{\rm{c+}}(L)$ and $r_{\rm{c-}}(L)$ are inferred in Appendix \ref{app:bdetection}.

We then integrate equation \eqref{eqn:action} numerically through the Simpson rule and by concentrating integration points around the two discontinuities at $r_{\rm{p}}$ and $r_{\rm{a}}$. We have tested our action integration scheme against the \textsc{galpy} implementation for an NFW profile and we found that with only $50$ integration steps, we can reach a relative accuracy better than $10^{-3}$. However, our implementation is highly vectorized so that we can evaluate e.g. the actions for $10^5$ different orbits on a single processor in $0.38$ seconds, whereas  \textsc{galpy} needs for the same calculation $811$ seconds so more than a factor $10^3$ longer. Note that the relative accuracy of our implementation can easily be increased to $5 \cdot 10^{-7}$ when using $1000$ integration steps, which still only take about $5.5$ seconds for the same calculation. However, we stick to the $50$ steps, since they are faster and good enough to achieve the accuracy that we require.

\subsection{The distribution function as a function of the actions}
To parameterize the initial phase space distribution of the NFW profile as a function of the actions $f_0(J_r,L)$, we set up a grid of energy and angular momentum vectors and calculate the action for each of them. The grid consists of $1000$ non-uniformly spaced energy values times $100$ angular momentum values, where the angular momentum values range each from 0 to the maximal permitted angular momentum for a given energy $E$ (i.e. the angular momentum of a circular orbit with energy $E$). With the resulting points $(E,J_r,L)$ we then set up a mesh-less linear 2D interpolator (using the function \textsc{LinearNDInterpolator} from the \textsc{scipy.interpolate} library) to interpolate from $(J_r,L)$ space to the corresponding energy $E$. 

By evaluating $E(J_r,L)$ for randomly sampled orbits, we have tested that this interpolation has a worst-case accuracy better than $10^{-4} |\phi_0|$. The distribution function can then be evaluated through $f_0(J_r,L) = f_0(E(J_r,L))$. We have checked that the resulting distribution function $f(J_r,L)$ has a relative accuracy better than $10^{-3}$ when evaluated for randomly sampled orbits.

This interpolation table has only to be set up once, since the initial NFW profile does not change in the iteration steps of Young's method. However, all of the remaining steps have to be repeated for each iteration of Young's method.

\subsection{Poisson solver}
We start with a density profile $\rho_i$ that is given on a grid of radii $r_i$ which is spaced uniformly in logarithmic space. We typically found that sampling about 200 radii on the range $\left[10^{-10}, 10^{3}\right] \rvir$ provides sufficient accuracy in all steps of the procedure. For the first iteration we simply assume $\rho_i = \rho_{\rm{NFW}}(r_i)$. In later iterations we always use the result of the previous iteration for the densities. 

We then solve for the cumulative mass by integrating
\begin{align}
    M(r) = M_0 + \int_{r_{\rm{0}}}^{r} 4 \pi \rho(r) r^2 \rm{d}r
\end{align}
and for the potential by integrating
\begin{align}
    \phi(r) = \phi_0 + 4 \pi \int_{r_{\rm{0}}}^{r} \frac{G M(r)}{r^2} \rm{d}r
\end{align}
both through the Simpson rule on the $r_i$ grid. For $M_0$ and $\phi_0$ we choose the fiducial values of the NFW below the smallest sampled radius $r_0 = 10^{-10} \rvir$. When we need to evaluate any of the functions later at other radii, we interpolate them through third order log-log-space interpolation.

\subsection{Boundary detection} \label{app:bdetection}
To detect the tidal boundary of a given potential profile, we search for the radius $r_{\rm{tid}}$ where $\phi(r)$ (which includes the contributions of the self-potential and the external tidal field) has a maximum through a binary search method. Further, we calculate for a grid of 500 radii that are spaced uniformly in logarithmic space between $r_0$ and $r_{\rm{tid}}$ the angular momenta and energies of circular orbits as in equations \eqref{eqn:lcirc} and \eqref{eqn:ecirc}. We use these to set up interpolators $r_{\rm{c+}}(L)$, $r_{\rm{c-}}(L)$, $E_{\rm{max}}(r) = E_c(r)$, $L_{\rm{min}}(E)$ and $L_{\rm{max}}(E)$ -- again using third order interpolation in log-log-space -- to paremeterise the integration boundaries of the space of possible orbits. 

\subsection{Action interpolator}
In each iteration we need to parameterize the phase space distribution as a function of energy and angular momentum by adiabatically matching it through the the actions to the initial phase space distribution. This requires knowledge of the radial action for each energy and angular momentum pair in the final profile. For this we set up in each iteration a grid in energy-angular momentum space, where energies range from the lowest ($\phi(r \rightarrow 0)$) to highest ($E_{\rm{max}}$) permitted energy and the angular momenta range from the lowest to highest angular momentum as a function of energy ($L_{\rm{min}}(E)$ to $L_{\rm{max}}(E)$) -- as shown in the right panel of Figure~\ref{fig:intboundaries}.  We have found that 250 steps in energy direction times 100 steps in angular momentum direction are sufficient. For each of these points we evaluate the action numerically and we then set up a third order regular grid interpolator (\textsc{RectBivariateSpline} from \textsc{scipy.interpolate}). By testing at random sampling locations we have checked that the $J_r(E,L)$ interpolation has close to $10^{-4}$ relative accuracy and that the distribution function $f(E,L)$ is reconstructed to even better than $10^{-4}$ relative accuracy.

\subsection{Density projection}
The final and most important step of each iteration is the calculation of the revised density estimate through equation \eqref{eqn:f_to_dens}. For each radius we evaluate the integrand from equation \eqref{eqn:f_to_dens} at 200 energy times 100 angular momentum locations. Since all functions are parameterized through interpolators the evaluations are very quick and highly vectorized. The sampling points are spaced non-uniformly to each put additional emphasis on the lower and upper integration boundary. 
We then evaluate the 2D integral by summing up all contributions through the 2D Simpson rule. 

We have benchmarked the whole pipeline by checking whether we can reconstruct the NFW profile when using $\lambda = 0$. Note that each of the numerical steps needs to be accurate for this to work. We found that we can reconstruct the initial NFW density profile better than $5 \cdot 10^{-3}$ relative accuracy on the range $\left[10^{-10}\rvir, 10^3\rvir\right]$. The remaining inaccuracy originates mostly from the inaccuracy in the description of the initial NFW phase space distribution. We expect that the numerical error in tidally truncated profiles should also be of that order. While we cannot strictly prove that this level of accuracy also holds in that case, the tests against N-body simulations in Section \ref{sec:atides_sim} show that the implementation seems at least perfect within the accuracy that N-body simulations can test.

\section{Simulations and convergence to the adiabatic limit} \label{app:adsimnumerics}
In Section \ref{sec:atides_sim} we have presented the final results of N-body simulations that applied a tidal field adiabatically. However, there are some numerical details of these simulations that we have omitted in the main text. These are not important for the interpretation or the evaluation of our results, but they are important to enable other research to reproduce such simulations. We will describe them in more detail here.

\subsection{Adaptive frame of reference}

To run simulations with adiabatically applied tidal fields, we have to add an analytic term as in equation \eqref{eqn:tidpot} to the force calculation in the \textsc{DASH} code, but can otherwise use the same algorithms for evaluating the self-gravity, as already discussed in \cite{Ogiya_2013} and \cite{ogiya_2019}. 

However, there is another complication, since the location $\myvec{x} = 0$ is an unstable equilibrium point of the tidal potential. If a particle was placed at $\myvec{x} = 0$ (with no self-gravity contributions), any small perturbation to its position would grow exponentially in time\footnote{at least if any eigenvalue of the tidal tensor is positive, which is the case for any scenario of interest}. Now, if we place a halo at that location, its centre of mass will follow a similar behaviour, i.e. the halo will start drifting away from $\myvec{x} = 0$ exponentially. It takes some time until the drift is significant, but after that time it doesn't take very long until the centre of mass position takes very large values. Physically this drift does not matter at all, since the internal dynamics of the halo are completely unaffected by it. However, when the components of the position vector become large, the round-off errors of the internal dynamics may become very large, due to the way floating point numbers are represented. 

Therefore, our simulations apply a shift in the coordinate frame in every time-step to ensure that the halo lies always at the origin. In each step we estimate the position and velocity of the halo as
\begin{align}
    \myvec{v}_0 &= \langle v_i \rangle \\
    \myvec{x}_0 &= \langle x_i \rangle
\end{align}
where the average goes over the $1\%$ most bound particles of the halo. Then we subtract $\myvec{x}_0$ and $\myvec{v}_0$ from every particle's position and velocity. At first sight it might seem that we are modifying the potential when we introduce such a spatial shift:
\begin{align}
    \phi(\myvec{x} - \myvec{x}_0) &= \phi(\myvec{x}) +  \myvec{x_0} \Tid \myvec{x}  - \frac{1}{2} \myvec{x}_0 \Tid \myvec{x}_0 ,
\end{align}
which corresponds to a shift in acceleration
\begin{align}
    \myvec{a}(\myvec{x} - \myvec{x}_0) &= \myvec{a}(\myvec{x}) -   \Tid \myvec{x_0} .
\end{align}
However, this modification of the potential corresponds simply to the boost to a uniformly accelerated frame of reference with acceleration $\Tid \myvec{x_0}$. Such `boost' operations were extensively discussed in \citet{stuecker_2021_bp}, and they do not affect the internal dynamics of the system. 

\subsection{Convergence to the adiabatic Limit} \label{app:adsimconvergence}
While we have only presented the final converged simulations of the adiabatic limit in the main text in Section \ref{sec:atides_sim}, we have also run several convergence tests to make sure that the presented simulations are converged to the adiabatic limit $\tau \rightarrow \infty$ and do not exhibit any relaxation effects $N \rightarrow \infty$. Here, we present convergence tests for three of the six cases that were presented in Section \ref{sec:atides_sim}, although we have of course checked convergence for all presented simulations.

\begin{figure}
    \centering
    \includegraphics[width=\columnwidth]{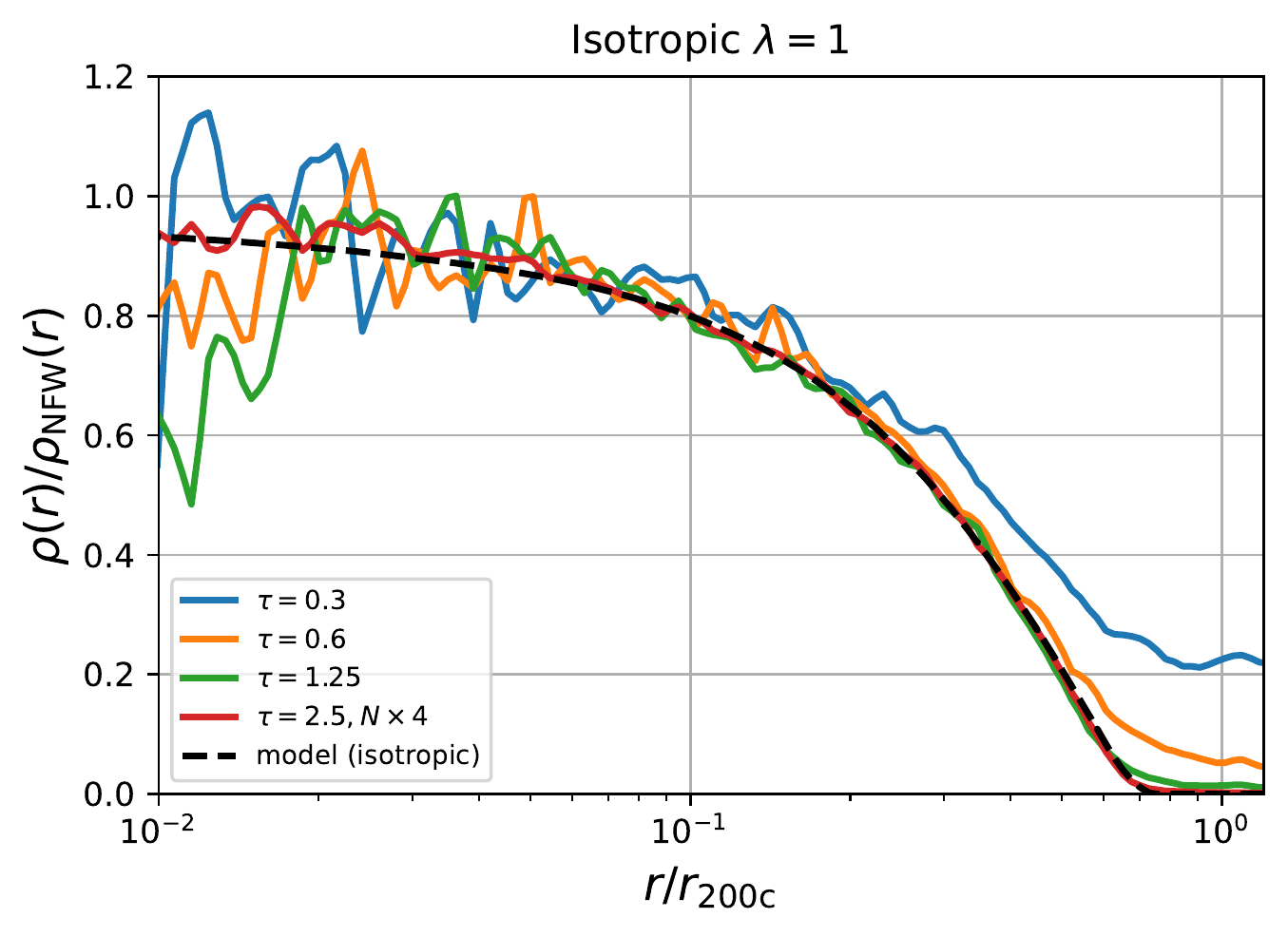}
    \includegraphics[width=\columnwidth]{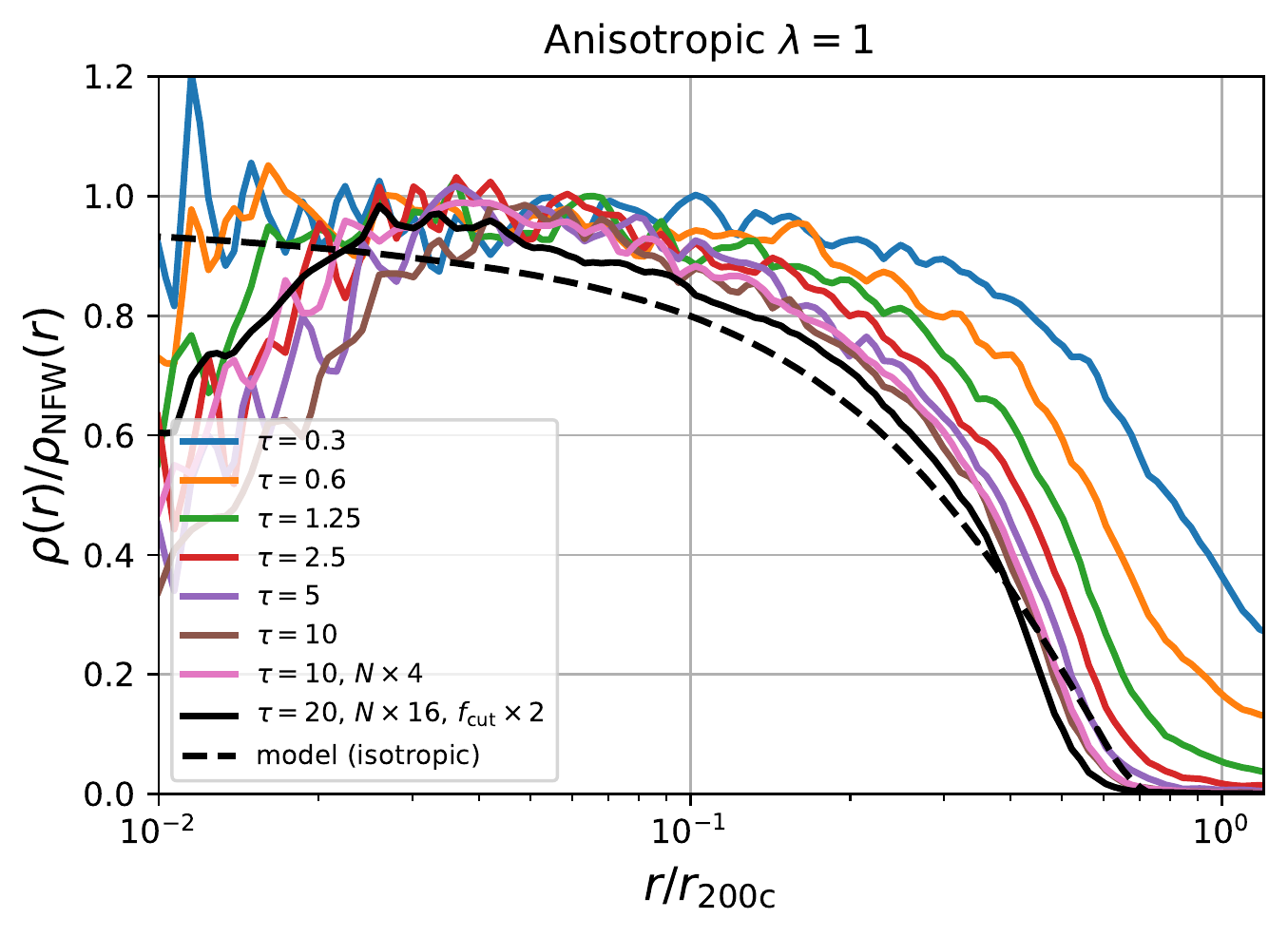}
    \includegraphics[width=\columnwidth]{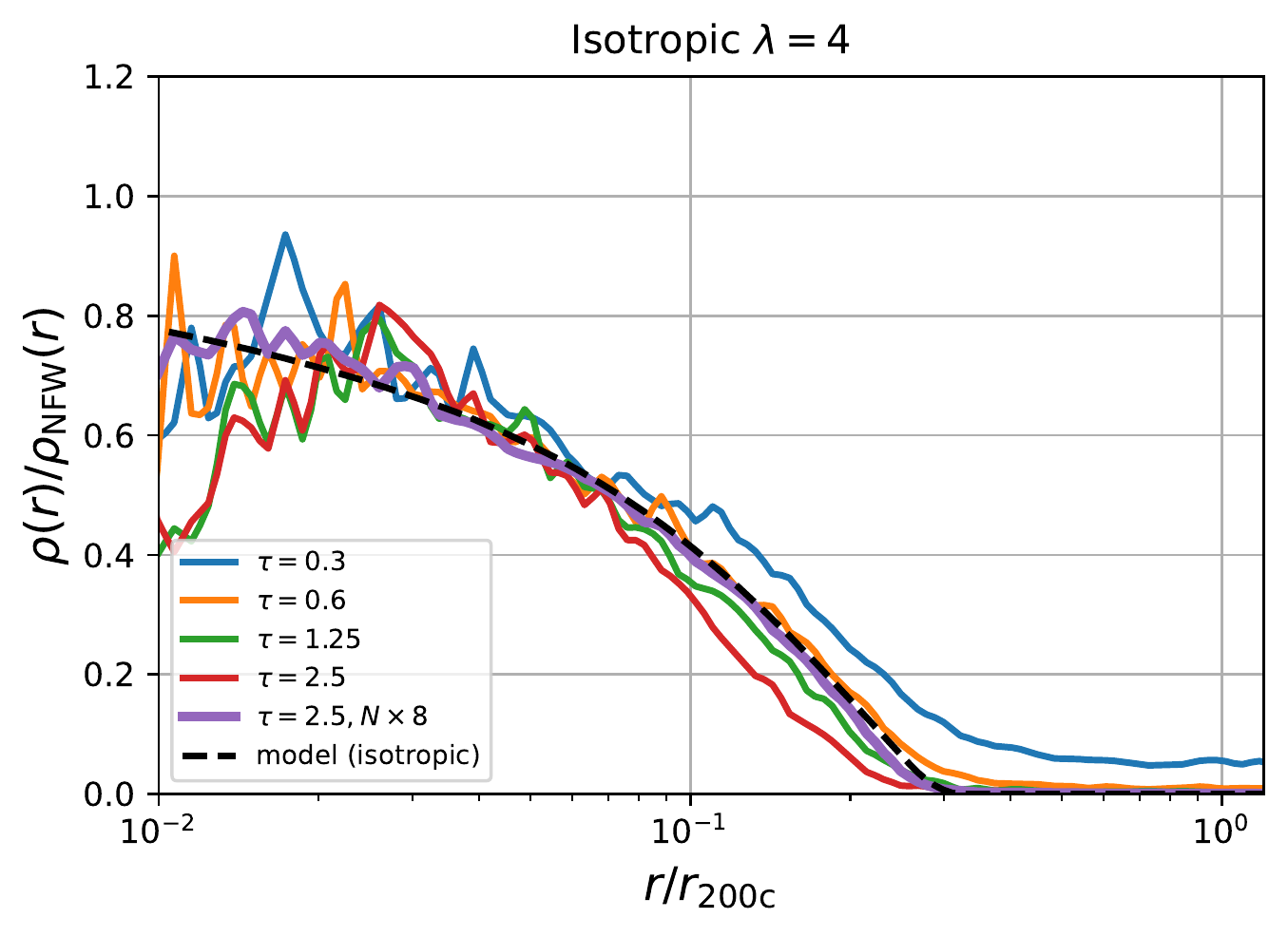}
    \caption{Convergence tests to infer the adiabatic limit. For large timescales $\tau$ the simulations become independent from the speed at which the tidal field was applied -- reaching the adiabatic limit. Reaching the adiabatic limit is quicker for isotropic than for anisotropic cases (top panel versus central panel). Further, the adiabatic limit is reached quicker for stronger tidal fields, since the orbital timescale at the tidal radius is smaller (top versus bottom panel). For increasing timescales it can be important to adjust the particle number to counteract the increasing relaxation effects (e.g. bottom panel). In some runs the central density has decreased due to two-body relaxation. }
    \label{fig:convergence}
\end{figure}

\begin{table}
    \centering
    \begin{tabular}{c|c|c|c|c|c}
         Type & $\lambda / \lvir$ & label & $N$ & $\tau / \tvir$ & $F_{\rm{cut}}$  \\
         \hline
         iso & 1 & $\tau = 0.3$ & $2^{17}$ & 0.3 & 4 \\
         iso & 1 & $\tau = 0.6$ & $2^{17}$ & 0.6 & 4 \\
         iso & 1 & $\tau = 1.25$ & $2^{17}$ & 1.25 & 4 \\
         iso & 1 & $\tau = 2.5$, Nx16 & $2^{21}$ & 2.5 & 4 \\
         \hline
         aniso & 1 & $\tau = 0.3$ & $2^{17}$ & 0.3 & 1 \\
         aniso & 1 & $\tau = 0.6$ & $2^{17}$ & 0.6 & 1 \\
         aniso & 1 & $\tau = 1.25$ & $2^{17}$ & 1.25 & 1 \\
         aniso & 1 & $\tau = 2.5$ & $2^{17}$ & 2.5 & 1 \\
         aniso & 1 & $\tau = 5$ & $2^{17}$ & 5 & 1 \\
         aniso & 1 & $\tau = 10$ & $2^{17}$ & 10 & 1 \\
         aniso & 1 & $\tau = 10$, $N \times 4$ & $2^{19}$ & 10 & 1 \\
         aniso & 1 & $\tau = 20$, $N \times 16$, $f_{\rm{cut}} \times 2$ & $2^{21}$ & 20 & 2 \\
         \hline
         iso & 4 & $\tau = 0.3$ & $2^{17}$ & 0.3 & 4 \\
         iso & 4 & $\tau = 0.6$ & $2^{17}$ & 0.6 & 4 \\
         iso & 4 & $\tau = 1.25$ & $2^{17}$ & 1.25 & 4 \\
         iso & 4 & $\tau = 2.5$ & $2^{17}$ & 2.5 & 4 \\
         iso & 4 & $\tau = 2.5$, $N \times 8$ & $2^{20}$ & 2.5 & 4
    \end{tabular}
    \caption{Simulation parameters that were used for the convergence tests in Figure~\ref{fig:convergence}. The first column indicates whether an isotropic or anisotropic tidal tensor was used, $\lambda$ indicates the strength of the tidal field, $N$ the particle number, $F_{\rm{cut}}$ the initial truncation radius in units of $\rvir$. The horizontal lines separate simulations that were presented in different panels of Figure~\ref{fig:convergence}.}
    \label{tab:convergencetests}
\end{table}

In Figure~\ref{fig:convergence} we present the transfer functions for the simulations with parameters listed in Table \ref{tab:convergencetests}. Each of the panels shows the convergence to the adiabatic limit for a different choice of the tidal field. The top panel shows the isotropic case with $\lambda=1$, the central panel the anisotropic case with $\lambda=1$ and the bottom panel the isotropic case with $\lambda=4$. 

We can see in the top panel that the isotropic $\lambda=1$ simulation converges rapidly to the adiabatic limit. It seems that beyond $\tau = 1.25 \tvir$ the density profile does not change any more. The red line shows the simulation which was presented in the main text which uses for safety an even larger $\tau = 2.5 \tvir$ and a four times larger number of particles. This shows that the adiabatic limit is indeed a well defined unique solution to the tidal stripping problem.

The situation is quite different for the anisotropic tidal field with $\lambda = 1$ which is shown in the central panel. At $\tau=1.25 \tvir$ this case is still far from converged. It reaches the adiabatic limit at around $\tau = 5-10 \tvir$. The solid black line indicates the simulation which is presented in the main text which uses for safety $\tau = 20 \tvir$ and an eight times larger particle number. The anisotropic case needs much longer to converge to the adiabatic limit, because the spatial region which particles can escape through is much smaller in this case than in the isotropic one (compare Figure~\ref{fig:tidalexperiment}). Note that the solid black line is also slightly offset from the (already converged) pink line, since we used a two times larger cut-off radius $f_{\rm{cut}}$ for this simulation. We had run the corresponding convergence tests before we realized that it is necessary to use $f_{\rm{cut}} \gtrsim 2$ to have results that are independent of the truncation radius.

The final example in the bottom panel of Figure~\ref{fig:convergence} is an isotropic case with larger tidal field $\lambda = 4$. We see that this case converges even quicker to the adiabatic limit than the $\lambda=1$ isotropic case and already reaches the limit around $\tau = 0.6 \tvir$. This is so, because the orbital time-scale at its tidal radius is much smaller than in the $\lambda = 1$ case. Additionally we note that the lower resolution cases for $\tau = 1.25 \tvir$ and $\tau = 2.5 \tvir$ seem to exhibit an additional suppression. However, we identified that this must be related to two body relaxation effects (which get more important for longer simulation times), since the simulation with eight times higher number of particles (purple line), which is also presented in the main text, agrees perfectly with our adiabatic model as well as the lower resolution $\tau = 0.6 \tvir$ simulation.

\section{The centrifugal contribution to the effective tidal field} \label{app:centrifugal}

As explained in Section \ref{sec:simcircular}, the centrifugal effect can  enhance the effective radial eigenvalue of the tidal tensor. For circular orbits this effect is given by equation \eqref{eqn:lameff}, as can easily be understood from the Jacobi-potential. However, we want to derive a generalization for the tidal field of non-circular orbits at the pericentre here.

At pericentre the centripetal force (gravity) forces a given subhalo onto a curved trajectory. Its instantaneous curvature radius $R$ can be calculated through
\begin{align}
    a_r &= \frac{v_{\rm{p}}^2}{R} \\
    R &= r_{\rm{p}} \left(\frac{v_{\rm{p}}}{v_{\rm{circ}}} \right)^2
\end{align}
where we have used that the circular velocity at pericentre is $v_{\rm{circ}} = \sqrt{r a_r}$ where $a_r$ is the absolute value of the radial acceleration. Now, if we want to estimate the forces that different particles experience due to the rotation of the reference frame, we have to consider the radial dependence of the centrifugal force at fixed angular frequency $\omega = v_{\rm{p}} / R$:
\begin{align}
    a_{\rm{cent}} &= \omega^2 R^*
\end{align}
where $R^*$ is a particle's distance to the instantaneous centre of rotation which is  $R^* = r + (R - r_{\rm{p}})$ along the radial line passing through the subhalo centre.
\begin{align}
    T_{\rm{cent}, rr} &=\frac{\partial a_{\rm{cent}}}{\partial r} \\
                      &= \omega^2 \\
                      &= \frac{v_{\rm{circ}}^2}{r_{\rm{p}}^2} \left(\frac{v_{\rm{circ}}}{v_{\rm{p}}} \right)^2 \\
                      &= \omega_{\rm{circ}}^2 \left(\frac{v_{\rm{circ}}}{v_{\rm{p}}} \right)^2
\end{align}
 We can see that a subhalo on a circular orbit experiences a centrifugal contribution to the effective pericentre tidal field of amplitude $\omega_{\rm{circ}}^2$, whereas non-circular orbits have the centrifugal term down-weighted by a factor $(v_{\rm{circ}}/v_{\rm{p}})^2$. 
 
In Sections \ref{sec:simcircular} and \ref{sec:orbiting_sim} it gets quite clear that this contribution is quite relevant for circular orbits. However, we want to briefly estimate how strong this effect is for non-circular cases.

\subsection{Relevance of the centrifugal contribution for generic orbits} \label{app:centnoncirc}
\begin{figure}
    \centering
    \includegraphics[width=\columnwidth]{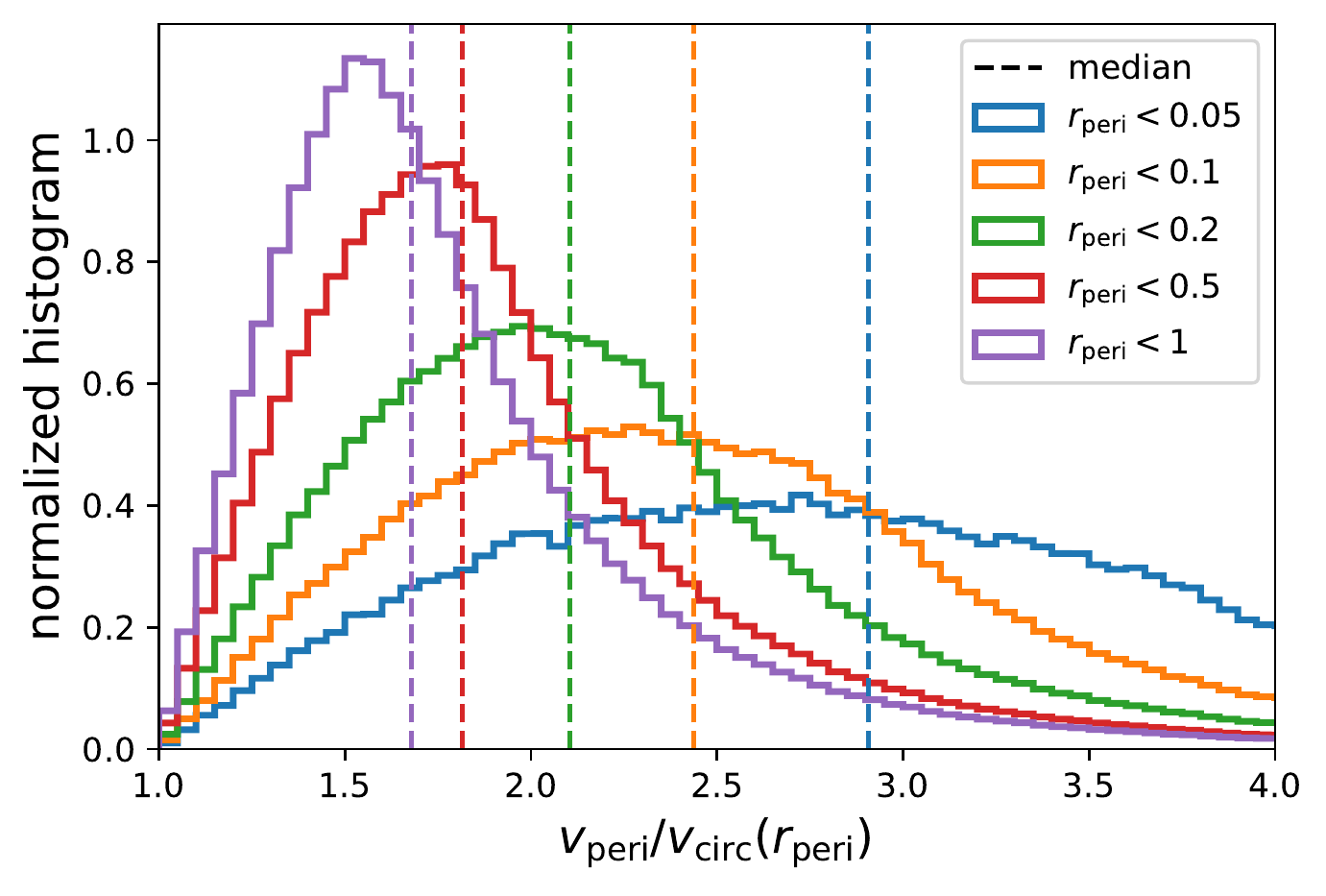}
    \caption{The distribution of the ratio between pericentre velocity and circular velocity at pericentre for the particles of an isotropic NFW. Different lines show particles on different pericentre radii (in units of the virial radius $\rvir$). In the inner regions of an NFW halo typical orbits have velocity ratios of order $2$ or larger.}
    \label{fig:vperi_over_vcirc}
\end{figure}

\begin{figure}
    \centering
    \includegraphics[width=\columnwidth]{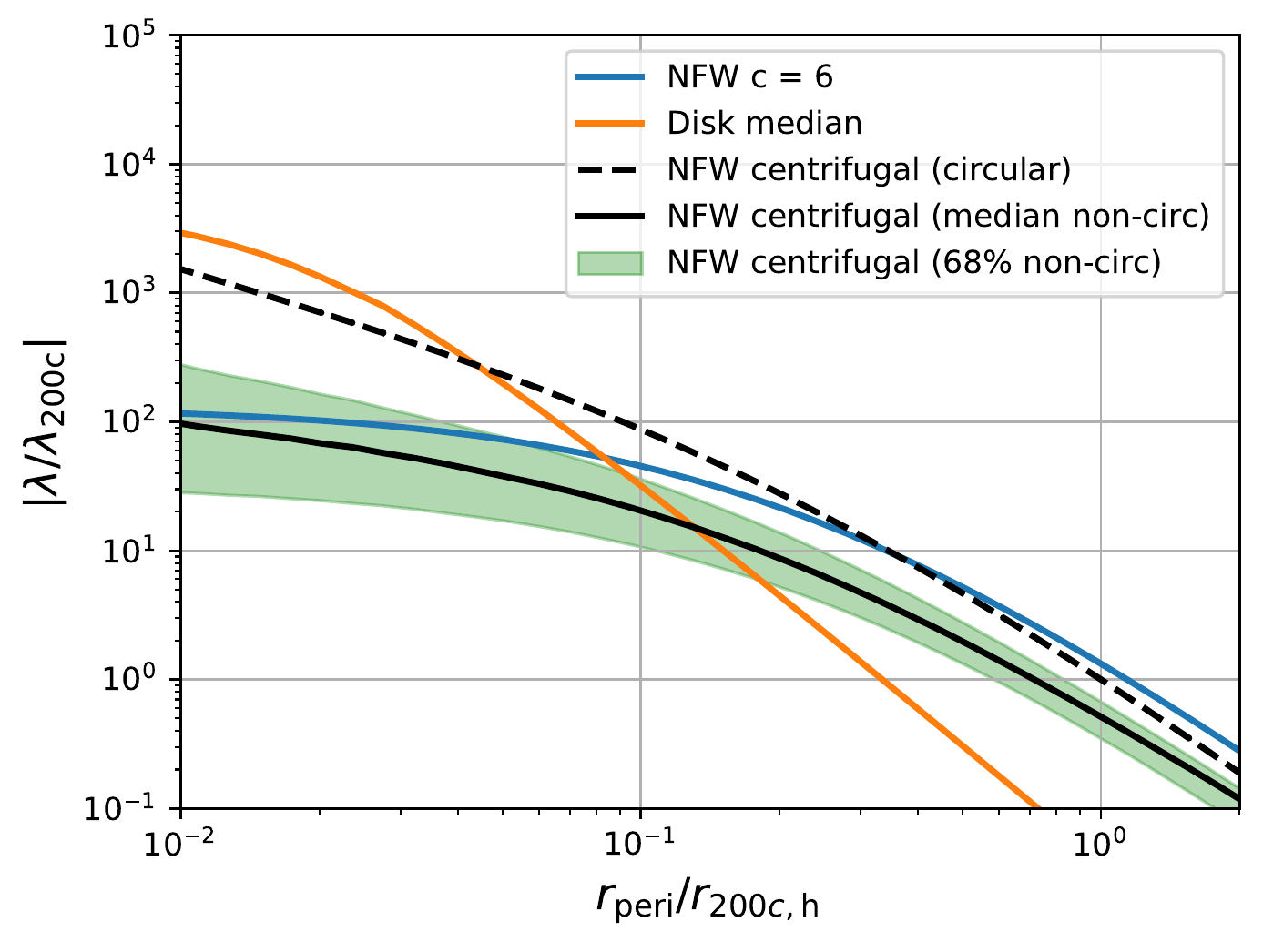}
    \caption{The tidal field of an NFW and a disk in comparison to the amplitude of the centrifugal contribution at pericentre. The black solid line and the green contours show the median and the 68 $\%$ region of the centrifugal field for NFW orbits which have their pericentre at the given radius. For circular orbits one can find situations where the centrifugal term poses a dominant contribution. However, for typical eccentric orbits in realistic scenarios which include the NFW host potential and baryonic components the centrifugal correction is small.}
    \label{fig:centrifugalfield}
\end{figure}

For typical non-circular orbits the centrifugal contribution is much lower than for circular ones. To evaluate this quantitatively we have calculated the velocity ratio at pericentre $v_{\rm{p}} / v_{\rm{circ}}(r_{\rm{p}})$  for a large number of orbits that were sampled from an isotropic NFW halo with concentration $c=6$. While the distribution of actual subhalo orbits might not exactly follow the distribution of dark matter particles, we think that this should at least give a reasonable impression of what typical  values for the velocity ratio should look like. A more sophisticated quantitative investigation could consider the actual distribution of infalling subhaloes \citep{Jiang_2015, Li_2020}. In Figure~\ref{fig:vperi_over_vcirc} we show histograms of this velocity ratio. Different lines show selections on different subpopulations of particles depending on their pericentre radius and the dashed lines indicate the median of each distribution. For example a subhalo that can be found at a radius smaller than $0.1 \rvir$ might approximately follow the orange distribution of velocity ratios.

We see that the median ratio is around $1.7$ when considering the whole halo, but it can even go up to values like $2.9$ when only orbits with pericentres within $0.05 \rvir$ are considered. Since the suppression in the centrifugal contribution to the effective tidal tensor scales as this value squared, we can expect it to be smaller by a factor $4-9$ for typical subhaloes that might be found in the inner regions of dark matter haloes. 

To estimate whether the centrifugal contribution might still be relevant when reduced by such large factors we show in Figure~\ref{fig:centrifugalfield} the different contributions to the tidal field by different components of the host halo. The NFW and the Disk lines in this figure have been calculated in the same way as in Figure~\ref{fig:host_tides}. The dashed line shows the centrifugal contribution for circular orbits. The solid black line shows the median and the green area the $68\%$ region of the distribution of the pericentre centrifugal term for orbits that were sampled from an NFW and have their pericentre at the given radius. 

We note that for the special case of circular orbits in pure NFWs the centrifugal term can be a significant contribution at all radii and can even become the dominant contribution at small radii $r < 0.2 \rhvir$. However, for typical non-circular orbits the typical centrifugal contribution is much smaller and subdominant at all radii. Its value only becomes quantitatively comparable to the pure NFW host potential at small radii.  However, if we additionally consider the tidal field of the disk, the typical centrifugal contribution is well below the gravitational tidal field at all radii. 

We conclude that including the centrifugal contribution can be relevant if one is interested in a theoretical understanding of the exact behaviour of circular orbits. However, typical orbits are not circular and for these the centrifugal effect at pericentre is much smaller. If additionally the effect of baryonic components is considered the centrifugal force will typically only contribute a $5-30\%$ enhancement to the effective tidal field. This is why we focused in the main-text of this paper on the limiting case of a vanishing centrifugal contribution.

\section{Powerlaw Profiles} \label{app:powerlaw}
\begin{table*}
    \centering
    \begin{tabular}{c|c|c|c|c|c||c|c|c|c}
         $\alpha$ & $r_{\rm{tid}}^* / r_0$ & $r_{\rm{max}}^* / r_0$ & $v_{\rm{max}}^* / v_{\rm{circ}}(r_0)$ & $M_{\rm{b}}^* / M_0$ & $L^* / L_0$ & $M(<r_{\rm{max}}) / M$ & $r_{\rm{tid}} / r_{\rm{max}}$ & $C_L$ & $v_{\rm{max}}^2 / r_{\rm{max}}^2 / \lambda$  \\
         \hline 
-0.4 & 4.374e-09 & 1.721e-09 & 5.568e-09 & 8.371e-26 & 1.862e-24 & 63.7\% & 2.54 & 0.815 & 10.5\\
-0.5 & 1.467e-06 & 5.655e-07 & 1.874e-06 & 3.159e-18 & 7.478e-17 & 62.9\% & 2.59 & 0.852 & 11.0\\
-0.75 & 1.367e-03 & 4.694e-04 & 1.782e-03 & 2.557e-09 & 7.314e-08 & 58.3\% & 2.91 & 0.914 & 14.4\\
-1 & 2.475e-02 & 7.784e-03 & 3.257e-02 & 1.517e-05 & 5.196e-04 & 54.4\% & 3.18 & 1.144 & 17.5\\
-1.25 & 1.061e-01 & 2.918e-02 & 1.433e-01 & 1.194e-03 & 5.491e-02 & 50.2\% & 3.64 & 1.850 & 24.1\\
-1.5 & 2.448e-01 & 5.445e-02 & 3.430e-01 & 1.467e-02 & - & 43.7\% & 4.50 & - & 39.7
    \end{tabular}
    \caption{Summary statistics for the adiabatic tidal remnants of powerlaw profiles. The first six columns indicate the slope of the profile $\alpha$, final tidal radius, the final radius where the circular velocity is maximal $r_{\rm{max}}$, the maximal circular velocity $v_{\rm{max}}$, the final mass in units of the initial mass contained within $r_0$ and the final luminosity versus the initial luminosity from material inside $r_0$. The remaining columns indicate quantities measured inside the radius of maximum circular velocity $r_{\rm{max}}$, i.e. the fraction of mass contained within $r_{\rm{max}}$, the tidal radius in units of $r_{\rm{max}}$, the $C_L$ factor as defined in equation \eqref{eqn:nfwluminosity} and the tidal ratio as in  equation \eqref{eqn:tidalratio}. All of these quantities can easily be rescaled to arbitrary combinations of tidal field, characteristic density and scale radius of the powerlaw profile. Luminosities are omitted for $\alpha \leq -1.5$, since they are divergent.}
    \label{tab:powerlaw}
\end{table*}

To understand how important the central slope of haloes is before they get exposed to tidal fields we define a set of powerlaw profiles and use the \textsc{adiabatic-tides} code to estimate their tidal remnant after being exposed adiabatically to a tidal field. We consider a density profile
\begin{align}
    \rho(r) &= \rho_0 \left( \frac{r}{r_0} \right)^\alpha
\end{align}
where $\rho_0$ is the characteristic density of a powerlaw profile, $r_0$ is its scale radius and $\alpha$ is its slope. Note that $\rho_0$ is degenerate with $r_0$ so that the density profile has only two independent parameters. The enclosed mass and the potential of this profile are given by
\begin{align}
    M(r) &= \frac{4 \pi \rho_0 r_0^3}{3 + \alpha} \left( \frac{r}{r_0} \right)^{3+\alpha} \\
    %a(r) &= \frac{4 \pi G \rho_0 r_0}{3 + \alpha} \left( \frac{r}{r_0} \right)^{1+\alpha} \\
    \phi(r) &= \frac{4 \pi G \rho_0 r_0^2}{(3 + \alpha)(2 + \alpha)} \left( \frac{r}{r_0} \right)^{2+\alpha} := \phi_0 \left( \frac{r}{r_0} \right)^{2+\alpha}
\end{align}
Further, we use Eddington inversion to determine the phase space distribution function, which is given for $\alpha > -2$ by
\begin{align}
    f(E) &= f_0 E^\beta \\
    \beta &= - \frac{6+\alpha}{4+2 \alpha} \\
    f_0 &= \frac{\Gamma(-\beta) \rho_0}{\sqrt{8 \pi^3} \Gamma(-\beta-\frac{3}{2}) \phi_0^{\beta+\frac{3}{2}}}
\end{align}
where $\Gamma$ is the Gamma-function. We note that the tidal truncation of a powerlaw profile with a given slope is an effective 0 parameter model where all other dependencies can be rescaled easily. We define the effective tidal field parameter
\begin{align}
    \lambda_0 &= \left| \frac{\partial_r \phi(r_0)}{r_0} \right|
\end{align}
which is so that if the tidal field is $\lambda = \lambda_0$, the (initial) potential saddle-point will lie at $r_0$. For a set of powerlaw profiles with different slopes $\alpha$ we calculate for $\lambda = \lambda_0$ the adiabatic remnant for which the iterative procedure described in Section \ref{sec:youngsmethod} is converged. Then we calculate different summary statistics for each of these cases and list them in Table \ref{tab:powerlaw}. These statistics include the final tidal radius (in units of the initial scale radius $r_0$), the final radius where the circular velocity is maximal $r_{\rm{max}}$ in units of $r_0$, the maximal circular velocity $v_{\rm{max}}$ in units of the initial circular velocity at $r_0$ and the final mass in units of the initial mass contained within $r_0$. Note that all these profiles have the same initial tidal radius at $r_0$ (and the same attractive force and mass contained within that radius), and therefore they can be compared fairly in this manner. 

\begin{figure}
    \centering
    \includegraphics[width=\columnwidth]{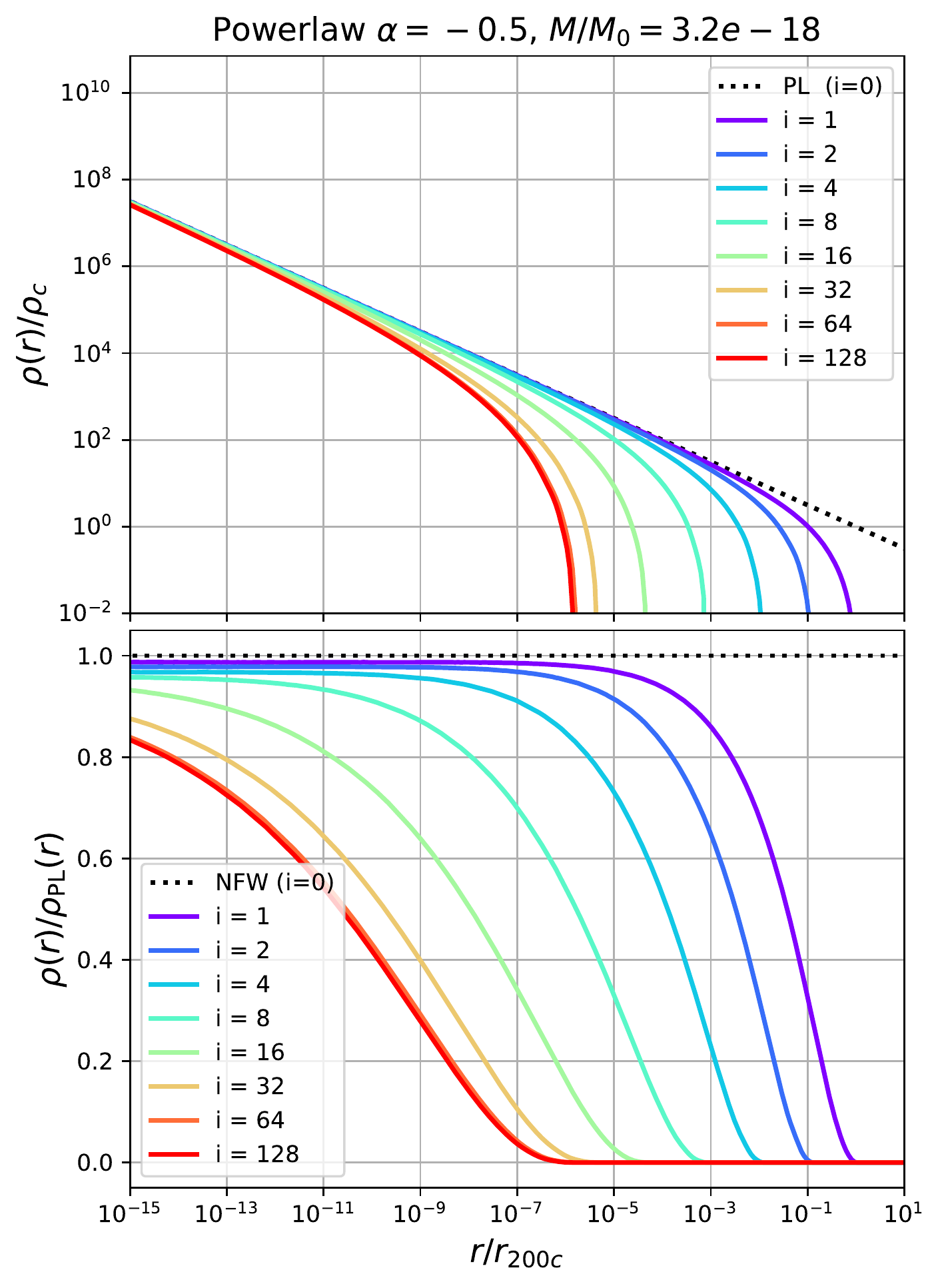}
    \caption{The convergence of Young's method for a powerlaw profile with slope $\alpha=-0.5$. For such a shallow slope gigantic amounts of mass are lost, but the iterative procedure still converges to a finite, stable remnant. }
    \label{fig:powerlawconvergence}
\end{figure}

We note that the amount of mass that remains in the adiabatic limit varies dramatically with the slope $\alpha$. For example, a powerlaw with slope $-1.5$ may retain about one percent of the mass that is contained within the initial tidal radius, while a powerlaw with $\alpha=-0.5$ reduces its mass by 18 orders of magnitude! However, we clearly find stable remnants even in such absurdly disrupted cases. That the adiabatic iterations converge, but do not form a runaway process for such shallow slopes is a non-trivial result. We show in Figure~\ref{fig:powerlawconvergence} an example for $\alpha=-0.5$ to exemplify how an equilibrium is reached even after such an extreme disruption. We speculate that no centrally divergent density profile can disrupt completely through tidal fields. However, we want to stress here that this point is very academic considering that the spatial extend of the remnant is typically many orders of magnitude smaller than the initial object.

All these quantities can be rescaled by considering how the initial tidal radius rescales under changes of the tidal field. We find the following scaling relations:
\begingroup
\allowdisplaybreaks  % These don't need to be all on the same page, by default align was requiring this and it created very weird effects...
\begin{align}
  \frac{r_{\rm{tid}}}{r_0} &= \left( \frac{r_{\rm{tid}}^*}{r_0} \right) \left(\frac{\lambda}{\lambda_0}\right)^{\frac{1}{\alpha}}\\
  \frac{r_{\rm{max}}}{r_0} &= \left( \frac{r_{\rm{max}}^*}{r_0} \right) \left(\frac{\lambda}{\lambda_0}\right)^{\frac{1}{\alpha}}\\
  \frac{v_{\rm{max}}}{v_{\rm{circ,0}}} &= \left(\frac{v_{\rm{max}}^*}{v_{\rm{circ,0}}} \right) \left(\frac{\lambda}{\lambda_0}\right)^{\frac{2 + \alpha}{2 \alpha}}\\
                                       &= \left(\frac{v_{\rm{max}}^*}{v_{\rm{circ,0}}} \right) \left( \frac{r_{\rm{max}}}{r_{\rm{max}}^*} \right)^{\frac{2 + \alpha}{2}}\\
  \frac{M_{\rm{b}}}{M_0} &= \left(\frac{M_{\rm{b}}^*}{M_0} \right) \left(\frac{\lambda}{\lambda_0}\right)^{\frac{3 + \alpha}{\alpha}}\\
  \frac{L}{L_0} &= \left(\frac{L^*}{L_0} \right) \left(\frac{\lambda}{\lambda_0}\right)^{\frac{3 + 2\alpha}{\alpha}} \\
  \left( \frac{v_{\rm{max}}}{r_{\rm{max}}} \right)^2 &= \left(\frac{v_{\rm{max}}^*}{v_{\rm{circ,0}}} \right)^2\left( \frac{r_{\rm{max}}^*}{r_0} \right)^{-2} \left( \frac{\lambda}{\lambda_0} \right)
\end{align}
\endgroup
where the quantities marked by a star correspond to the values in Table \ref{tab:powerlaw} and the ones without star to the values when rescaled to different amplitudes of the tidal field. Note that there are several trivial, but important consequences for the strong mass-loss regime of an NFW halo with central slope $\alpha=-1$: The $v_{\rm{max}}(r_{\rm{max}})$ relation needs to have an asymptotic slope of $-0.5$ in the limit of very strong mass-loss. In the limit of strong mass-loss, doubling the tidal field (or halving the characteristic density) halves the tidal radius, quarters the bound mass and halves $v_{\rm{max}}^2$. Further we note that powerlaw profiles with different slopes respond quite differently to tidal fields. For example a powerlaw profile with $\alpha=-1.5$ would only half its mass when doubling the tidal field, but a profile with $\alpha=-0.5$ would reduce its mass to $1/32$ when doubling the value of the tidal field!

%\begin{align}
%    \rho(\phi) &=  \rho_0 \left(\frac{\phi}{\phi_0}\right)^{\frac{\alpha}{2 + \alpha}} \\
%    \rho'(\phi) &= \frac{\rho_0}{\phi_0} \frac{\alpha}{2 + \alpha}  \left(\frac{\phi}{\phi_0}\right)^{\frac{-2}{2 + \alpha}} \\
%    \rho''(\phi) &= \frac{\rho_0}{\phi_0^2}  \frac{- 2 \alpha}{(2 + \alpha)^2}  \left(\frac{\phi}{\phi_0}\right)^{\frac{-4 - \alpha}{2 + \alpha}} \\
%                 &= N \phi^\beta
%\end{align}

%\begin{align}
%    f_0(E) &= \frac{1}{\sqrt{8} \pi^2} \int_E^{E_{\rm{max}}} \frac{d^2 \nu}{d^2 \Phi} (\Phi - E)^{-1/2} d \Phi \\
%           &= \frac{N}{\sqrt{8} \pi^2} \int_E^{E_{\rm{max}}} \Phi^\beta (\Phi - E)^{-1/2} d \Phi \\
%           &= \frac{N}{\sqrt{8} \pi^2} \left[2 \sqrt{\Phi - E} \Phi^\beta \left(\frac{\phi}{E} \right)^{-\beta} F_{21} \left(\frac{1}{2}, -\beta, \frac{3}{2}, 1 - \frac{\phi}{E}\right) \right]^{E_{\rm{max}}}_E \\
%           &= \frac{2 N }{\sqrt{8} \pi^2} E_m^\beta \sqrt{E_m - E}  \left(\frac{E_m}{E} \right)^{-\beta} F_{21} \left(\frac{1}{2}, -\beta, \frac{3}{2}, 1 - \frac{E_m}{E}\right)
%\end{align}
%\begin{align}
%    f(E) &= 
%\end{align}

\section{A worst-case estimate of tidal shocks} \label{app:tidal_heating}

One of the main criticisms that could be brought forward to the adiabatic limit calculations in this article is that a time-varying tidal field may redistribute energy and therefore additionally unbind particles. This effect is completely neglected in the adiabatic limit approximation. In this appendix,  we try to evaluate how important this effect is. 

Here, we only attempt to estimate the effect tidal shocks would have onto an \textsc{adiabatic-tides} remnant, but we do not attempt to estimate its relevance for reaching such a long term limit. We note that energy redistribution will still be relevant for understanding the progression of mass loss even if it has little impact on the final `equilibrium structure' of a subhalo. The effects of tidal heating are often treated in the impulsive limit where particles are approximated to not move during a shock \citep[e.g.][]{spitzer_1958,gnedin_ostriker_1999}. However, the impulsive limit is always a poor approximation for the particles contained inside \textsc{adiabatic-tides} remnants, since all of these are inside the instantaneous pericentre tidal radius and therefore the tidal force is at most as large as the internal attractive force, so that shocks that have a significant impact on particles' velocities would also be long enough that those particles would have moved a significant fraction of their orbit.

In general, the effects of tidal heating depend on the time-scale the tidal field is applied over in comparison to the internal time-scales of the particles in the halo. The dependency works in two different counter-acting ways: If a tidal field is applied over longer time-scales, then it has more time to accelerate particles and change their energy. However, the longer it takes to increase (or decrease) the tidal field, the more particles will experience the change of the tidal field adiabatically and therefore not redistribute their energy. These two counter-acting effects make it somewhat difficult to make precise quantitative estimates of how the energy-levels of particles change without specializing to specific scenarios and explicitly running full dynamical simulations. 

We want to give a worst-case estimate of energy-redistribution, by constructing a reference case where no particles are adiabatically shielded and particles may be accelerated due to the tidal field for arbitrary long time-scales. Let us imagine that a subhalo would be exposed to a tidal field with the following time-dependence
\begin{align}
\Tid(t)=
\begin{cases}
 0   &\text{for } t < t_1\\
 f \left(\frac{t - t_1}{\tau_{\rm{grow}}} \right) \Tid_0   &\text{for } t_1 < t < t_1 + \tau_{\rm{grow}}\\
 \Tid_0 &\text{for } t_2 < t < t_2 + \tau_{\rm{shock}}\\
 f \left(\frac{t_4 - t}{\tau_{\rm{grow}}} \right) \Tid_0   &\text{for } t_3 < t < t_3+ \tau_{\rm{grow}} \\
 0   &\text{for } t > t_4\\
\end{cases}
\end{align}
where $T_0$ is the peak tidal field, $f(x)$ is an arbitrary function that smoothly goes from 0 at $x=0$ to 1 at $x=1$, $t_2 = t_1 + \tau_{\rm{grow}}$, $t_3 = t_2 + \tau_{\rm{shock}}$ and $t_4 = t_3 + \tau_{\rm{grow}}$. Here we have defined two time-scales $\tau_{\rm{grow}}$ and $\tau_{\rm{shock}}$ which control independently the time that is needed to grow the tidal field and the time it is applied over. Now, the worst case scenario of a tidal shock is the one where $\tau_{\rm{grow}} \rightarrow 0$  and $\tau_{\rm{shock}}$ is large, because then no particle is adiabatically shielded. In this case we can simplify
\begin{align}
\Tid(t)=
\begin{cases}
 0   &\text{for } t < t_1\\
 \Tid_0 &\text{for } t_1 < t < t_1 + \tau_{\rm{shock}}\\
 0   &\text{for } t > t_2\\
\end{cases}
\end{align}
where $t_2 = t_1 + \tau_{\rm{shock}}$. In this case a particle experiences the following change in energy
\begin{align}
    \Delta E_{\rm{tid}} &= \phi_{\rm{tid}}(\myvec{x}_1) - \phi_{\rm{tid}}(\myvec{x}_2) \\
    \phi_{\rm{tid}}(\myvec{x}) &= \frac{1}{2} \myvec{x} \Tid_0 \myvec{x}
\end{align}
where $\myvec{x}_1$ is the particles location relative to the subhalo's centre at $t_1$ and  $\myvec{x}_2$ is its location at $t_2$.\footnote{Here we have neglected the effect of changes in the self-potential of the halo between $t_1$ and $t_2$, which are a separate effect, but should be rather weak for the \textsc{adiabatic-tides} remnant.} In this picture, the particle is lifted to a new energy level at time $t_1$, then orbits with conserved energy at that energy level in the joint potential -- given by the sum of self-potential and $\phi_{\rm{tid}}$ -- and then is 'lowered' back down to a different energy level at $t_2$, since it has moved to a different location $\myvec{x}_2$. If $\tau_{\rm{shock}}$ is very small then $\Delta E_{\rm{tid}}$ is also very small, since $\myvec{x}_1 \approx \myvec{x}_2$. However, we are interested in the limit of an arbitrary long shock $\tau_{\rm{shock}} \rightarrow \infty$, so that we can set an upper limit to $\Delta E$ which is independent of the details of the shock. For $\tau_{\rm{shock}} \rightarrow \infty$ there are two groups of particles that need to be distinguished. The first group consists of particles that can escape the self-gravity of the halo while the shock is ongoing ($t_1 < t < t_2$). The trajectories of those particles may diverge during an arbitrary long shock $\myvec{x}_2 \rightarrow \infty$ as $\tau_{\rm{shock}} \rightarrow \infty$ and therefore $\Delta E$ diverges as $\tau_{\rm{shock}} \rightarrow \infty$. Therefore, for particles which are outside the tidal boundary (or such particles which have enough energy that they could move outside the boundary) the change in energy can become arbitrary large in this limit.

\begin{figure}
    \centering
    \includegraphics[width=\columnwidth]{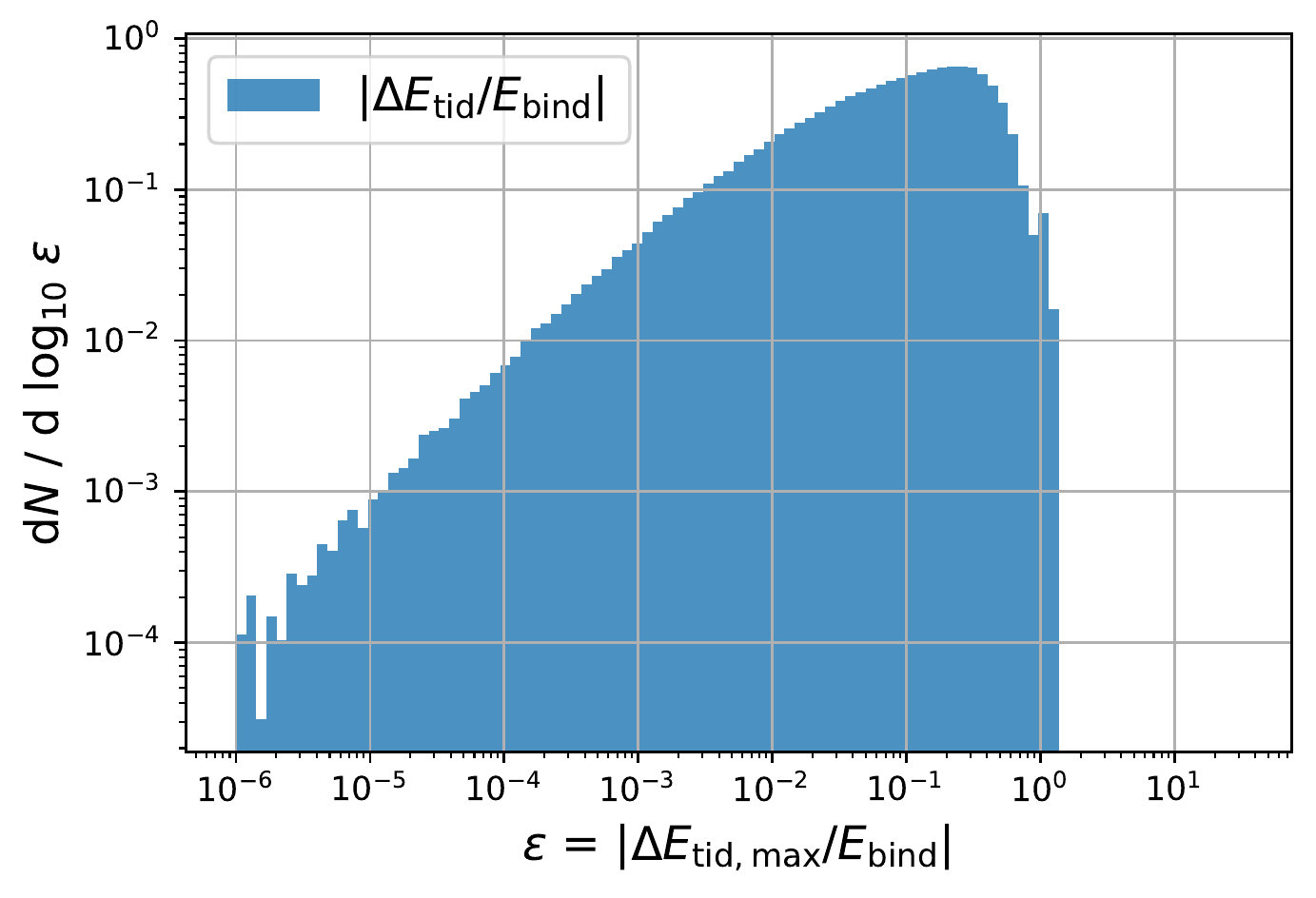}
    \caption{Worst-case estimate of the energy that may be injected to particles of an \textsc{adiabatic-tides} remnant in a single tidal shock in comparison to their binding energy. The remnant of an adiabatically applied tidal field $\lambda$ is also robust to instantaneoous changes in the tidal field of the same amplitude.}
    \label{fig:tidal_heating_worst_case}
\end{figure}

However, by construction, the halo remnants that have been calculated with the \textsc{adiabatic-tides} model have (almost) exclusively particles which are on bound orbits when the tidal field is applied. Therefore, for these particles $\Delta E$ will not diverge for a worst-case scenario of a tidal shock $\tau_{\rm{grow}} \rightarrow 0$ and $\tau_{\rm{shock}} \rightarrow \infty$, as long as the peak tidal field $T_0$ is limited. We can put an upper bound to the energy change that such bound particles may experience as
\begin{align}
    \Delta E_{\rm{tid}} < \left|\lambda_{|\rm{max}|}\right| r_{\rm{a}}^2 = \Delta E_{\rm{tid, max}} \label{eqn:thworstcase}
\end{align}
where $\lambda_{|\rm{max}|}$ is the eigenvalue of $\Tid_0$ which has the largest absolute value and $r_{\rm{a}}$ is the largest radius the particle can possibly reach (in the presence of a tidal field).  

In most cases it will be $\lambda_{|\rm{max}|} = \lambda_1$ i.e. the largest eigenvalue $\lambda_1$ is also the eigenvalue with the largest absolute value. However, for the specific case of the NFW host (presented in the main-text), the second and third eigenvalue of the tidal tensor (which are negative) have a larger absolute value at small radii $r \lesssim 0.1 \rhvir$. This may explain the stronger deviations from the approximate structure-tide degeneracy for non-circular cases at small radii found in Section \ref{sec:structide_sim}, since these two eigenvalues may not be completely neglected for the progression of mass loss. In the limit where the subhalo's pericenter lies arbitrary far inside of the hosts scale radius $r_{\rm{p}} \rightarrow 0$ -- as it was investigated e.g. by \citet{Delos_2019} -- $\Delta E_{\rm{tid}}$ can become arbitrary large (since $\lambda_{2,3} \rightarrow - \infty$) whereas the \textsc{adiabatic-tides} prediction would be limited by a saturating central eigenvalue ($\lambda_1 \rightarrow \text{const.}$). Therefore, energy redistribution would become arbitrary important in the limit $r \rightarrow 0$ and we cannot expect the \textsc{adiabatic-tides} prediction to be good at arbitrarily small radii. However, it is rather unrealistic to assume the NFW potential that far in the center of a Milky Way like galaxy where the actual potential is dominated by baryons. In the case with baryons the largest eigenvalue of the tidal field is usually also the eigenvalue with the largest absolute value. This is is also the case e.g. for the isothermal sphere potential assumed by \citet{errani_2021}. Therefore, for simplicity we assume here that $\lambda_{|\rm{max}|} = \lambda_1$, i.e. that the eigenvalue of the tidal field that limits the energy redistribution is the same as the eigenvalue that we have used for evaluating the \textsc{adiabatic-tides} model.

We remind the reader that \eqref{eqn:thworstcase} is a worst-case estimate and that in realistic scenarios the energy change due to a tidal shock will be much lower. However, it is instructional to compare this worst-case estimate to the change in energy that is required to unbind these particles.

We set up an \textsc{adiabatic-tides} remnant with $\lambda = \lambda_{|\rm{max}|} = 0.1 \lambda_{\rm{s}}$. An estimate of the energy needed to remove the particle from the potential-well is given by the self-energy
\begin{align}
    E_{\rm{bind}} \approx \left|\phi_{\rm{s}} (\myvec{x}) + \frac{1}{2} \myvec{v}^2 \right|
\end{align}
with the self-potential $\phi_{\rm{s}}$ normalized so that $\phi_{\rm{s}} \rightarrow 0$ for $r \rightarrow \infty$. We then show the histogram of $\epsilon = |\Delta E_{\rm{tid, max}} / E_{\rm{bind}}|$ in Figure~\ref{fig:tidal_heating_worst_case}. Here, we have sampled particles from \textsc{adiabatic-tides} remnant and calculated $r_{\rm{a}}$ in the presence of the tidal field. We see that in this worst-case estimate the energy that is injected in a shock reaches the binding energy for a small fraction of particles. For typical particles we have $\epsilon \sim 0.1$ and we note that particles can be found at arbitrary small values of $\epsilon$. Note that one could estimate the number of orbits that are needed until a particle may become unbound due to tidal-shocks as $1 / \epsilon$. However, since all estimates here are worst-case estimates, it is likely that most particles stay much longer.

Therefore, we conclude that most particles in the \textsc{adiabatic-tides} remnants are robustly protected from instantaneous changes in the tidal field for a few shocks even in the worst case scenario. In realistic scenarios the injected energy will be much lower and the remnant probably stable for tens of orbits. Further, we note that there are particles in the remnant with arbitrary small values of $\epsilon$ and therefore we expect that any finite tidal field cannot completely disrupt an NFW halo, even if shocks are applied in such ways that no part of the system is adiabatically shielded and the remnant has gone through any large finite number of orbits.

%%%%%%%%%%%%%%%%%%%%%%%%%%%%%%%%%%%%%%%%%%%%%%%%%%

% Don't change these lines
\bsp	% typesetting comment
\label{lastpage}
\end{document}